\documentclass[footinbib,a4paper,aps,prx,superscriptaddress,reprint,twocolumn,preprintnumbers,amsmath,amssymb,nobalancelastpage,10pt,accepted=2023-06-13]{quantumarticle}
\pdfoutput=1
\usepackage{amsmath}
\usepackage{verbatim}
\usepackage{graphicx}
\usepackage{color}
\usepackage{tikz}
\usepackage{subfigure}
\usepackage{float}
\usepackage{amssymb}
\usepackage{amsmath}
\usepackage{amsfonts}
\usepackage{array}
\usepackage{bm}
\usepackage{xcolor}
\graphicspath{{figures/}}

\usepackage{braket}

\definecolor{mygrey}{gray}{0.35}
\definecolor{myblue}{rgb}{0.2,0.2,0.8}
\definecolor{myzard}{cmyk}{0,0,0.05,0}
\definecolor{mywhite}{rgb}{1,1,1}
\definecolor{myred}{rgb}{1,0.,0.3}

 \def\ee{\mathord{\rm e}}
 
 \def\ii{\mathord{\rm i}}

\def\half{\textstyle\frac{1}{2}}

\def\fourth{\textstyle\frac{1}{4}}

\renewcommand{\ii}{{\rm i}}
\renewcommand{\ee}{{\rm e}}
\def\beq{\begin{equation}}
\def\eeq{\end{equation}}
\def\barray{\begin{eqnarray}}
\def\earray{\end{eqnarray}}
\usepackage{braket}

\definecolor{mygrey}{gray}{0.35}
\definecolor{myblue}{rgb}{0.2,0.2,0.8}
\definecolor{myzard}{cmyk}{0,0,0.05,0}

\definecolor{myred}{rgb}{1,0.,0.3}

 \def\ee{\mathord{\rm e}}
 \def\ii{\mathord{\rm i}}

\def\half{\textstyle\frac{1}{2}}

\def\fourth{\textstyle\frac{1}{4}}
\renewcommand{\ii}{{\rm i}}
\renewcommand{\ee}{{\rm e}}
\def\beq{\begin{equation}}
\def\eeq{\end{equation}}
\def\barray{\begin{eqnarray}}
\def\earray{\end{eqnarray}}

\begin{document}

\preprint{IFT-UAM/CSIC-23-11}
\title{Fermion production at the boundary of an expanding universe: a cold-atom gravitational analogue}

\author{C. Fulgado-Claudio}
\affiliation{Instituto de F\'isica Te\'orica, UAM-CSIC, Universidad Aut\'onoma de Madrid, Cantoblanco, 28049 Madrid, Spain. }

\author{J. M. S\'anchez Vel\'azquez}
\affiliation{Instituto de F\'isica Te\'orica, UAM-CSIC, Universidad Aut\'onoma de Madrid, Cantoblanco, 28049 Madrid, Spain. }

\author{A. Bermudez}
\affiliation{Instituto de F\'isica Te\'orica, UAM-CSIC, Universidad Aut\'onoma de Madrid, Cantoblanco, 28049 Madrid, Spain. }

\begin{abstract}
We study the phenomenon of cosmological particle production of  Dirac fermions in a Friedmann-Robertson-Walker  spacetime,  focusing on a (1+1)-dimensional case in which  the evolution of the scale factor is set  by the equations of  Jackiw-Teitelboim gravity.  As a first step towards a quantum simulation of this phenomenon, we consider two possible lattice regularizations,  which allow us to explore  the interplay of  particle production and  topological phenomena  in spacetimes with a boundary. In particular, for a Wilson-type discretization of the Dirac field,  the asymptotic Minkowski vacua  connected by the intermediate expansion correspond to  symmetry-protected topological groundstates, and have a boundary manifestation in the form of zero-modes exponentially localized to the spatial boundaries. We show that  particle production  can also populate these zero modes, which contrasts with the situation with a na\"{i}ve-fermion discretization, in which conformal zero-mass fields do not allow for particle production. We  present a  scheme for the quantum simulation  of this gravitational analogue by means of ultra-cold atoms in Raman optical lattices, which require real-time control of the Raman-beam detuning according to the scale factor of the simulated spacetime, as well as band-mapping measurements.
\end{abstract}

\maketitle

\setcounter{tocdepth}{2}
\begingroup
\hypersetup{linkcolor=black}
\tableofcontents
\endgroup

  \section{\bf Introduction}
  
  Quantum field theory (QFT) provides a unifying language to describe  quantum many-body systems at widely different scales. For instance,  observed phenomena in  high-energy physics can be accounted for by the standard model of particle physics~\cite{Peskin:1995ev}, a QFT of fermions
coupled to scalar and vector bosons. Here, Poincar\'e invariance determines the arena for such fields: the flat Minkowski spacetime of special relativity. At much smaller energy scales, within the realm of condensed-matter systems, non-relativistic  QFTs are routinely used to explain various collective phenomena~\cite{fradkin_2013}. Interestingly, relativistic QFTs analogous to
those of particle physics  also  arise   in  coarse-grained descriptions of certain phase transitions~\cite{sachdev_2011}, or in materials such as graphene~\cite{RevModPhys.81.109}, Weyl semimetals~\cite{RevModPhys.90.015001}, and topological insulators and superconductors~\cite{RevModPhys.83.1057}. Here, Poincaré invariance  and   an effective speed of light  emerge at long wavelengths~\cite{Anderson393}, such that the  low-energy excitations can be described by quantum fields  in an effective Minkowski spacetime. As first realized in the context of the propagation of quantized sound waves in fluids~\cite{PhysRevLett.46.1351}, there are also situations in which the emergent invariance is related to  general coordinate transformations, i.e. diffeomorphisms~\cite{carroll_2019}. One then obtains emergent QFTs in a curved spacetime, leading to condensed-matter analogues of phenomena studied within the realm of general relativity~\cite{Barcelo2005}.  

Note that, in this so-called {\it analogue gravity}~\cite{doi:10.1098/rsta.2019.0239}, the  emerging spacetime metric $g_{\mu\nu}(x)$   is typically a classical field corresponding to a particular solution of Einstein's field equations~\cite{carroll_2019}. Accordingly, these condensed-matter analogue systems do not aim at mimicking a  full quantum theory of  gravity in the laboratory, but rather at reproducing characteristic phenomena of  QFTs under   classical background gravitational fields~\cite{Parker}, exploring the interplay of gravitation and quantum physics well below the Planck scale. The study of {\it QFTs in curved spacetimes} has lead to important predictions of this interplay,  with paradigmatic examples  being {\it (i)} the evaporation of black holes due to quantum effects~\cite{HAWKING1974}, which elucidates on the thermodynamic nature of such objects; and {\it (ii)}  particle production  during inflation, which is crucial to understand the large-scale behaviour of the  universe~\cite{PhysRevD.23.347}. One of the attractive features of analogue gravity  is that one can  mimic  these phenomena, which are notoriously difficult to observe in a real gravitational context,  in a controlled tabletop experiment. In fact, the range of most applications of QFTs in curved spacetimes is  believed to lie far away from any experimental probe~\cite{carroll_2019}. One such example is the aforementioned evaporation of black holes~\cite{HAWKING1974}, where the emission of thermal Hawking radiation from stellar-size black holes leads to vanishingly-small temperatures in comparison to the cosmic microwave background (i.e. $10n{\rm K}$ versus $T_{\rm CMB}\sim 2.7$K, and its observed inhomogeneities $\delta T_{\rm CMB}\sim 10\mu$K). In the context of analogue systems, on the contrary, Hawking radiation of bosonic fields has already been observed in the propagation of either light in non-linear media~\cite{PhysRevD.83.024015,doi:10.1126/science.1153625,PhysRevLett.105.203901,PhysRevLett.122.010404}, or sound in Bose-Einstein condensates~\cite{PhysRevLett.85.4643,PhysRevLett.105.240401,Steinhauer2016,Kolobov2021}. The related phenomenon of the Unruh effect~\cite{PhysRevD.14.870} has also been observed with Bose-Einstein condensates~\cite{Hu2019}. These experiments are leading to a paradigmatic shift: while, for many decades, gedanken experiments have been crucial to understand the interplay of general relativity and quantum mechanics, it is nowadays possible to turn them into real experiments in analogue-gravity labs.

In this manuscript, we will focus on cosmological  particle production  in an expanding universe~\cite{PhysRevLett.21.562,PhysRev.183.1057}, which can find analogues in the quantized sound waves of Bose-Einstein condensates~\cite{PhysRevA.68.053613,PhysRevA.69.033602} and in trapped-ion crystals~\cite{PhysRevA.98.033407}. In fact, the essence of quantum fields in an expanding spacetime has been recently observed in these two experimental platforms~\cite{PhysRevX.8.021021,PhysRevLett.123.180502}. Further progress along these lines has allowed to implement various specific metrics of expanding curved  spacetimes in the lab~\cite{Viermann2022,tolosasimeon2022curved,S_nchez_Kuntz_2022}, opening the route to very promising future advances.  
Although most of the recent progress has focused on bosonic fields, these atomic experiments  can also be performed with Fermi gases~\cite{RevModPhys.80.885}, and it would be very interesting to observe gravity analogues of the more elusive  {\it fermion production in expanding spacetimes}~\cite{parkerfermions}. In fact, one could go a step further and realise emerging spacetimes with {\it exotic geometries and topologies}  that, despite being  allowed by the theory, do not have any  clear observational pathway in a real gravitational context.  In fact, some of the above analogues have been realised in ring-shaped condensates~\cite{PhysRevX.8.021021}, such that the bosonic fields are defined on spacetimes with a non-trivial spatial topology $\mathbb{R}\times\mathbb{R}\to \mathbb{R}\times S^1$, where $S^1$ is a circle. In spacetimes with $D=(d+1)$ dimensions, one may consider $\mathbb{R}^d\to\mathbb{R}^{d-n}\times S^n$, where there are $n$ compactified spatial dimensions. Such toroidal topologies, originally addressed in the context of  Kaluza-Klein compactification of extra dimensions~\cite{doi:10.1142/S0218271818700017,Klein1926},  can lead to  interesting consequences for the quantum fields, such as topological mass generation and topological symmetry restoration~\cite{KHANNA2014135}. Other boundary conditions  can also play a role in general relativity, such as  in the context of black hole thermodynamics~\cite{Wald2001}. In QFTs, one may consider base spaces $\mathbb{R}^d\to\mathbb{R}^{d-1}\times I$, where $I$ is a finite  interval  of the real line in which one imposes Dirichlet boundary conditions on the fields,  leading to analogues of the Casimir effect in  both its static~\cite{zbMATH03048490} and 
 dynamical~\cite{doi:10.1063/1.1665432} incarnations. We note that  gauge field theories in manifolds with a boundary have also been explored~\cite{cmp/1104178138, ELITZUR1989108,doi:10.1142/S0217751X95000966}, and can lead to an interesting bulk-boundary correspondence.
 
  Coming back to the notion of extra dimensions, rather than compactifying them, one may instead interpret them as the bulk of certain lattice models  displaying non-trivial {\it topology in reciprocal space}~\cite{GOLTERMAN1993219}. This bulk topology has a boundary manifestation in the form of field solutions that are exponentially localized within the boundaries. Remarkably, the    lattice field theories describing these boundary degrees of freedom  can display properties that cannot exist in the absence of a bulk, such as certain quantum anomalies~\cite{CALLAN1985427,Kaplan:2009yg,KAPLAN1992342}. This connects directly to the aforementioned condensed-matter experiments with topological insulators and superconductors~\cite{RevModPhys.83.1057}. We believe it would be  interesting to explore the interplay of these effects with gravitation in analogue systems, where the corresponding QFTs with non-trivial topologies evolve in real time under a background curved spacetime. 

Let us finally note that the essence of most phenomena of QFTs in curved spacetimes is  exemplified by studying free quantum fields evolving in a background curved metric, although  interesting effects can also arise when exploring such {\it real-time dynamics for interacting quantum fields}. In the spirit of quantum simulations~\cite{Feynman_1982}, one could exploit some of these analogue gravity systems, those in which one can prepare various initial states and measure relevant observables after a controllable real-time evolution, as quantum simulators that  address dynamical and non-perturbative problems of QFTs in curved spacetimes, going beyond the capabilities of current numerical simulations based on classical computers.

The goal of our work is to explore a specific system where all of the above points can be addressed. First of all, we will focus on analogue gravity for fermions, exploring the cosmological production of Dirac fermions in an expanding Friedmann-Robertson-Walker spacetime in $D=(1+1)$ dimensions. We will show that specific lattice regularizations of these QFTs lead to the aforementioned non-trivial topologies in reciprocal space, which are manifested by the existence of zero modes localized at the spatial boundaries of the expanding spacetime. We show   how bulk fermions reproduce exactly the continuum QFT prediction for the production of particle-antiparticle pairs as a consequence of the accelerated cosmological expansion. In particular, in the zero-mass limit, one reaches a conformally-invariant situation where the fermion production in the bulk is zero. This contrasts with the situation within the spacetime boundaries, where fermions bound to the spatial edges can be created at finite rates despite having strictly zero energy due to a protecting symmetry. We discuss how, by working in conformal time, this phenomenon could be observed in experiments of ultra-cold fermions in Raman optical lattices. This experimental realization brings a very interesting perspective for future work, as one could explore how this real-time dynamics is affected by non-perturbative phenomena such as dynamical mass generation  in a Gross-Neveu QFT~\cite{PhysRevD.10.3235}.

This article is organised as follows. In Sec.~\ref{sec:dirac_fields_curved}, we revise the theoretical background of particle creation in a (1+1)-dimensional Friedmann-Robertson-Walker universe during a de Sitter expansion phase. This expansion is characterized by an exponentially-growing scale factor. We obtain both analytical and numerical results for the spectrum of created particles, and  its number density. In Sec.~\ref{sec:discrete}, we start by reviewing some topics related to symmetry-protected topological phases, and their  relevance in  lattice  models within condensed matter and high-energy physics. We then introduce two different discretization schemes for the theory of Dirac fields in a FRW background, discussing how they deal with the phenomenon of fermion doubling, and how this can affect  the description of particle production for periodic boundary conditions. We then move to study the effect of imposing open boundary conditions in Wilson's scheme of discretization, and observe the appearance of topological phases connected to the asymptotic vacua of the lattice field theory. These vacua are characterized by a non-zero topological invariant in reciprocal space, which has a boundary correspondence in the form of zero-modes exponentially localized to the spatial edges of the system. We conclude that these topological modes are also produced as a consequence of the expansion of the universe. In Sec.~\ref{sec:implementation}, we propose a detailed experimental scheme for the quantum simulation of this phenomenon in systems of ultra-cold fermionic atoms in Raman optical lattices. Finally, in Sec.~\ref{sec:conclusions}, we present the conclusions and outlook of the presented results.  

\vspace{0.5cm}
\begin{figure}
\includegraphics[scale=0.4]{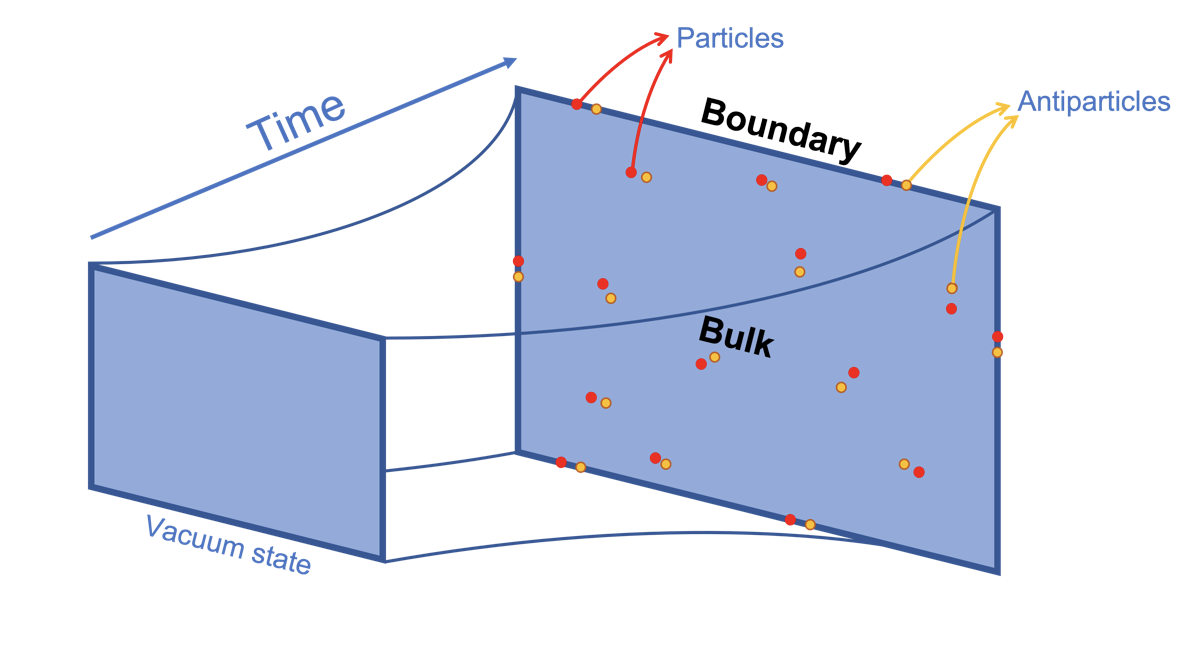}
\caption{{\bf Particle creation for expanding universes with boundary:} Pictorial representation of the phenomenon of particle production in an expanding spacetime. The dynamics of the background metric produces excitations of the field in pairs, which are interpreted as particles (represented in red) and antiparticles (represented in orange). For certain lattice discretizations of the field, the spatial boundaries of the spacetime can actually host  zero-energy modes that are exponentially localized to the boundaries, and thus propagate in a lower effective dimension. We explore the phenomenon of particle production for such zero modes, which is a landmark of the interplay of topological vacua and the  curved spacetime.}
\label{fig:cartoon}
\end{figure}

\section{\bf Fermion production in an expanding universe}
\label{sec:dirac_fields_curved}

\subsection{\label{sec:level1}Dirac fermions in Friedmann-Robertson-Walker spacetimes in  (1+1) dimensions}

We consider Dirac fermions evolving in a spatially-homogeneous and  spatially--isotropic Friedmann-Robertson-Walker (FRW) universe. The details on how to describe a QFT of Dirac fermions in curved spacetimes can be found in Appendix \ref{app:dirac_qftcs}. The FRW metric forms the basis of the standard  model of cosmology,  capturing the large-scale structure of the universe. The line element   is 
\beq
\label{eq:line_element}
    {\rm d}s^2=g_{\mu\nu}(x){\rm d}x^{\mu}{\rm d}x^{\nu}=-{\rm d}t^2+\mathsf{a}^2\!(t){\rm d}\boldsymbol{\Sigma}^2,
    \eeq
     where $\mathsf{a}(t)$ is a dimensionless scale factor, and $d\boldsymbol{\Sigma}^2$  depends on  the metric of the spatial slices, which corresponds to that of   a maximally-symmetric manifold~\cite{carroll_2019}. Accordingly, the spatial metric has spherical symmetry, and  a   uniform curvature that can be either positive, negative, or zero.  In FRW spacetimes, the scale factor $\mathsf{a}(t)$ determines how big the spatial slices are and, thus,  how the universe  expands, being its specific time dependence  determined by the nature of the stress-energy tensor  that  sources  Einstein's field equations. Assuming that this source corresponds to a perfect fluid, which is homogeneous and isotropic, one can derive the so-called Friedmann equations that determine how $\mathsf{a}(t)$ evolves in time.  For instance, when the Einstein field equations are sourced by the vacuum energy (i.e. a positive cosmological constant), one obtains an exponentially-growing  scale factor,  leading to a de Sitter expansion, which will be discussed in more detail below.
    
    In $D=1+1$, the  symmetric manifold would correspond to the line ${\rm d}\boldsymbol{\Sigma}={\rm d}\mathrm{x}$,  such that the  spatial curvature vanishes and the discussion is, in principle, greatly simplified. However, regardless of the nature of the stress-energy tensor,  the Einstein tensor  vanishes identically for such reduced  dimensionality~\cite{Brown:1988am} and, thus,   matter cannot act as a source of curvature following Einstein's theory. Moreover,  including a positive cosmological constant in Einstein's equations, which leads to the aforementioned de Sitter expansion in $D=3+1$ dimensions, now implies that the volume element of the metric vanishes, which is incompatible  with the desired FRW spacetime~\eqref{eq:line_element}. The details on how to overcome these problems in $D=1+1$ can be found in Appendix \ref{app:jt_gravity}.
   
      For the moment, we consider an arbitrary time dependence for the scale factor, and develop the formalism by introducing the conformal time 
    \beq
    \label{eq:conformal_time}
    \eta=\int_0^t\frac{\rm dt'}{\mathsf{a}(t')}.
    \eeq The curved metric can be related to that of a flat Minkowski spacetime by a conformal transformation
\begin{equation}
g_{\mu\nu}(x)=\mathsf{a}^2(\eta)\eta_{\mu\nu},\hspace{1ex} g^{\mu\nu}(x)=\mathsf{a}^{-2}(\eta)\eta^{\mu\nu},
\label{eqmetric}
\end{equation}
which shows that both spacetimes  share the same causal structure.
It is then straightforward  to compute the \textit{zweibein}~\eqref{eq:vielbein}, and the curved gamma matrices~\eqref{eq:curved_gammas}, which lead to 
\beq
e_{\mu}^{{a}}(\eta)=\mathsf{a}(\eta)\delta_{\mu}^{a},\hspace{2ex}  \tilde{\gamma}^\mu\!(\eta)=\mathsf{a}^{-1}\!(\eta)\gamma^\mu.
\eeq
We see that both  depend on the conformal time through the  scale factor, but remain spatially homogeneous. The same occurs for the Christoffel symbols and the spin connection. Using the expressions \eqref{eq:christoffel}-\eqref{eq:spin_connection}, one finds  that the non-vanishing components are $\Gamma^0_{00}(\eta)=\Gamma^0_{11}(\eta)=\Gamma^1_{01}(\eta)=\Gamma^1_{10}(\eta)=\partial_\eta\log(\mathsf{a}(\eta))$, and $\omega^{10}_1(\eta)=-\omega^{01}_1(\eta)=\omega^{10}_0(\eta)=-\omega^{01}_0(\eta)=-\half\partial_\eta\log(\mathsf{a}(\eta))$.

Let us also note  that, in $D=1+1$ dimensions, the flat gamma matrices are proportional to the Pauli matrices, such that one can work with two-components spinors as mentioned previously. In this manuscript, we make the following choice
\beq
\label{eq:flat_gammas}
\gamma^0=-\mathrm{i}\sigma^z,\,\,\,\,\gamma^1=-\sigma^x.
\eeq
In this reduced dimensionality, there is  a single Lorentz transformation $\Lambda$ with a boost of speed $\mathsf{v}={\rm atanh}\xi$, which acts on the spinors via Eq.~\eqref{eq:lorentz_spinors} with  the generators $S_{01}=-S_{10}=\sigma^y/2$, and  depends on the boost rapidity  via $\Omega_{01}=-\Omega_{10}=\xi$.
The covariant derivative~\eqref{eq:cov_derivative} also depends on these generators  via the  connection field~\eqref{eq:connection_field} which, in this case, only depends on conformal time
\beq
\label{eq:spin_connection_2d}
\omega_0\!(\eta)=0,\hspace{2ex}\omega_1\!(\eta)=-\half\partial_\eta \text{log}\left(\mathsf{a}(\eta)\right)\sigma^y.
\eeq
We then substitute all these  expressions in the generic action~\eqref{eq:Dirac_action_curved}, and find that the resulting dynamics amounts to that of Dirac fermions with a time-dependent mass in a static Minkowski spacetime with coordinates $x=(\eta,{\rm x})$, and is dictated by the following action
\begin{equation}
\label{eq:Dirac_action_curved_rescaled}
    S=\!\!\int\!\!{\rm d}^2x\,\overline{\chi}(x)\!\left(- \eta^{\mu\nu} {\gamma}_\mu\!{\partial}_{\!\nu}+m\mathsf{a}(\eta)\right)\!\!\chi(x).
\end{equation}
Here, we have rescaled the spinor field as $\psi(x)\mapsto\chi(x)=\sqrt{\mathsf{a}(\eta)}\psi(x)$, such that the principle of stationary actions yields 
the following Dirac equation
\begin{equation}
     \left(\gamma^\mu\partial_\mu-m\mathsf{a}(\eta)\right)\chi(x)=0.
\label{eqn:DiracEquation}
\end{equation}

 Therefore, all the effects of the expanding spacetime are encoded in a dynamical multiplicative renormalization of the bare mass, which will  depend on the specific time dependence of the scale factor that is itself determined by the underlying classical field equations of the gravity model. 
 Let us note that, regardless of the particular scale factor, one can  see that, for conformal QFTs in which the bare fermion mass vanishes, $m=0$, there is no effect of the expanding spacetime apart from a trivial rescaling, and thus     no particle production. 

Before moving on, let us emphasize  again that the fermionic fields are not the source of the specific  expansion of the scale factor, which should have some other  origin as will be discussed in Sec.~\ref{sec:level3}. Before moving there, we discuss the physics of particle production for a generic scale factor in the following subsection, recalling that  we do not consider any back-action from the field onto the metric, and thus we work in the QFT in curved spacetime formalism. In any case, this effect should  be negligible in light of the Fermi-Dirac statistics of the fields. 

\subsection{\label{particlecreation} Bogoliubov transformations and particle creation in asymptotic Minkowski vacua}

Let us momentarily go back to QFT in flat spacetimes. Upon quantization, the vacuum of a Dirac field in a static Minkowski spacetime can be uniquely defined. In fact, the vacuum remains the same at any   instant of time, such that  the notion of particle and antiparticle excitations is unambiguous~\cite{Peskin:1995ev}. Introducing additional  interactions in the Dirac action~\eqref{eq:Dirac_action} brings in interesting dynamical effects, since  these excitations can  scatter,  and particle-antiparticle pairs can be created from the vacuum. This situation changes for curved spacetimes since, as advanced in the introduction, there are specific situations where the curvature/dynamics of the universe can lead to particle  creation~\cite{HAWKING1974,PhysRevD.23.347} even in the absence of interactions with other quantum fields. For the problem at hand, where the fermions are coupled to an  expanding  FRW universe, this occurs when the field  absorbs the required energy to create excitations from the gravitational background~\cite{PhysRev.183.1057,parkerfermions,Ford_2021}. This is particularly transparent in the  conformal-time description of the action, Eq.~\eqref{eq:Dirac_action_curved_rescaled}, where the effect of the expanding universe is encoded in a time-dependent mass, suggesting  that  the energy of the Dirac fields shall not be conserved and may result  in particle creation.

However, a precise interpretation  of particle production in curved spacetimes is complicated by the fact that, in contrast to the Minkowski spacetime, the vacuum   is coordinate-dependent and, thus, not uniquely defined~\cite{Ford_2021,Mukhanov}. Therefore,  the typical  notion of particles and antiparticles that arises in the canonical quantization of field theories in flat spacetimes~\cite{Peskin:1995ev} must be re-addressed with some care. Let us discuss the canonical quantization of the Dirac theory~\eqref{eq:Dirac_action_curved_rescaled} at a fixed instant of time $x=(\eta_\star,\mathrm{x})$, assuming thus  a specific inertial frame~\cite{Peskin:1995ev}. One    imposes canonical anti-commutation relations for  $\chi(x)$ and its conjugate momentum $\Pi_\chi(x)=\bar{\chi}(x)\gamma_0=\ii\chi^{\dagger}\!(x)$, which are then upgraded  to  field operators and  denoted using  a hat  $\chi(x),\Pi_\chi(x)\mapsto\hat{\chi}(x),\hat{\Pi}_\chi(x)$ fulfilling equal-time canonical anti-commutation relations. Since the Dirac equation~\eqref{eqn:DiracEquation} is linear, we can expand the field operators in the complete basis of its positive- and negative-frequency solutions, so-called  mode functions. Accounting for the previous rescaling with the scale factor, we find that the Dirac field at a fixed instant of time can be expanded as 
\begin{equation}
    \hat{\psi}(x)=\!\!\left.{\int}\!\frac{{\rm d}\mathrm{k}}{2\pi}\frac{1}{ \sqrt{2\omega_{\rm k}\mathsf{a}(\eta)}}\left(\hat{a}_\mathrm{k} \tilde{u}_{\rm k}(x)+\hat{b}^\dag_\mathrm{k} \tilde{v}_{\rm k}(x)\right)\right..
    \label{eqn:decompositionmodefunctions}
\end{equation}
Here,    the mode functions are  obtained by the product of   spinor solutions with  Fourier components  $\ee^{\pm\ii {\rm k}\,{\rm  x}}$
\beq
\label{eq:mode_solutions}
\tilde{u}_{\rm k}(x)=\tilde{u}_{\rm k}(\eta_\star)\ee^{\ii {\rm k} \,{\rm x}},\hspace{1ex}  \tilde{v}_{\rm k}(x)=\tilde{v}_{\rm k}(\eta_\star)\ee^{-\ii {\rm k} \,{\rm x}},
\eeq 
 In the expansion~\eqref{eqn:decompositionmodefunctions}, we have chosen a normalization that depends on $\omega_{\rm k}=(\mathrm{k}^2+m^2\mathsf{a}^2(\eta_\star))^{1/2}$, and is consistent with the anti-commutation algebra  of the creation-annihilation operators
\beq
\label{eq:ferm_algebra}
\left\{\hat{a}_\mathrm{k},\hat{a}^{\dagger}_{\mathrm{k}'}\right\}=\left\{\hat{b}_{\mathrm{k}},\hat{b}^{\dagger}_{\mathrm{k}'}\right\}=2\pi \delta(\mathrm{k}-\mathrm{k}'),
\eeq
 while the remaining anti-commutators  vanish. The spinor solutions are  normalized according to a standard choice~\cite{Peskin:1995ev}, namely $\tilde{u}_{\rm k}^{\dagger}\!(\eta_\star) \cdot \tilde{u}_{\rm k}^{\phantom{\dagger}}(\eta_\star)=\tilde{v}_{\rm k}^{\dagger}(\eta_\star)\cdot \tilde{v}_{\rm k}^{\phantom{^{\dagger}}}(\eta_\star)=2\omega_{\rm k}$, where the dot here represents the  matrix-vector multiplication. In addition, they fulfill $\tilde{v}_{\rm k}^{\dagger}\!(\eta_\star) \cdot \tilde{u}_{-\rm k}(\eta_\star)=\tilde{u}_{\rm k}^{\dagger}\!(\eta_\star) \cdot \tilde{v}_{-\rm k}(\eta_\star)=0$ in order to recover the equal-time anti-commutation relations of the conjugate field operators $\{\hat{\chi}(\eta_\star,{\rm x}),\hat{\Pi}_\chi(\eta_\star,{\rm y})\}=\ii\delta({\rm x-y})$. 

In the case of a   flat and static spacetime,  the scale factor is trivial $\mathsf{a}(\eta)=1$, and $\tilde{u}_{\rm k}(\eta_\star)=\tilde{u}_{\rm k}\ee^{-\ii \omega_{\rm k}\eta_{\star}},\tilde{v}_{\rm k}(\eta_\star)=\tilde{v}_{\rm k}\ee^{+\ii \omega_{\rm k}\eta_{\star}}$, such that the mode functions~\eqref{eq:mode_solutions} can be expressed in terms of plane waves $\ee^{\pm\ii kx}$, where $kx=\eta^{\mu\nu}k_\mu x_\nu$ and  the 2-momentum     is defined on mass shell $k=(\omega_{\rm k},{\rm k})$. One can then find the single-particle states $\ket{k}$ as unitary irreducible representations of the Poincar\'e group  $x\mapsto x'=\Lambda x+d$. Accordingly, these states  transform as $\ket{ k}\mapsto U(\Lambda,d)\ket{ k}=\ee^{-\ii kd}\ket{\Lambda k}$, where the operators yield a representation of the group fulfilling $U^{\dagger}(\Lambda,d)U(\Lambda,d)=\mathbb{I}$ and $U(\Lambda_1,d_1)U(\Lambda_2,d_2)=U(\Lambda_1\Lambda_2,\Lambda_1d_2+d_1)$, and  are generated by acting with rescaled creation operators on the QFT vacuum  $\ket{ k}_a=\sqrt{2\omega_{\rm k}}\hat{a}^{\dagger}_{\rm k}\ket{0}$, $\ket{ k}_b=\sqrt{2\omega_{\rm k}}\hat{b}^{\dagger}_{\rm k}\ket{0}$. We recall that the vacuum  fulfills  $\hat{a}_{\mathrm{k}}\ket{0}=\hat{b}_{\mathrm{k}}\ket{0}=0$, and is the only state   left invariant under  transformations  within the Poincar\'e group~\cite{preskill_notes_QFT_curved}. Using this  invariance, one can   change the  inertial frame used to define the canonical momenta at any other instant of time $\eta=\eta_\star$, and extend the notion of the vacuum to any other instant of time,  arriving in this way to an unambiguous notion of particles and antiparticles. In particular, the modes do not mix under   Poincar\'e  transformations and, moreover, 
 the individual
 number operators can be shown to be Poincar\'e invariant
\begin{equation}
    \hat{N}_{a}=\int\!\frac{{\rm d}\mathrm{k}}{2\pi}\hat{a}^{\dagger}_{\mathrm{k}}\hat{a}_{\mathrm{k}},\hspace{1.5ex}     \hat{N}_{b}=\int\!\frac{{\rm d}\mathrm{k}}{2\pi}\hat{b}^{\dagger}_{\mathrm{k}}\hat{b}_{\mathrm{k}},
\end{equation}
 such that  any inertial observer would agree on the specific  particle/antiantiparticleparticle content of the state. In fact, Eq.~\eqref{eqn:decompositionmodefunctions} is valid for any  instant of time letting $x=(\eta_\star,\mathrm{x})\mapsto(\eta,\mathrm{x})$, which shows that  the evolution of the  operators is trivial $\hat{a}_{\rm k}(\eta)=\hat{a}_{\rm k}\ee^{-\ii\omega_{\rm k}\eta}, \hat{b}_{\rm k}(\eta)=\hat{b}_{\rm k}\ee^{-\ii\omega_{\rm k}\eta}$, and there is no particle/antiparticle creation from the initial vacuum, nor scattering between different particles, unless additional interactions are incorporated in the QFT. 

A similar philosophy can be followed for curved spacetimes, although, as advanced previously,  more profound conceptual challenges arise.    In the case of an expanding spacetime $\mathsf{a}(\eta)\neq1$, Poincar\'e invariance is superseded by diffeomorphism invariance, the vacuum becomes coordinate dependent, and particles can no longer be associated with unitary irreducible representations of the Poincar\'e group~\cite{preskill_notes_QFT_curved}. Since there is, in principle, no preferable coordinate system, the notion of  vacuum  and particle becomes ambiguous for curved spacetimes. As discussed in~\cite{Birrell, Mukhanov}, a reasonable approximation  to discuss particle production is that of  adiabatic vacua, which connect to the so-called Bunch-Davies vacuum in de Sitter spacetimes~\cite{Bunch:1978yq} after  an extended period of inflation~\cite{Ford_2021}. In the case of Dirac fields, recursive methods to construct such adiabatic vacua  have been recently discussed in~\cite{PhysRevD.98.025016}. Ultimately, one may adopt  an operational philosophy, where the notion of   particles/antiparticles is related to specific local detectors that click by absorbing  energy from the field, the so-called Unruh-DeWitt detectors~\cite{Birrell}. In this article, however, we will be interested in situations amenable to analogue-gravity experiments, in which the simple and unambiguous notions of the   vacuum and particles  in flat spacetimes are still useful.
Paralleling  the in-out formalism of scattering in interacting QFTs  mentioned above, we  consider scale factors that tend adiabatically to constant values in the remote past and distant future. Therefore, the metric~\eqref{eqmetric} tends  to asymptotically-flat Minkowski spacetimes, where the vacuum and  particle/antiparticle states have a well-defined meaning. Accordingly, we can use the field decomposition in Eq.~\eqref{eq:mode_solutions} for those distant  times $\eta\approx\eta_{0}$ and $\eta\approx\eta_{\rm f}$, using   their corresponding mode functions and creation-annihilation operators. However, although customary in flat spacetime, there are some problems with the mode expansion as defined in Eq.~\eqref{eqn:decompositionmodefunctions} when $\mathsf{a}(\eta)\neq1$. In particular, the above normalization of the modes $\tilde{u}_{\rm k}(\eta)$ and $\tilde{v}_{\rm k}(\eta)$ would imply that their norm  is time-dependent, and thus their evolution  cannot be unitary. We thus define new modes normalized to 1 as $u_{\rm k}(\eta)=\tilde{u}_{\rm k}(\eta)/{\sqrt{2\omega_{\rm k}(\eta)}}$, $v_{\rm k}(\eta)=\tilde{v}_{\rm k}(\eta)/{\sqrt{2\omega_{\rm k}(\eta)}}$, so that the normalization now reads $u^{\dagger}_{\rm k}(\eta)\cdot u_{\rm k}(\eta)=1$, $v_{\rm k}^{\dagger}(\eta)\cdot v_{\rm k}(\eta)=1$, while the orthonormal conditions $u_{\rm k}^{\dagger}(\eta)\cdot v_{\rm -k}(\eta)=v_{\rm k}^{\dagger}(\eta)\cdot u_{\rm -k}(\eta)=0$ remain the same. Using these normalized modes, the expansion of the Dirac field~\eqref{eqn:decompositionmodefunctions} becomes
\begin{equation}
    \hat{\psi}(x)=\!\!\left.{\int}\!\frac{{\rm d}\mathrm{k}}{2\pi}\frac{1}{\sqrt{\mathsf{a}(\eta)}}\left(\hat{a}_\mathrm{k} u_{\rm k}(x)+\hat{b}^\dag_\mathrm{k} v_{\rm k}(x)\right)\right.,
  \label{eq:normalizedmodeexpansion}
\end{equation}
where the mode functions are defined in analogy to  Eq.~\eqref{eq:mode_solutions}, and evolve in conformal time following the Dirac equation~\eqref{eqn:DiracEquation} in momentum space
 \begin{align}
 &(\gamma^0\partial_\eta +\ii{\rm k}\gamma^1-m\mathsf{a}(\eta))u_{\rm k}(\eta)=0,
  \label{umodeevolution}\\
  &(\gamma^0\partial_\eta -\ii{\rm k}\gamma^1-m\mathsf{a}(\eta))v_{\rm k}(\eta)=0.
   \label{vmodeevolution}
  \end{align}
This convention is discussed in more detail in Appendix~\ref{app_c}. 

Let us then define the asymptotic vacua $\ket{0_{0}}$ and $\ket{0_{\rm f}}$,  which are annihilated by the corresponding operators $\hat{a}_{\mathrm{k}}(\eta_{0}),\,\hat{a}_{\mathrm{k}}(\eta_{\rm f})$, and $\hat{b}_{\mathrm{k}}(\eta_{0}),\,\hat{b}_{\mathrm{k}}(\eta_{\rm f})$. Note that, due to the intermediate expansion, the initial vacuum  $\ket{0_{0}}$ may not evolve into the instantaneous groundstate of the QFT at any later  time, nor to the asymptotic distant-future  vacuum $\ket{0_{\rm f}}$ after the complete expansion.
In contrast to the flat spacetime, in which the time-evolution of the creation-annihilation operators was a trivial complex phase, the corresponding evolution in the expanding spacetime leads to the following canonical, so-called Bogoliubov,  transformation 
\begin{align}
&\hat{a}_{\mathrm{k}}^{\phantom{\dagger}}(\eta_{\rm f})=\alpha^{\phantom{\dagger}}_{\mathrm{k}}(\eta_{\rm f})\hat{a}^{\phantom{\dagger}}_{\mathrm{k}}(\eta_{0})-\beta_{\mathrm{k}}(\eta_{\rm f})\hat{b}_{-\mathrm{k}}^{\dagger}(\eta_{0}),\label{eqn:evolutiona}\\
&\hat{b}^{\phantom{\dagger}}_{\mathrm{k}}(\eta_{\rm f})=\alpha^{\phantom{\dagger}}_{-\mathrm{k}}(\eta_{\rm f})\hat{b}^{\phantom{\dagger}}_{\mathrm{k}}(\eta_{0})+\beta_{-\mathrm{k}}(\eta_{\rm f})\hat{a}_{-\mathrm{k}}^{\dagger}(\eta_{0})\label{eqn:evolutionb},
\end{align}
where  $\alpha_{\rm k}(\eta_{\rm f})$ and $\beta_{\rm k}(\eta_{\rm f})$ are known as the Bogoliubov coefficients \cite{bogo1,bogo2}. To maintain the anti-commutation algebra~\eqref{eq:ferm_algebra}, these dimensionless coefficients fulfill
\beq
|\alpha_{\mathrm{k}}(\eta_{\rm f})|^2+|\beta_{\mathrm{k}}(\eta_{\rm f})|^2=1.
\eeq

Under such a Bogoliubov transformation, the annihilation operators in the distant future become a linear superposition of both the creation  and annihilation operators of the remote past, which  is the key to account for particle production. One can  readily see that the number of particles in the far future,   $N_{a}=\int{\rm d}\mathrm{k}\bra{0_{0}
}\hat{a}^{\dagger}_{\mathrm{k}}(\eta_{\rm f})\hat{a}_{\mathrm{k}}(\eta_{\rm f})\ket{0_{0}}/2\pi \mathsf{a}(\eta_{\rm f})
$, upon which all inertial observers agree, is given by $N_a=n_a\delta(0)$. Here, we have divided by the scale factor to take into account the rescaling of the field~\eqref{eqn:decompositionmodefunctions}.  The divergent factor $\delta(0)$ in the particle number corresponds to the infinite spatial volume of the FRW universe, such that the
mean  density of produced particles, in a fiducial volume cell, is
\begin{equation}
\label{eq:density_produced}
    n_a=\frac{1}{ \mathsf{a}(\eta_{\rm f})}\int{\rm d}\mathrm{k}\;|\beta_{\mathrm{k}}(\eta_{\rm f})|^2.
\end{equation}

Given that this particle density is obtained by integrating over the spatial  momentum, which has inverse length units, whereas the scale factor is dimensionless, one finds that the  Bogoliubov coefficient $|\beta_{\mathrm{k}}(\eta_{\rm f})|^2$ is proportional  to the number of produced particles for a specific mode. Additionally, according to Eq.~\eqref{eqn:evolutionb},
 the mean  density of  antiparticles is equal $ n_b=    n_a$, which is a consequence of the fact that the expansion conserves the total charge.

\subsection{\label{sec:level3} Fermion production for a de Sitter expansion}

Once the formalism to understand the phenomenon of particle creation has been discussed, we present the details on how to calculate these quantities for a specific  expansion of a FRW spacetime~\eqref{eqmetric}. In particular, we consider an exponentially-growing  scale factor 
\beq
\mathsf{a}(t)=\ee^{Ht}\mapsto \mathsf{a}(\eta)=-\frac{1}{H\eta},
\label{eq:deSitterPhase}
\eeq
where we note that the conformal and cosmological times~\eqref{eq:conformal_time} are related via  $\eta=-{\rm exp}\{-Ht\}/{H}$, such that $\eta\in(-\infty,0)$ for $t\in(-\infty,+\infty)$. This evolution displays   a constant rate of expansion $
\mathsf{a}^{-1}{\rm d}\mathsf{a}/{\rm d}t = H$ that is commonly known as the Hubble parameter. Let us now discuss how this specific expansion   arises in the two approaches to define a Dirac QFT  in a FRW spacetime of $D=(1+1)$ dimensions introduced in Sec.~\ref{sec:level1}. 

First of all, we consider  the situation in which this arises as an effective QFT when the fermions are forced to move along a single spatial section of the $D=(3+1)$-dimensional FRW spacetime. The scale factor~\eqref{eq:deSitterPhase} can be obtained  from the standard Einstein field equations 
\beq
\label{eq:Eisntein}
G_{\mu\nu}(x)=R^{\lambda}_{\;\mu\lambda\nu}(x)-\half R(x) g_{\mu\nu}(x)=\Lambda g_{\mu\nu}(x),
\eeq
which are expressed in terms of the Riemann curvature tensor
     \beq
     R^{\rho}_{\;\sigma\mu\nu}(x)=\partial_\mu\Gamma^\rho_{\;\nu\sigma}-\partial_\nu\Gamma^\rho_{\;\mu\sigma}+\Gamma^{\rho}_{\;\mu\lambda}\Gamma^\lambda_{\;\nu\sigma}-\Gamma^{\rho}_{\;\nu\lambda}\Gamma^\lambda_{\;\mu\sigma},
     \eeq
a positive cosmological constant $\Lambda$, and  the scalar curvature 
\beq
R(x)=g^{\mu\nu}(x)R^{\lambda}_{\;\mu\lambda\nu}(x).
\eeq
Using the specific expressions of these quantities for the $(3+1)$-dimensional FRW spacetime~\cite{carroll_2019}, one can then derive the so-called Friedmann equations for the scale factor from Eq.~\eqref{eq:Eisntein} which, in this case, have a simple exponentially-growing solution~\eqref{eq:deSitterPhase} with $H=\sqrt{\Lambda/3}$. This de Sitter expansion, which is an exact solution of Einstein's equations,   is actually used to model the inflationary epoch of the early universe in the standard model of cosmology, focusing on a slow-roll regime in which the Hubble parameter is approximately constant~\cite{carroll_2019}. By including a term proportional to the stress-energy tensor in the right-hand side of Eq.~\eqref{eq:Eisntein}, one can also obtain other evolutions of the scale factor associated to matter- and radiation-dominated universes. Indeed, from this perspective, the positive cosmological constant can be interpreted as the result of a vacuum energy acting as a source of Einstein's equations.

Let us now move to the second alternative, where the Dirac fields move in a $(1+1)$- dimensional FRW spacetime that evolves according to  JT gravity. As noted above, the problem with Einstein's equations~\eqref{eq:Eisntein} in this reduced dimensionality is that $G_{\mu\nu}(x)=0$. JT gravity constructs  an alternative  field equation using directly  the curvature scalar. In the presence of a positive cosmological constant,      the  constant-curvature Jackiw-Teitelboim  equation simply reads  
  \beq
  \label{eq:jt__eqs}
  R(x)=\Lambda,
  \eeq
For the FRW spacetime,    the scalar curvature is   $R=\frac{2}{\mathsf{a}}\frac{{\rm d}^2\mathsf{a}}{{\rm d}t^2}$, and one can easily obtain the simple exponentially-growing solution~\eqref{eq:deSitterPhase} with $H=\sqrt{\Lambda/2}$, which differs from  the dependence  in higher dimensions. 
In the  context of JT gravity, one can also derive the analogue of the Friedmann equation when a term proportional to the stress-energy scalar is added in the right-hand side of Eq.~\eqref{eq:jt__eqs}. This leads to other time evolutions of the scale factor $\mathsf{a}(t)$ for a matter- and radiation-dominated universes~\cite{AESikkema1991}, which  differ from those of Einstein's gravity.   In any case, since we are interested in a de Sitter expansion, and both approaches lead to the same exponential growth with the same Hubble constant, any of the interpretations of the origin of the $(1+1)$-dimensional Dirac QFT in the expanding FRW spacetime will be valid. We can thus carry on with the phenomenon of particle production.

To connect with our previous discussion of particle production, this scale factor must be connected to  the  asymptotic flat spacetimes limits, which requires adiabatically ramping up/down the scale factor. One way of doing this~\cite{Anderson:2017hts} is by means of the following factor of expansion 
\begin{align}
\label{eq:scalefactor}
\mathsf{a}(\eta)=\exp\left[\frac{1}{2\lambda}\log\frac{\cosh\left(\lambda\log\frac{\eta_{\rm in}}{\eta}\right)}{\cosh\left(\lambda\log\frac{\eta_{\rm out}}{\eta}\right)}-\frac{\log\left(\eta_{\rm in}\eta_{\rm out}\right)}{2}\right],
\end{align}
which is used to interpolate smoothly between each of the three following regimes
\beq
\mathsf{a}(\eta)=\left\{\begin{tabular}{ll }
    $-\frac{1}{H\eta_{\rm in}},$ & {\rm if} $\eta\ll\eta_{\rm in},$  \\
   $ -\frac{1}{H\eta},$ & {\rm if} $\eta_{\rm in}\ll\eta\ll\eta_{\rm out},$ \\
   $-\frac{1}{H\eta_{\rm out}},$ & {\rm if} $\eta\gg\eta_{\rm out}.$  \\
\end{tabular}
\right.
\eeq
This scale factor is regulated by a parameter $\lambda$. The greater $\lambda$ is, the flatter are the side regions and the better approximated is the de Sitter phase of expansion. We expect that, for $\lambda$ sufficiently large, the vacuum at $\eta_0\ll\eta_{\rm in}$ is adiabatically connected to the instantaneous groundstate at $\eta_{\rm in}$, such that no excitations are produced during the ramping-up phase. 
Then, the de Sitter expansion between $(\eta_{\rm in},\eta_{\rm out})$ will cause non-adiabatic effects, such that the time-evolved state shall no longer coincide with the  instantaneous groundstate at $\eta_{\rm out}$, leading to particle production. Conversely, after the de Sitter phase of expansion, the instantaneous groundstate will be adiabatically connected to the vacuum state of the asymptotic future $\eta_{\rm f}\gg\eta_{\rm out}$, such that no extra particles are produced during the ramping-down phase. In this way, the particle creation   has a well-defined interpretation considering the flat-spacetime asymptotic limits, and is essentially caused by the period of de Sitter expansion which, as will be shown below, admits a closed analytical expression. We note that other hyperbolic-tangent scale factors also allow for closed analytical expressions for the production of Dirac fermion in FRW spacetimes~\cite{PhysRevD.17.964,PhysRevD.82.045030,Martin_Martinez_2012}, although the specific expansions cannot be connected to the Einstein or JT gravity field equations sourced by a simple cosmological constant.

We now discuss how to calculate the Bogoliubov coefficients for this expansion. For reasons that will become clear when discussing the cold-atom analogue-gravity implementation, we
 follow the diagonalization method~\cite{Grib1994VacuumQE,Haro_2008}, which  uses the instantaneous eigenstates of the single-particle Hamiltonian as the mode functions, and consequently introducing at each time annihilation operators $\hat{a}_{\rm k}(\eta)$, $\hat{b}_{\rm k}(\eta)$ which are used to define the vacuum state. The details of this method can be found in Appendix \ref{app_c}. This approach  starts by noticing that the dynamics of the rescaled spinor field  in the FRW spacetime can be described by a single-particle Hamiltonian $H_{\mathrm{k}}(\eta)$ with instantaneous eigenvalues $\pm\omega_{\mathrm{k}}(\eta)=\pm\sqrt{\mathrm{k}^2+m^2\mathsf{a}^2(\eta)}$, and normalized eigenvectors ${\mathsf{v}_{{\rm k}}^\pm(\eta)}$. These eigenvectors correspond, up to a normalization factor, to  the  spinor solutions introduced above, Eq.~\eqref{eqn:decompositionmodefunctions}, for the specific time instant $\eta=\eta_\star$, namely ${\mathsf{v}_{\rm k}^+(\eta_\star)}\propto \tilde{u}_{\rm k}(\eta_\star),{\mathsf{v}_{\rm k}^-(\eta_\star)}\propto \tilde{v}_{\rm k}(\eta_\star)$, which will now be set to the asymptotic remote past $\eta_\star=\eta_{0}$. At later times,  the evolution of the Dirac field is  described by Eq.~\eqref{eq:normalizedmodeexpansion}, but the normalised mode functions $u_{\rm k}(\eta), v_{\rm k}(\eta)$  generally depart from these instantaneous spinor solutions ${\mathsf{v}_{{\rm k}}^\pm(\eta)}$. This departure arises as one leaves the asymptotic remote past $\eta\approx\eta_{\rm in}\gg\eta_0$ and enters in a region with non-adiabatic changes in the scale factor  $\mathsf{a}(\eta)$. In this period of expansion, the mode functions must be found by solving the set of coupled ordinary differential equations (ODEs) in Eq.~\eqref{umodeevolution}, which for each component and in our particular representation reads
 \begin{align}
\begin{split}
&\mathrm{i}\partial_\eta u_{k,1}=-m\mathsf{a}(\eta)u_{k,1}-\mathrm{i}\mathrm{k}u_{k,2},\\&\mathrm{i}\partial_\eta u_{k,2}=+\mathrm{i}\mathrm{k}u_{k,1}+m\mathsf{a}(\eta)u_{k,2},
\label{eoms}
\end{split}
\end{align}
where   $u_{k,1}$ ($u_{k,2}$) is the upper (lower) spinor component.
These ODEs can be expressed as $\mathrm{i}\partial_\eta{u_k}=H_\mathrm{k}(\eta)u_k$, where the aforementioned single-particle Dirac Hamiltonian reads
\begin{equation}
H_\mathrm{k} (\eta)= \begin{pmatrix}
-m\mathsf{a}(\eta) & -\mathrm{i}\mathrm{k} \\
\mathrm{i}\mathrm{k} &m\mathsf{a}(\eta)
\end{pmatrix}.
\label{hamdS}
\end{equation}

One finds that the mode solution $u_{\rm k}(\eta)$ at any instant of time is related to the instantaneous eigenstates of the single-particle Hamiltonian via
\begin{align}
    &u_{\rm k}(\eta)=\alpha_{\rm k}(\eta)\mathsf{v}_{\rm k}^{+}(\eta)+\beta_{\rm k}^*(\eta)\mathsf{v}_{\rm k}^{-}(\eta),\label{eqn:lambdabogo}\\
    &v_{\rm -k}(\eta)=-\beta_{\rm k}(\eta)\mathsf{v}_{\rm k}^{+}(\eta)+\alpha_{\rm k}^*(\eta)\mathsf{v}_{\rm k}^{-}(\eta),
    \label{eqn:taubogo}
\end{align}
which is the manifestation of the Bogoliubov transformation in Eqs.~\eqref{eqn:evolutiona}-\eqref{eqn:evolutionb} at the level of single-particle solutions. Neglecting the adiabatic changes in the asymptotic regions $(\eta_0,\eta_{\rm in})$ and $(\eta_{\rm out}, \eta_0)$, and assuming a purely de Sitter phase \eqref{eq:deSitterPhase}, we solve Eq.~\eqref{eoms} for $u_{\rm k}(\eta)$, choosing as the initial condition the  instantaneous eigenstate 
\begin{equation}
    u_{\rm k}(\eta_{\rm in})=\mathsf{v}_{\rm k}^+(\eta_{\rm in}),
    \label{eqn:initialconditionoriginal}
\end{equation}
so that $\beta_{\rm k}(\eta_{\rm in})=0$. Then, after the expansion, the  Bogoliubov coefficient can be obtained from the overlap of the evolved mode function with the negative-energy instantaneous eigenstate
\begin{equation}
\label{eq:beta_coeff}
|\beta_{\rm k}(\eta_{\rm out})|^2=| u^{\dagger}_{\rm k}(\eta_{\rm out})\cdot\mathsf{v}_{\rm k}^-(\eta_{\rm out})|^2.
\end{equation}
 \begin{figure*}[t]
	\centering
	\includegraphics[width=1.6\columnwidth]{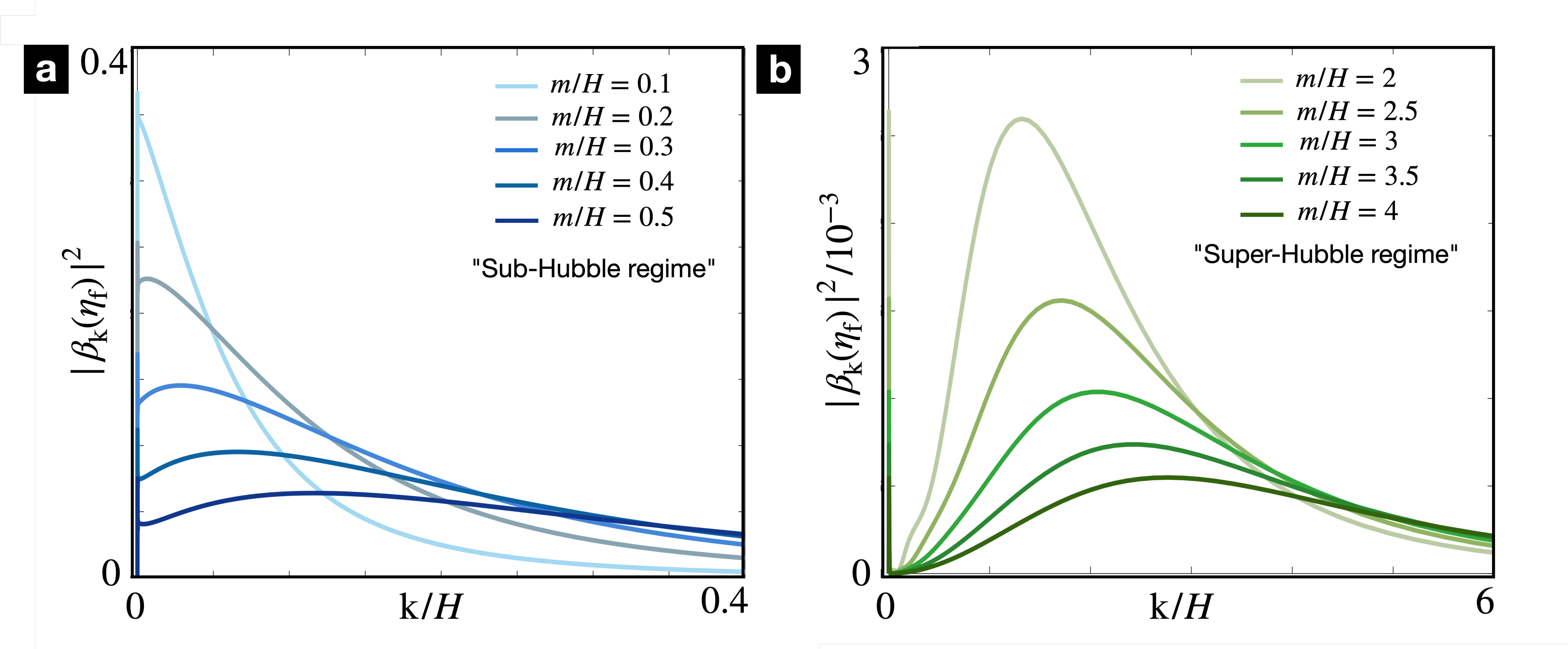}
	\caption{\label{fig:spec_sub} {\bf Spectral distribution for the produced fermions:} We represent the Bogoliubov parameter $|\beta_{\rm k}(\eta_{\rm f})|^2$ for fixed expansion time parameters $\eta_{0}=-10^{4}/H$, $\eta_{\rm in}=-10^{3}/H$, $\eta_{\rm out}=-1/H$ and $\eta_{\rm f}=-10^{-10}/H$, as a function of momentum. In light of Eq.~\eqref{eq:density_produced}, this can be interpreted as a spectral distribution  of the produced  particles, and we focus on the regime of {\bf (a)} sub-Hubble masses,  and  {\bf (b)} super-Hubble masses.  }
	\end{figure*}

From this expression, it becomes clear that the production of particles will be negligible if the adiabatic theorem \cite{Sakurai:2011zz} holds, since the mode solutions will remain instantaneous eigenstates of the single particle Hamiltonian at any latter times $\eta>\eta_{\rm in}$, and so the overlap in \eqref{eq:beta_coeff} will be zero. For this particular Hamiltonian and for a purely de Sitter expansion phase, the adiabatic theorem holds as long as the parameters satisfy
\begin{align}
\left|\frac{m{\rm k}H^2\eta}{\left(m^2+{\rm k}^2H^2\eta^2\right)^{3/2}}\right|\ll1, 
\label{eqn:adiabatic_condition}
\end{align}
which is consistent with previously proposed adiabatic parameters \cite{Chung:2011ck}. We are thus interested in situations where the expansion does not satisfy condition \eqref{eqn:adiabatic_condition}, because it is the non-adiabaticity in the expansion of the universe which induces particle production. 

As shown in Appendix~\ref{app:exact_solution}, the analytical solution for the mode functions is found  by decoupling the ODEs~\eqref{eoms} into a pair of  Bessel differential equations, whose solutions can be expressed in terms of Hankel functions \cite{1988AmJPh..56..958A}
\begin{align}
\begin{split}
    &u_{{\rm k},1}(\eta)=C_{1,1}\sqrt{\eta}H^{(1)}_{\nu_{+}}(\mathrm{k}\eta)+C_{1,2}\sqrt{\eta}H^{(2)}_{\nu_{+}}(\mathrm{k}\eta),\\&u_{{\rm k},2}(\eta)=C_{2,1}\sqrt{\eta}H^{(1)}_{\nu_{-}}(\mathrm{k}\eta)+C_{2,2}\sqrt{\eta}H^{(2)}_{\nu_{-}}(\mathrm{k}\eta),
\end{split}
\label{solutions_Hankel}
\end{align}
with $\nu_{\pm}=\half\pm \mathrm{i}\frac{m}{H}$. Here,  the four integration constants $C_{i,j}$ are  not independent, as we started from  two first-order  ODEs in Eq.~(\ref{eoms}). We thus  need  two initial conditions, which are given by Eq.~(\ref{eqn:initialconditionoriginal}).
We note that similar expressions in terms of Hankel functions can be found in the literature for the $(3+1)$-dimensional case~\cite{paperema}, where differences arise due to the helicity of the spinor solutions, and also for scalar fields~\cite{josederivative}, where  the order of the Hankel functions is real $\nu\in\mathbb{R}$. As discussed in the Appendix~\ref{app:exact_solution}, our solution in terms of Hankel functions can also be related to previously-found solutions that make use of the less-familiar cylinder functions~\cite{victorvillalba}. Before moving on, let us note that this analytical solutions rests on the assumption that no particles will be produced on the  adiabatic switching regions, the validity of which will be explored below numerically for specific switchings.

Let us now comment on  a simple analytical expression for the Bogoliubov coefficient~\eqref{eq:beta_coeff}, and thus  the density of produced particles~\eqref{eq:density_produced}  after an infinitely-long phase of expansion, namely   $\eta_{\rm out}\approx\eta_{\rm f}\rightarrow0^-$ (i.e. $t_{\rm out}\rightarrow\infty$), $\eta_{\rm in}\approx\eta_0\rightarrow-\infty$ (i.e. $t_{\rm in}\rightarrow-\infty)$, where  the only restriction is that of a non-vanishing bare mass $m\neq0$. The eigenvalues in those limits adopt the form $\mathsf{v}_{\rm k}^{\pm}(\eta_{\rm out}\rightarrow-\infty)=\left(\pm\mathrm{i},\;1\right)^{\rm t}/\sqrt{2}$, $\mathsf{v}_{\rm k}^{+}(\eta_{\rm out}\rightarrow0^-)=\left(\mathrm{i},\;0\right)^{\rm t}$ and $\mathsf{v}_{\rm k}^{-}(\eta_{\rm out}\rightarrow0^-)=\left(0,\;1\right)^{\rm t}$.
The constants $C_{i,j}$ in (\ref{solutions_Hankel}) are  determined by the initial  condition (\ref{eqn:initialconditionoriginal}), such that \begin{equation}
    u_{\rm k}(\eta)=C_{\mathrm{k}}\sqrt{\eta}\begin{pmatrix}-H_{\nu_{+}}^{(1)}(\mathrm{k}\eta)\\ \ee^{\ii\pi\nu_{-}}H_{\nu_{-}}^{(1)}(\mathrm{k}\eta)\end{pmatrix},
    \label{eqn:solutioninfpast}
\end{equation}
where $C_{\mathrm{k}}=\frac{\sqrt{\pi \mathrm{k}}}{2}\ee^{-\mathrm{i}\left(\mathrm{k}\eta_{\rm in}+\frac{\pi\nu_{-}}{2}-\frac{\pi}{4}\right)}$. As discussed in Appendix~\ref{app_b},  using the asymptotic behaviour of the Hankel functions for $z=\mathrm{k}\eta_{\rm out}\rightarrow0^-$, which assumes the existence of a cutoff for the spatial momenta $\mathrm{k}\leq\Lambda_{\rm c}$, we arrive at the result
$
\left|\beta_{{\rm k}}(0^-)\right|^2={1}/({\ee^{\frac{2\pi m}{H}}+1}),
$
which is reminiscent of a Dirac-Fermi distribution at an effective temperature $T=H/2\pi$. Let us note, however, that  the instantaneous energy dispersion $\omega_{\mathrm{k}}(\eta)=(\mathrm{k}^2+m^2\mathsf{a}^2(\eta))^{1/2}$ does not appear in the expression, which would thus yield an infinite density of produced particles when integrated. Altogether, our  result is
\begin{equation}
\left|\beta_{\mathrm{k}}(0^-)\right|^2=\frac{1}{\ee^{\frac{2\pi m}{H}}+1}\theta(m),
\label{eqn:thermic}
\end{equation}
where $\theta(m)$ is Heaviside's step function, $\theta(m)=1$ if $m>0$ and zero elsewhere. Note that, in the limit of large masses $m\gg H$,  particle production is exponentially suppressed.

\begin{figure}[t]
	\centering
	\includegraphics[width=0.95\columnwidth]{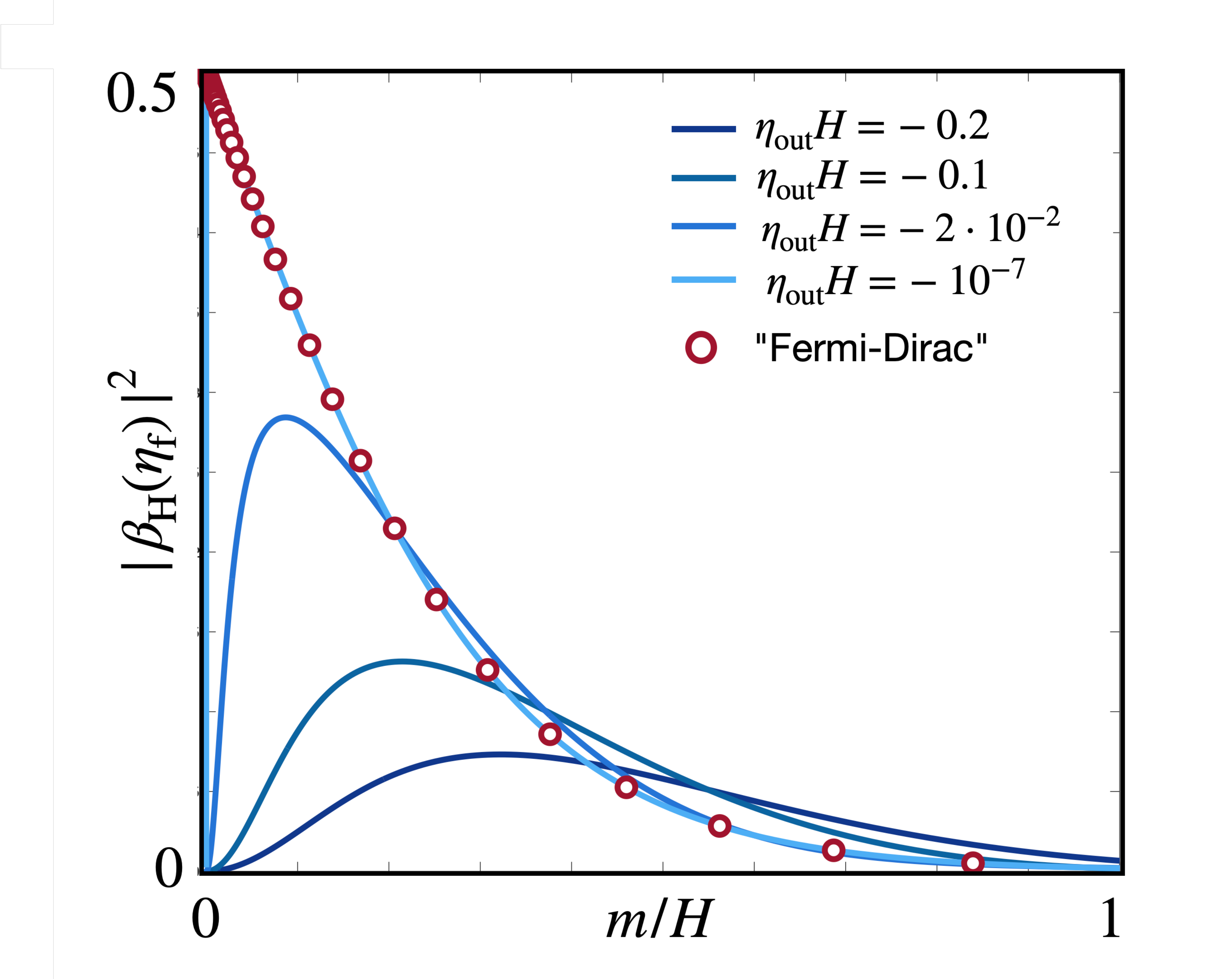}
	\caption{\label{fig:thermic} {\bf Number of created fermions at the Hubble scale:} We represent the Bogoliubov parameter $|\beta_{\rm k}(\eta_{\rm f})|^2$ for momentum ${\rm k}=H$ as a function of the bare mass $m$ for different expansion times $\eta_{\rm f}$ (solid lines) for $\eta_{0}=-10^3/H$, $\eta_{\rm in}=-10^2/H$ and $\eta_{\rm f}=-10^{-10}/H$. When the total time of expansion approaches $\eta_{\rm out}\rightarrow0$, the solid lines representing the numerical calculation of $|\beta_{\rm k}(\eta_{\rm f})|^2$ tend to the  Fermi-Dirac-like distribution~\eqref{eqn:thermic} (red circles). }
\end{figure}

Let us now assess the validity of these results by taking into account specific parameters for the adiabatic switching regions. We will no longer use the approximations $\eta_{\rm in}\rightarrow-\infty$ and $\eta_{\rm out}\rightarrow0^-$, and so we will not use the limiting forms that we stated before, but calculate numerically the instantaneous eigenstates in the corresponding adiabatic regions using the specific scale factor~\eqref{eq:scalefactor}. A numerical benchmark of our result can be found in Fig. \ref{fig:thermic}, where we present the results of a numerical calculation of $|\beta_{\rm k}(\eta_{\rm f})|^2$, which involves solving the system of ODEs~\eqref{eoms} with the specific scale factor~\eqref{eq:scalefactor}, as a function of the bare mass. We consider   various finite values of the final expansion time, and fix the parameter of the adiabatic switching to $\lambda=30$, which ensures a smooth and slow connection to the asymptotic Minkowski vacua. We observe that, as $\eta_{\rm out}$ approaches zero, $|\beta_{\rm k}(\eta_{\rm f})|^2$ indeed tends to the previous Fermi-Dirac-like function~\eqref{eqn:thermic}, confirming  the validity of our analytical treatment and, in particular, the adiabatic switching that leads to a faithful interpretation of the particle production. Let us also note that we have also implemented other switching profiles, which lead to a similar agreement with the analytical prediction.

We now  explore how particle production depends on the spatial momentum of the fermions. We see in  Fig.~\ref{fig:spec_sub} that, for masses within the sub-Hubble regime ($m<H$), there is a peak in the spectrum for $\mathrm{k}\ll H$, and also a non-trivial shape for $\mathrm{k}\approx0$. Again, the production is lower for heavier particles, although the distribution is broader. For super-Hubble masses ($m>H$), we see that the peak occurs at higher spatial momentum  $\mathrm{k}$. Analogously, its height lowers and its width broadens  as the mass increases, although not as much as in the sub-Hubble case.

\begin{figure}[t]
	\centering
	\includegraphics[width=0.95\columnwidth]{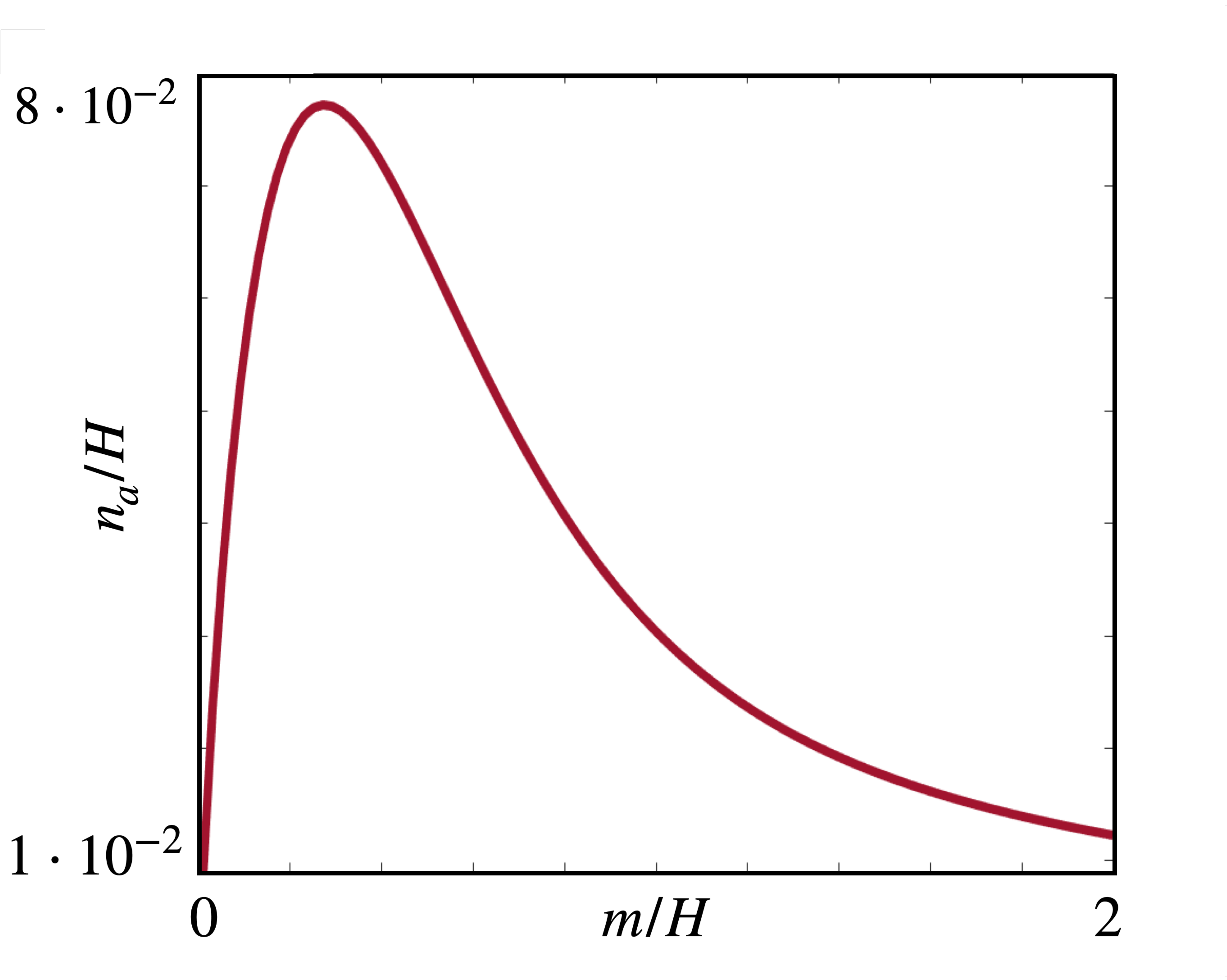}
	\caption{\label{fig:nVSm}  {\bf Density of produced fermions:} We represent the density of produced fermions $n_a$ as a  function of the  bare mass $m$, both in units of the Hubble constant $H$. Accordingly, by increasing the value of this parameter, which will be possible in the context of quantum simulations, the density of produced fermions can reach large values. The parameters of the de Sitter expansion and adiabatic switching times are $\eta_{0}=-10^4/H$, $\eta_{\rm in}=-10^3/H$, $\eta_{\rm out}=-1/H$,  and $\eta_{\rm f}=-10^{-10}/H$. A peak of production occurs for $m\approx 0.274 H$. }
\end{figure}

Finally, if we integrate this spectrum in momentum space, we can obtain the  density of created particles via Eq.~\eqref{eq:density_produced}. If we calculate this integral for different masses, we obtain the plot in Fig. \ref{fig:nVSm}. We can see that, for lighter fermions,  the total number of produced particles grows as the mass increases (i.e. although the height of the peak of the spectrum  lowers as mass increases, it also broadens in such a way that the total area increases, yielding a higher  density of produced particles). This occurs until a certain value of the mass is reached, where the broadening of the spectrum is not sufficient to compensate for the lowering of the peak, and the density of produced fermions starts to decrease with the bare mass. The physical reason for this decrease is that the creation of  heavy energetic fermions is suppressed as the gravitational background does not have enough energy to produce them. On the other hand, if the mass is very small, we are close to a conformal-invariant expansion where particle production is also suppressed, as displayed in the figure. Altogether, there is a maximum for the density of the produced fermions with an intermediate mass that  balances between these two effects. This occurs for fermions with a mass $m\approx0.274H$, which differs with respect to the scalar field case with $m=H$~\cite{josederivative}.

\section{\bf \label{sec:discrete}Lattice regularization and spacetime boundaries}

In this section, we consider two different  lattice discretizations of  the Dirac QFT in a  curved spacetime. In high-energy physics, the lattice is an artificial scaffolding for the fields that serves to regularize the QFT, allowing to treat interacting problems with ultraviolet divergences beyond the perturbative renormalization group~\cite{WILSON197475}. In the context of flat Minkowski spacetimes,  lattice field theories (LFTs)  are  routinely used for this purpose   in quantum chromodynamics~\cite{PhysRevD.10.2445,lattice_qcd, latticegaugetheory}. As advanced in the introduction, there are certain lattice  discretizations~\cite{Wilson1977,Kaplan:2009yg,KAPLAN1992342}  that   display  a non-trivial topology in reciprocal space~\cite{GOLTERMAN1993219,10.1143/PTP.73.528}, and connect these QFTs~\cite{PhysRevLett.105.190404,PhysRevLett.108.181807,PhysRevX.7.031057,BERMUDEZ2018149,PhysRevB.99.064105,PhysRevB.99.125106,PhysRevD.102.094520,https://doi.org/10.48550/arxiv.2112.06954} to the physics of topological insulators and superconductors in condensed matter~\cite{RevModPhys.83.1057}. In the condensed-matter context, the lattice is actually physical, and one is not only interested in recovering a continuum limit devoid of lattice artifacts that can appear around certain phase transitions, but also in charting the full phase diagram in which the specific lattice discretization can play a key role.

A celebrated example is that of the  quantum anomalous Hall  (QAH) effect~\cite{doi:10.1146/annurev-conmatphys-031115-011417} in both the honeycomb~\cite{PhysRevLett.61.2015,RevModPhys.89.040502}, and square~\cite{PhysRevB.78.195424}  lattices. These models can be connected to a Hamiltonian formulation of Dirac fermions in (2+1) dimensions with a Wilson-type discretization~\cite{Wilson1977}, and  host groundstates that cannot be understood from the paradigm of Landau's theory of spontaneous symmetry breaking~\cite{landau_37,Xiao:803748}. In fact, characterizing these states requires introducing an invariant known as the Chern number, which characterizes the topology in reciprocal space, and is responsible for the robustness of the quantized transverse conductance~\cite{PhysRevLett.49.405,nakahara_2017}.  The QAH phases
 are a specific type of the so-called symmetry-protected topological (SPT) phases~\cite{originaltopophases}. In general, different SPT phases can be found within the same symmetry class, and cannot thus be described by a local order parameter and connected via a symmetry-breaking mechanism. Characterizing these phases and the phase transitions requires instead the use of other   topological invariants that can only change via a gap-closing phase transition that is not associated to symmetry breaking. For fermionic models, there are various SPT phases for different symmetry classes and spatial dimensionalities~\cite{doi:10.1063/1.3149495,doi:10.1063/1.3149495,classification_spt}, which can also be connected to discretizations of Dirac QFTs~\cite{Ryu_2010}, and turn out to be robust to perturbations that do not explicitly break the specific symmetry.

In $(1+1)$ dimensions, an archetype of SPT physics is the so-called Su-Schrieffer–Heeger  model in the limit of a classical  lattice dimerization~\cite{PhysRevLett.42.1698, 2016LNP...919.....A,shankar}. Here, one can distinguish a topological phase in the symmetry class $\mathsf{BDI}$ from a trivial band insulator using a  topological invariant known as the Zak's phase $\varphi_{\text{Zak}}$~\cite{Zak}, which is defined as  the  integral  of the  Berry connection $\mathcal{A}_n(\mathrm{k})$~\cite{doi:10.1098/rspa.1984.0023,berry_review} in reciprocal space. The model allows for a topological phase transition as one changes the microscopic couplings, through which the energy gap to excitations vanishes, and the Zak's phase changes abruptly $\varphi_{\text{Zak}}:0\to\pi$. This marks the onset of  topological effects, which have a boundary manifestation: the existence of charge fractionalization in zero-energy excitations that are exponentially localized to the edges of the lattice, i.e. boundary zero modes. If the lattice dimerization is dynamical, such boundary modes can be localized to solitons interpolating between two different groundstates, paralleling the Jackiw-Rebbi mechanism of charge fractionalization in QFTs~\cite{PhysRevD.13.3398}.

Although the Su-Schrieffer–Heeger  model can be rigorously connected to a specific regularization of a continuum Dirac QFT with boundary zero modes~\cite{alejandroTopologicalQED,PhysRevB.100.115152}, it is not  a standard discretization in the Hamiltonian formulation of LFTs~\cite{SusskindwilsonHamiltonian,latticefermions}. As discussed in~\cite{BERMUDEZ2018149}, it can be related to the aforementioned Wilson discretization~\cite{Wilson1977} for a specific choice of gamma matrices and microscopic parameters. This Wilson discretization can also be depicted as a lattice model of fermions hopping on a cross-link two-leg ladder and subjected to an external $\pi$-flux, the so-called Creutz ladder~\cite{PhysRevLett.83.2636,RevModPhys.73.119}. This model can also host  boundary zero modes exponentially localized to the left- and right-most boundaries of the ladder, signaling the occurrence of an SPT phase  in the symmetry class $\mathsf{AIII}$~\cite{Mazza_2012,1907.11460,Zurita2021tunablezeromodes}. The role of  finite temperatures~\cite{PhysRevB.86.155140,PhysRevLett.112.130401},  dynamical quenches~\cite{PhysRevLett.102.135702,PhysRevB.99.054302}, and charge pumping~\cite{PhysRevB.96.035139}  has also been discussed.  The interplay of  interactions and topology has  been explored recently, including repulsive and attractive Hubbard-type interactions~\cite{PhysRevX.7.031057,PhysRevB.99.125106,BERMUDEZ2018149, PhysRevResearch.2.023058,PhysRevB.105.024502,PhysRevB.105.094201}, many-body~\cite{PhysRevB.103.L060301} and disorder~\cite{PhysRevB.104.094202,PhysRevB.104.L161118} induced localization, and interactions mediated by a discrete gauge field~\cite{PhysRevX.10.041007}. Additionally, by exploring regimes away from the external $\pi$-flux limit, the continuum limit connects to Lorentz-violating QFTs, allowing for the characterization of  topological phenomena  via persistent groundstate currents~\cite{PhysRevB.106.045147}.

Let us note that all of the above phenomena correspond to discretizations in a flat Minkowski spacetime. To the best of our knowledge, the introduction of a curved metric remains largely unexplored in the context of SPT phases. 
 Consequently, it is very natural to ask ourselves if the topological properties can have an interplay with the typical phenomena studied within the realm of QFTs in curved spacetimes. This is not only interesting from the theoretical perspective, but might also be fruitful for a dynamical manifestation of topological effects in possible quantum simulation experiments, as discussed below.

\subsection{Na\"ive discretization and fermion doubling}

Let us start by discussing the lattice regularization of the Dirac theory in a curved metric~\eqref{eq:Dirac_action_curved}, focusing on the $(1+1)$-dimensional case. We begin from the Hamiltonian density of the theory, which can be obtained from the Lagrangian density associated to  the action~\eqref{eq:Dirac_action_curved_rescaled}, namely
\begin{equation}
\label{eq:Dirac_conformal_time_lagrangian}
    \mathcal{L}=\overline{\chi}(x)\!\left(-\eta^{\mu\nu} {\gamma}_\mu\!{\partial}_{\!\nu}+m\mathsf{a}(\eta)\right)\!\!\chi(x).
\end{equation}
The conjugate momentum for the spinor fields is $\Pi_\chi(x)=\mathrm{i}\chi^{\dagger}\!(x)$, yielding thus the Hamiltonian density
\begin{align}
\label{eq:Dirac_hamiltonian_curved_rescaled}
\mathcal{H}&=\overline{\chi}(x)\left(\gamma^1{\partial}_{\!1}-m\rm{\mathsf{a}(\eta)}\right)\chi\!(x).
\end{align}

Following the usual canonical quantization approach, we upgrade the fields to operators ${\chi}\!(x),{\Pi}_\chi\!(x)\mapsto\hat{\chi}\!(x),\hat{\Pi}_\chi\!(x)$, imposing  canonical anti-commutation relations between the fields and their conjugate momenta at equal times, namely  $\{\hat{\chi}\!(\eta,{\rm x}),\hat{\Pi}_\chi\!(\eta,{\rm y})\}=\ii\delta({\rm x}-{\rm y})$, $\{\hat{\chi}(\eta,{\rm x}),\hat{\chi}\!(\eta,{\rm y})\}=\{\hat{\Pi}_{\chi}\!(\eta,{\rm x}),\hat{\Pi}_\chi\!(\eta,{\rm y})\}=0$, which  correspond to 
\begin{align}
    &\{\hat{\chi}\!(\eta,{\rm x}),\hat{\chi}^{\dagger}\!(\eta,{\rm y})\}=\delta({\rm x}-{\rm y}),\\
    &\{\hat{\chi}\!(\eta,{\rm x}),\hat{\chi}\!(\eta,{\rm y})\}=\{\hat{\chi}^{\dagger}\!(\eta,{\rm x}),\hat{\chi}^{\dagger}\!(\eta,{\rm y})\}=0.
\end{align}

The most intuitive way to discretize this theory is by defining a chain of $N$ evenly-spaced sites, separated by a length $a$, and defining fermionic creation and annihilation operators on those sites. 
The spatial derivatives appearing in the Hamiltonian must be replaced by a finite difference, such that 
\begin{align}
\label{eq:spinor_lattice_field}
    \hat{\chi}\!(\eta,{\rm x})&\longrightarrow\hat{\chi}_n(\eta):=\hat{\chi}\!(\eta,\,na),\\
    \partial_1\hat{\chi}\!(\eta,{\rm x})&\longrightarrow\frac{\hat{\chi}_{n+1}\!(\eta)-\hat{\chi}_{n-1}\!(\eta)}{2a}.
\end{align}

The discretized Hamiltonian  is then obtained by direct substitution on Eq.(\ref{eq:Dirac_hamiltonian_curved_rescaled}), and by discretizing the spatial integral
\begin{align}
\label{eq:naive_hamiltonian}
{\hat{H}}_{\text{N}}=a\sum_n\hat{\chi}_n^{\dagger}\left(\gamma^5\frac{\hat{\chi}_{n+1}-\hat{\chi}_{n-1}}{2a}-\ii m\mathsf{a}(\eta)\gamma^0\hat{\chi}_n\right),
\end{align}
where we have introduced the chiral gamma matrix 
\beq
\gamma^5=\ii\gamma^0\gamma^1=-\ii\sigma^y,
\eeq
 which is anti-Hermitian for the mostly-plus metric.

In LFTs, one usually confines the fields in a box, and then allows its size to diverge. Since one expects the fields' amplitudes to decay sufficiently fast, boundary effects are usually neglected, and so periodic boundary conditions (PBC) and a basis of plane waves are generally used. We follow this approach now, and relegate the study of other boundary conditions to the following subsection, which will allow us to study boundary effects  related to the aforementioned SPT phases of matter. For now, we assume PBC, and use $\hat{\chi}_n=\frac{1}{\sqrt{N}}\sum_{\rm k}\hat{\chi}_{\rm k}e^{i{\rm k}an}$. Since this is a discrete theory with spatial periodicity $a$, only those values of the crystal momentum within the Brillouin zone  ${\rm k}\in{\rm BZ}=\{-\frac{\pi}{a}+\frac{2\pi n}{Na}$, $n\in\mathbb{Z}_N\}$ will be allowed, leading to the Hamiltonian  in momentum space
\begin{align}
\label{eq:naive_hamiltonian_momentum_space}
{\hat{H}}_{\rm N}=\sum_{{\rm k}\in{\rm BZ}}\hat{\chi}_{\rm k}^{\dagger}\left(\ii\gamma^5\frac{\sin({\rm k}a)}{a}-\ii m\mathsf{a}(\eta)\gamma^0\right)\hat{\chi}_{\rm k}.
\end{align}

This way of discretizing the theory is usually known as the \textit{na\"ive discretization}, which is afflicted by the so-called fermion doubling~\cite{latticefermions, lattice_qcd, latticegaugetheory}. In $d=1$ spatial dimension, this implies that, when taking the continuum limit $a\rightarrow0$, one recovers  twice as many fermions as there were in the original continuous theory. In general, for Hamiltonian field theories in $D=d+1$ dimensional spacetimes, one would   encounter $2^d$ doublers. This presents a problem when one is interested in particle production, since it means that the discrete theory has additional low frequency excitations. For our particular model~\eqref{eq:naive_hamiltonian_momentum_space}, this will result in an overproduction of  particles. In order to prove that,  we need to calculate the Bogoliubov coefficients $\beta_{\rm k}(\eta)$, noting that the previous system of differential equations~\eqref{eoms} gets modified due to the discretization 
\begin{align}
\begin{split}
&\mathrm{i}\partial_\eta u_{{\rm k},1}=-m\mathsf{a}(\eta)u_{{\rm k},1}-\mathrm{i}\frac{\sin(\mathrm{k}a)}{a}u_{{\rm k},2},\\&{\ii}\partial_\eta u_{{\rm k,}2}=+{\ii}\frac{\sin(\mathrm{k}a)}{a}u_{{\rm k},1}+m\mathsf{a}(\eta)u_{{\rm k},2}.
\label{eoms_naive}
\end{split}
\end{align}
The instantaneous eigenvectors ${\mathsf{v}_{{\rm k}}^\pm(\eta)}$, which  we used to  impose the initial condition ~\eqref{eqn:initialconditionoriginal}
 and calculate the final density of produced particles~\eqref{eq:beta_coeff}, are also modified, since the  single-particle Dirac Hamiltonian~\eqref{hamdS} changes due to the discretization. This can be directly read from Eq.~\eqref{eq:naive_hamiltonian_momentum_space}, yielding for our representation
\begin{equation}
\label{ham_SP_naive}
H_{\rm k} (\eta)= \begin{pmatrix}
-m\mathsf{a}(\eta) & -{\ii}\frac{\sin(\mathrm{k}a)}{a} \\
{\ii}\frac{\sin(\mathrm{k}a)}{a} &m\mathsf{a}(\eta)
\end{pmatrix}.
\end{equation}

\begin{figure}[t]
	\centering
	\includegraphics[width=0.95\columnwidth]{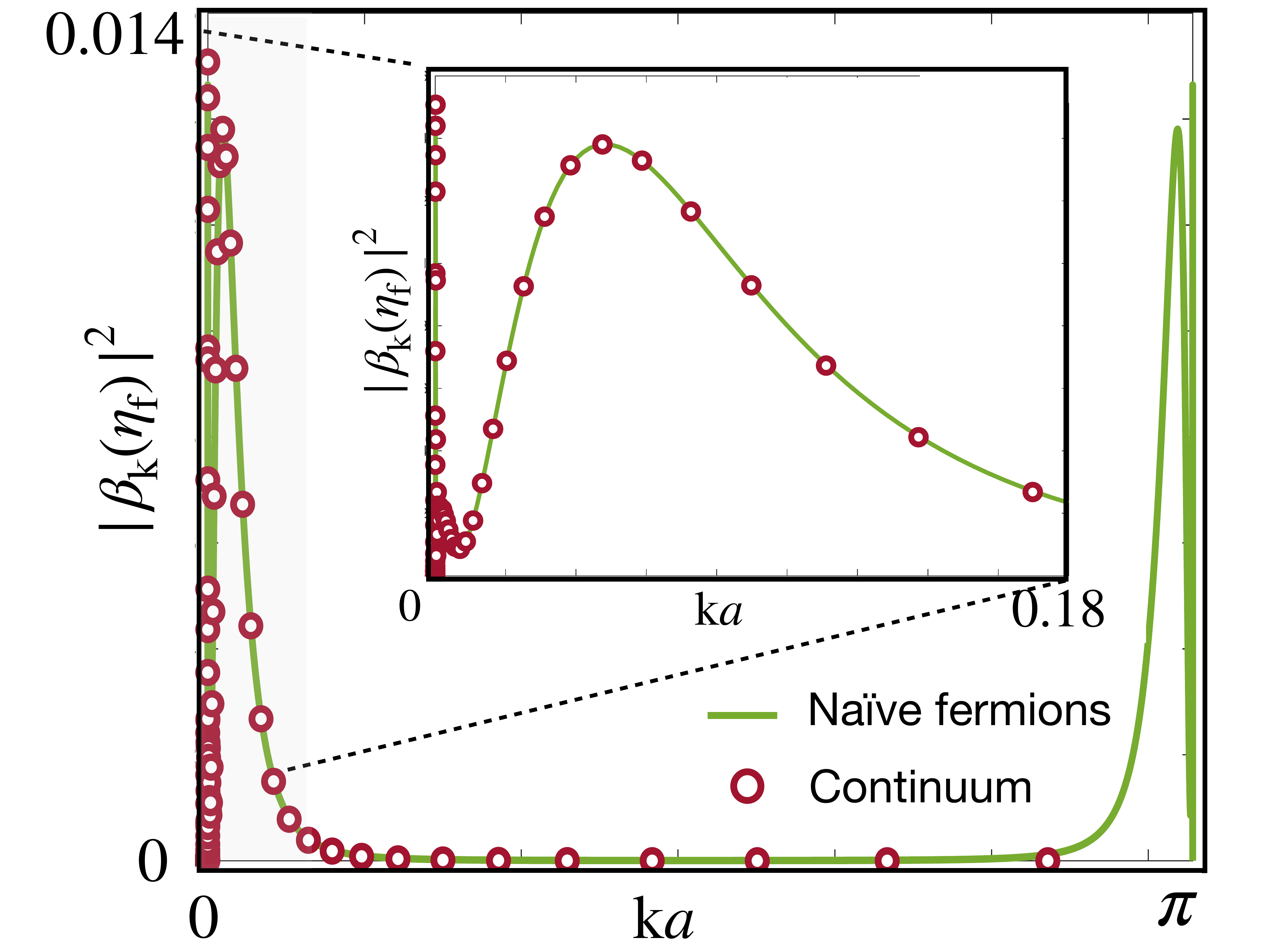}
	\caption{\label{fig:naivediscretization} {\bf Spectral distribution  for  na\"ive fermions:} We represent the spectrum of fermion production  $|\beta_{\rm k}(\eta_{\rm f})|^2$ as a function of the momentum ${\rm k}\in{\rm BZ}$ for an expansion time parameters $\eta_{\rm 0}=-10^5a$, $\eta_{\rm in}=-10^4a$, $\eta_{\rm out}=-10a$, and $\eta_{\rm f}=-10^{-9}a$, where $a$ is the lattice spacing. This is obtained by solving numerically the ODEs~\eqref{eoms_naive} for $m=0.1/a$, and $H=0.1/a$. Only the absolute value of ${\rm k}$ is represented, since there is a symmetry ${\rm k}\rightarrow-{\rm k}$ with respect of the center  of the BZ. The solid lines represent the lattice results, and the red circles stand for the  continuum QFT.}
\label{naive_spectrum}
\end{figure}

With these new equations of motion and instantaneous eigenstates, we calculate numerically the fermion production. One would expect to recover a good approximation to the continuum results for $\mathrm{k}\ll1/a$, where the dispersion relation becomes similar to that of Dirac fermions. As the momentum increases, however,  the differences between the continuum equations and the na\"ively-discretized ones becomes more important, until reaching  the edge of the Brillouin zone $\mathrm{k}\approx\pi/a$, where the fermion doubler lies. Since the dispersion relation there is again similar to that of a Dirac fermion,  a high contribution to the production of particles can again take place.  As can be seen in Fig. \ref{fig:naivediscretization}, the spectrum of particle production at small momenta (solid line) reproduces accurately the continuum result (dots), but we observe spurious creation of particles caused by the fermion doubling when approaching the boundary of the Brillouin zone.  One can clearly see that the distortion of the particle-production spectrum is fully symmetric, as both low-energy fermions are equally affected by the time-dependent mass.

\begin{figure}[t]
	\centering
	\includegraphics[width=0.95\columnwidth]{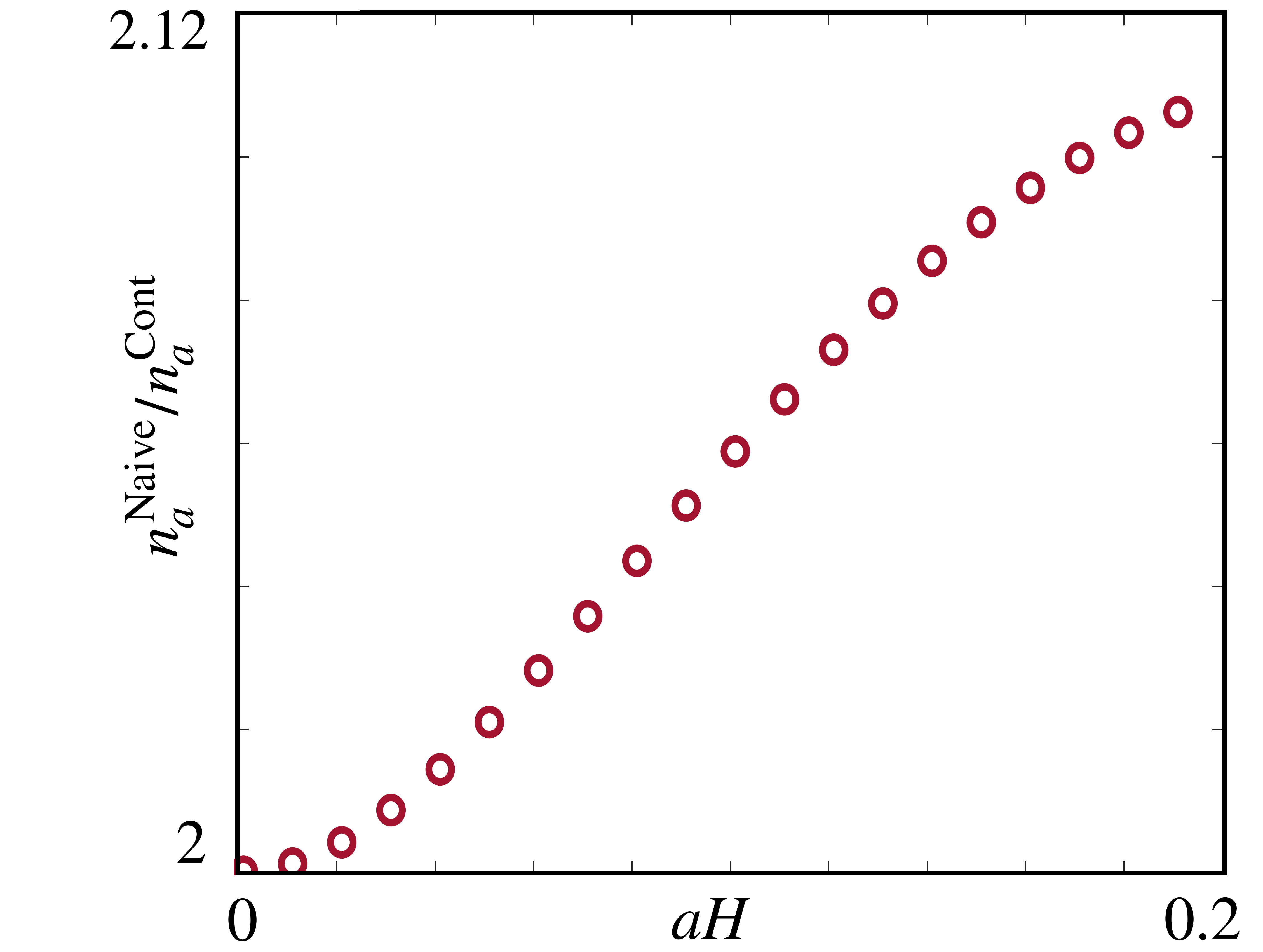}
	\caption{{\bf Continuum limit for the na\"ive-fermion  density: } We represent the ratio between the density of produced particles in the continuum theory and the one with  na\"ive fermions as a function of the lattice spacing. We see that, as the spacing lowers, the rate approaches $2$, since the BZ broadens and the most important contributions of the spectra to the integral (\ref{eq:density_produced}) are sent to the sides of the BZ zone where the dispersion is just slightly modified.The specific time parameters are $\eta_{\rm 0}=-10^3/H$, $\eta_{\rm in}=-2\cdot10^2/H$, $\eta_{\rm out}=-0.1/H$, and $\eta_{\rm f}=-10^{-10}/H$, and the mass is $m=0.2H$.}
\label{fig:rateNvsSpacing}
\end{figure}

Let us now discuss how the total number of produced fermions in this discretized theory is related to that in the continuum theory.  Let us recall that, in the original continuum QFT,  the    spectra in Figs. \ref{fig:spec_sub} {\bf (a)} and {\bf (b)} peak at a certain momentum $\mathrm{k}$ that decreases when the bare mass $m$ is lowered
 with respect to the Hubble parameter $H$. If such  peaks correspond to the region in the Brillouin zone in which the dispersion relation of the discretized model faithfully approximates the continuum one,  we expect to get twice the density of produced particles. We can check this numerically by calculating the continuum density of produced particles, and then comparing that value with the one obtained after the discretization is done. We show the results of doing this in Fig. \ref{fig:rateNvsSpacing}, where the  ratio between the continuum density of created particles and the one obtained after the discretization $\frac{n_{\rm Naive}}{n_{\rm Cont}}$ is presented. As shown in this figure, as $aH=H/\Lambda_{\rm c}\ll1$, which corresponds to a regime where the peak of particle production lies well-below the lattice cutoff $\Lambda_{\rm c}=1/a$, the total density of produced fermions in the discretized model is twice the one estimated from the continuum QFT.

There are several proposals to get rid of these doublers. For example, in the \textit{staggered fermion} approach, due to Kogut and Susskind \cite{SusskindwilsonHamiltonian}, it is proposed to reduce the number of degrees of freedom by using a single component field in each site of the lattice, which  halves the doublers. An alternative that makes connection to topological phases of matter is the so-called \textit{Wilson's fermion} approach \cite{wilsonfermionsoriginal}, which we develop below in the context  of gravitational particle creation.

\subsection{Wilson fermions and reciprocal-space topology}
The idea proposed by Wilson~\cite{wilsonfermionsoriginal} is to include an additional term in the Hamiltonian that acts as a momentum-dependent mass, known as the Wilson mass. This mass sends all of the spurious doublers to the cutoff of the theory, apparently removing all of  their effects in the long wavelength properties of the continuum limit. The objective is to leave the mass equal to the bare mass around ${\rm k}=0$, but making it very heavy around the edges of the Brillouin zone, where the doublers lie. The  Wilson term that must be added to the {na\"ive} Hamiltonian {(\ref{eq:naive_hamiltonian})  is
\begin{align}
\label{eq:Wilson_H}
    {\hat{H}}_{\text{W}}=-a\sum_{n}\hat{\chi}_n^{\dagger}\ii\gamma^0\left(\frac{{r}}{a}\hat{\chi}_n-\frac{{r}}{2a}\hat{\chi}_{n+1}-\frac{{r}}{2a}\hat{\chi}_{n-1}\right),
\end{align}
where $r$ is the so-called Wilson parameter, a dimensionless parameter that is typically set to $r=1$, although it can take other real values for more generality.

This term is the discretized version of a second derivative, which is  a priori irrelevant in a renormalization-group (RG) sense for the continuum QFT~\cite{wilsonfermionsoriginal}. However, allowing for negative bare masses can change the physics considerably, as it can lead to topological phases with boundary zero modes that would not be affected by the coarse-graining and rescaling of the RG, and also a non-zero topological invariant that is  preserved under the RG flow~\cite{PhysRevB.99.125106}. Let us discuss these properties  by considering the total Hamiltonian ${\hat{H}}={\hat{H}}_{\text{N}}+{\hat{H}}_{\text{W}}$.

\begin{figure}[t]
	\centering
	\includegraphics[width=0.95\columnwidth]{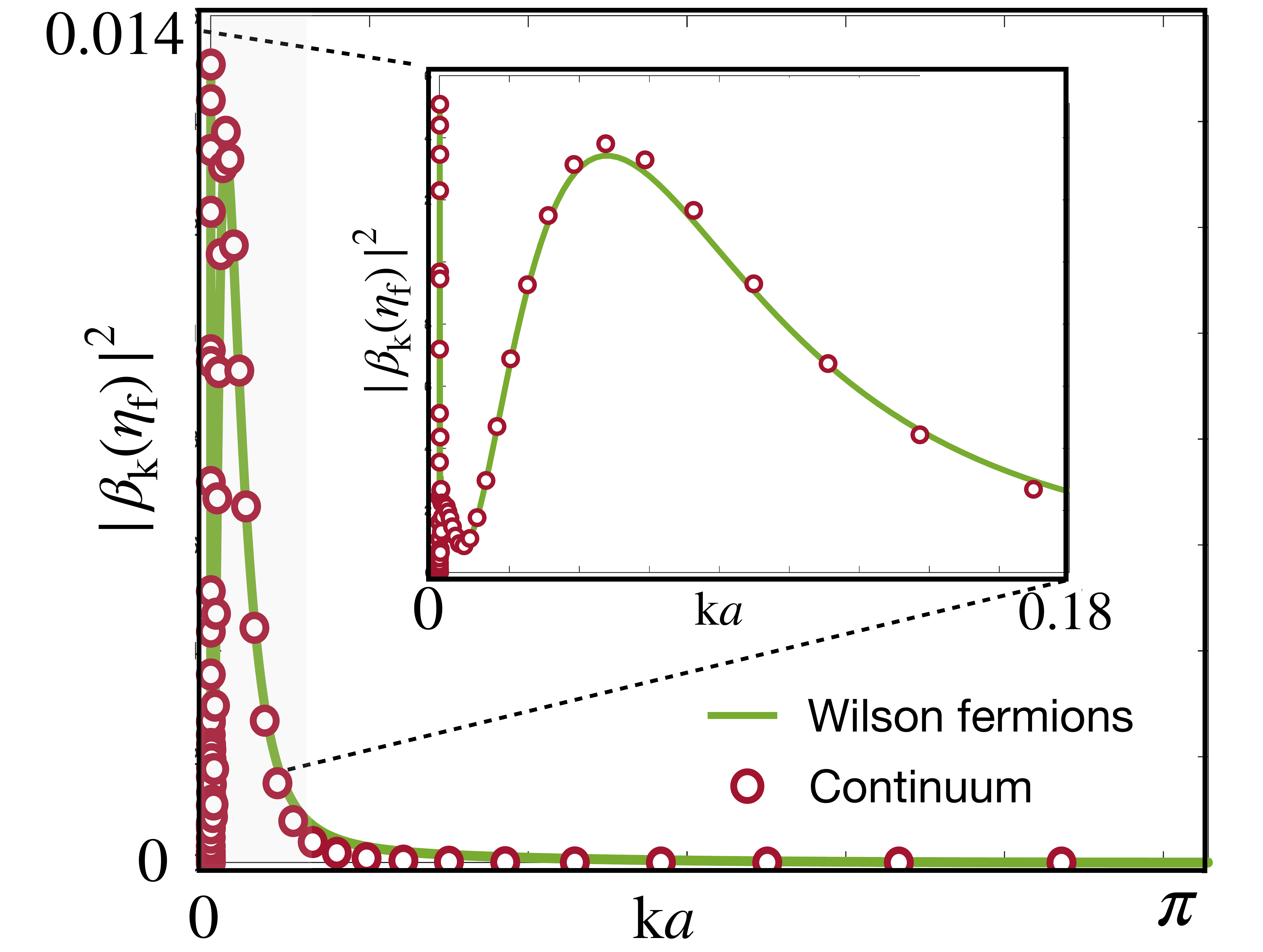}
	\caption{\label{fig:wilson_spectrum}  {\bf Spectral distribution  for  Wilson fermions:} We represent the spectrum of fermion production  $|\beta_{\rm k}(\eta_{\rm f})|^2$ as a function of the momentum ${\rm k}\in{\rm BZ}$ for a time parameters $\eta_{0}=-10^5a$, $\eta_{\rm in}=-10^4a$, $\eta_{\rm out}=-10a$, and $\eta_{\rm f}=-10^{-9}a$, where $a$ is the lattice spacing. This is obtained by solving numerically the ODEs obtained from~\eqref{ham_SP_wilson} for $m=0.1/a$,  $H=0.1/a$, and $r=1$. The solid lines represent the lattice results, and the red circles stand for the  continuum QFT.}
\end{figure}

\begin{figure}[t]
	\centering
	\includegraphics[width=0.95\columnwidth]{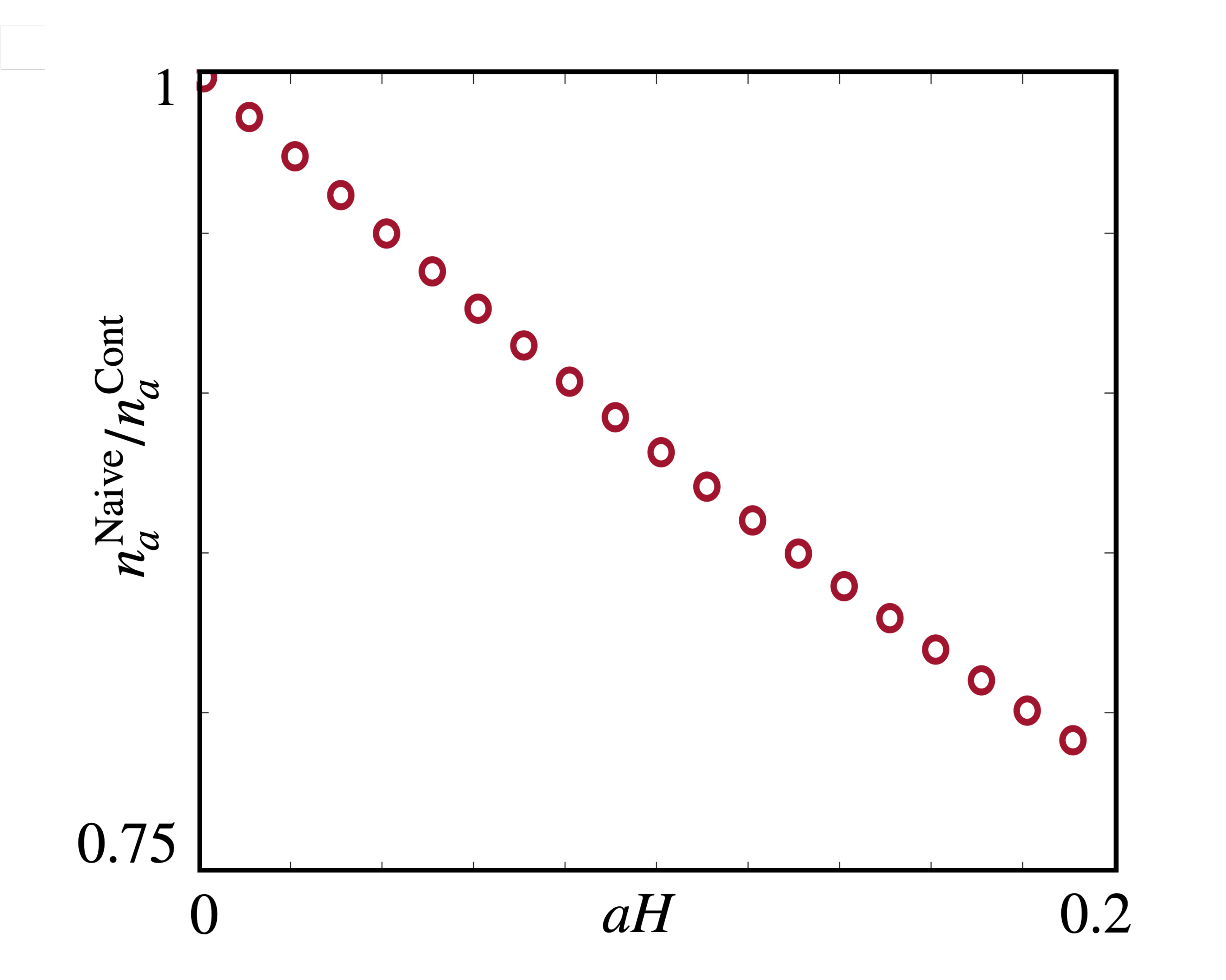}
	\caption{{\bf Continuum limit for the Wilson-fermion  density: } We represent the ratio between the density of produced particles in the continuum theory and the one with  Wilson fermions as a function of the lattice spacing. We see that, as spacing lowers, the rate approaches $1$, since the BZ broadens and the most important contributions of the spectra to the integral (\ref{eq:density_produced}) are sent to the sides of the BZ zone where the dispersion is just slightly modified. The specific time parameters are $\eta_{\rm 0}=-10^3/H$, $\eta_{\rm in}=-2\cdot10^2/H$, $\eta_{\rm out}=-0.1/H$, and $\eta_{\rm f}=-10^{-10}/H$, and the mass is $m=0.2H$.}
\label{fig:rateWilsonSpacing}
\end{figure}

Again,  working with PBC and a basis of plane waves, the new single-particle Hamiltonian is
\begin{align}
\nonumber
&H_{\rm k} (\eta)= \\&=\begin{pmatrix}
-m\mathsf{a}(\eta)-\frac{r(1-\cos({\rm k}a))}{a} & -{\ii}\frac{\sin({\rm k}a)}{a} \\
{\ii}\frac{\sin(\mathrm{k}a)}{a} &m\mathsf{a}(\eta)+\frac{r(1-\cos({\rm k}a))}{a}
\end{pmatrix},
\label{ham_SP_wilson}
\end{align}
 which corresponds to a substitution of the bare mass term for a momentum-dependent one, the so-called Wilson mass
 \begin{equation}
 \label{eq:WIlson_mass}
    m_{\rm W}({\rm k},\eta)=m\mathsf{a}(\eta)+\frac{r(1-\cos({\rm k}a))}{a}.
\end{equation}
This momentum-dependent mass behaves as it should: one recovers the original mass around ${\rm k}\approx0$, whereas the mass of the doubler  becomes very heavy in the continuum limit $a\rightarrow0$ 
\begin{align}
    m_\text{W}(0,\eta)=m\mathsf{a}(\eta),\hspace{2ex} m_{\text{W}}\left(\pm\frac{\pi}{a},\eta\right)=m\mathsf{a}(\eta)+\frac{2r}{a}.
\end{align}

When calculating the particle-production spectrum of  the discrete theory in  Wilson's approach, we use this new single-particle Hamiltonian to obtain the instantaneous eigenstates, and modify the equations of motion~\eqref{eoms_naive}
by substituting the bare mass with the Wilson one $m\mathsf{a}(\eta)\mapsto m_{\rm W}({\rm k},\eta)$. The numerical solution leads to 
 the spectral distribution of  Fig.~\ref{fig:wilson_spectrum}, where the solid line corresponds to Wilson's discretization and the dots stand for the continuum predictions. Comparing this figure with the na\"{i}ve-fermion case in Fig. \ref{fig:naivediscretization}, we see that  the peak of the contribution of the doubler on the edge of the Brillouin zone has become negligible.  For the parameters chosen, this is consistent with our previous results, as we concluded in Sec.~\ref{sec:level3}  that production of very heavy particles is highly suppressed. Accordingly, in order to recover the correct continuum results, we need to impose $aH\ll 1$ such that the dynamical change of the bare mass does not interfere with Wilson's prescription:  the doublers must remain at the cutoff of the QFT.  We must consider  the combined effects of the modification of the dispersion relation away from the doublers,  which can be  neglected by choosing values of $m$ and $a$ such that the tail of the spectrum of created particles is negligible around ${\rm k}\approx\pm\frac{\pi}{2a}$, and the effect of the new Wilson mass term around the center of the BZ, whose correction to the original mass is of order $O({\rm k}^2a^2)$. Once more, we should choose small values of the mass $m$, so that its spectrum presents negligible amplitudes for large values of the momentum, and values of $a$ which enlarge the Brillouin zone, making the correction of the mass term smaller. In Fig.~\ref{fig:rateWilsonSpacing} we represent the ratio between the continuum density of created particles and the one obtained after the Wilson's discretization. We can see that the regime $aH=H/\Lambda_{\rm C}\ll1$ corresponds to a situation where the peak of production lies well-below the lattice cutoff $\Lambda_{\rm c}=1/a$, and one recovers the continuum results without the effects of the fermion doubler.

\subsection{Topological fermion production at the boundary}

Now that we have our discrete Hamiltonian field theory in Eqs.~\eqref{eq:naive_hamiltonian} and~\eqref{eq:Wilson_H}, we  discuss how its vacuum is related to zero-temperature SPT phases, and how this can modify the spectrum of particle production depending on the boundary conditions. 
We stated above that the characterization of SPT phases is based on  topological invariants which, in the $(1+1)$-dimensional case correspond to the Zak's phase $\varphi_{\text{Zak}}$~\cite{Zak}. This topological invariant is defined as the integral of the Berry connection~\cite{doi:10.1098/rspa.1984.0023,berry_review}  over the Brillouin zone, $\varphi_{\text{Zak}}(\eta)=\int_{\text{BZ}}d{\rm k}\mathcal{A}_{\rm k}(\eta)$. The Berry connection is defined as
\beq
\mathcal{A}_{\rm k}(\eta)=\ii({\mathsf{v}_{\rm k}^-(\eta)})^{\dagger}\cdot \partial_{\rm k}{\mathsf{v}_{\rm k}^-(\eta)},
\eeq 
where we recall that ${\mathsf{v}_{\rm k}^-}(\eta)$ is the single-particle negative-energy instantaneous eigenstate. Since we have a dynamical mass in our problem, the dependence with $\eta$ is here treated as parametric, and we consider the topological invariant associated to each one of these instantaneous groundstates. The Zak's phase will differentiate SPT phases ($\frac{\varphi_{\text{Zak}}}{\pi}\in\mathbb{Z}$) from topologically-trivial ones ($\frac{\varphi_{\text{Zak}}}{2\pi}\in\mathbb{Z}$), which can be rephrased in terms of a $\mathbb{Z}_2$-valued gauge-invariant Wilson loop $W_{\rm Zak}=\ee^{\ii\varphi_{\rm Zak}}\in\{-1,+1\}$. In  Wilson's approach to the discretized Dirac QFT~\cite{BERMUDEZ2018149}, the Zak's phase is given by 
\begin{align}
    \varphi_{\text{Zak}}(\eta)=\frac{1}{2}\pi\bigl[\text{sgn}\left(m_{\text{W}}(0,\eta)\right)-\text{sgn}\left(m_{\text{W}}(\pi/a,\eta)\right)\bigl],
\end{align}
which amounts to a sign difference of the mass of the Dirac fermions, those at the center and edges of the BZ.
Accordingly, the instantaneous groundstate displays a non-trivial Zak's phase $\varphi_{\text{Zak}}(\eta)=\pm\pi$ when the Dirac fermion and its spatial doubler have opposite masses. 
Since the expansion of the spacetime is embodied in a time-dependent bare mass, there may be certain parameters of the theory for which the evolution of the spacetime itself will induce a topological phase transition of the instantaneous groundstates.

 As an example, let us suppose a bare mass $m=-2/a$, and  a Wilson parameter that will henceforth be set to  $r=1$.  We also consider a Hubble constant $H=1/a$, and  assume that the spacetime expansion begins at $\eta_{\rm in}=-2a$ and finishes at $\eta_{\rm out}=-0.5a$. We consider a purely de Sitter expansion ${\rm a}(\eta)=-1/H\eta$, neglecting the asymptotic flat regions for this example. Then, at $\eta=\eta_{\rm in}$, we have $m_{\text{W}}(0,\eta_{\rm in})=-1/a$ and $m_{\text{W}}(\pi/a,\eta_{\rm in})=+1/a$, i.e.  the sign of the masses is opposite  and, thus, we are initially  in a SPT phase with $\varphi_{\text{Zak}}=-\pi$. At $\eta=\eta_{\rm out}$, after the spacetime has undergone an expansion, the instantaneous Wilson masses evolve into $m_{\text{W}}(0,\eta_{\rm out})=-\frac{4}{a}$ and $m_{\text{W}}(\pi/a,\eta_{\rm out})=-\frac{2}{a}$, acquiring the same sign, such that the Zak's phase vanishes in this case $\varphi_{\text{Zak}}(\eta_{\rm out})=0$. Thus, the system would start in a topological phase and end in a trivial one, which can only occur through an intermediate gap-vanishing phase transition.  This situation reminds of the adiabatic dynamical quenches mentioned in the introduction~\cite{PhysRevLett.102.135702,PhysRevB.99.054302}, where a quantum system crosses a critical point by the external modification of a microscopic parameter. Interestingly,   gravity is responsible  for such external modification of the parameters, leading to a  topological phase transition via the de Sitter expansion. Note that, due to the Kibble-Zurek mechanism~\cite{KIBBLE1980183,ZUREK1996177,PhysRevLett.95.105701}, the crossing of the phase transition requires a breakdown of the adiabatic approximation, and can lead to  excitations that are not connected to the particle production of the continuous QFT described in the previous section. We will thus avoid this situation and explore those parameters for which the instantaneous groundstates remain in a SPT phase during the whole de Sitter expansion. The whole particle-antiparticle production will thus be a  consequence of the breakdown of Poincar\'e invariance in the expanding spacetime, and the change of the notion of vacua in the asymptotically-flat spacetimes. 

So far, we have only looked at bulk properties by imposing PBC. However, we know that SPT phases have a bulk-boundary correspondence~\cite{RevModPhys.83.1057} manifested in the appearance of zero-energy states, which are exponentially localized to the boundaries of the chain, i.e. to the spatial boundaries of  spacetime. 
The first step to do this is to change our boundary conditions from periodic (PBC) to open (OBC). As a consequence, we will no longer be able to use momentum as a good quantum number~\eqref{ham_SP_wilson}. In turn, our strategy will be to directly diagonalize the total Hamiltonian in position space. It is very intuitive to understand this Hamiltonian as a tight-binding-like model \cite{creutz}, with tunnelings and self-energies shown in Figs.~\ref{fig:diractightbinding} {\bf (a)} and {\bf (b)}. 

\begin{figure}[t]
	\centering
	\includegraphics[width=0.9\columnwidth]{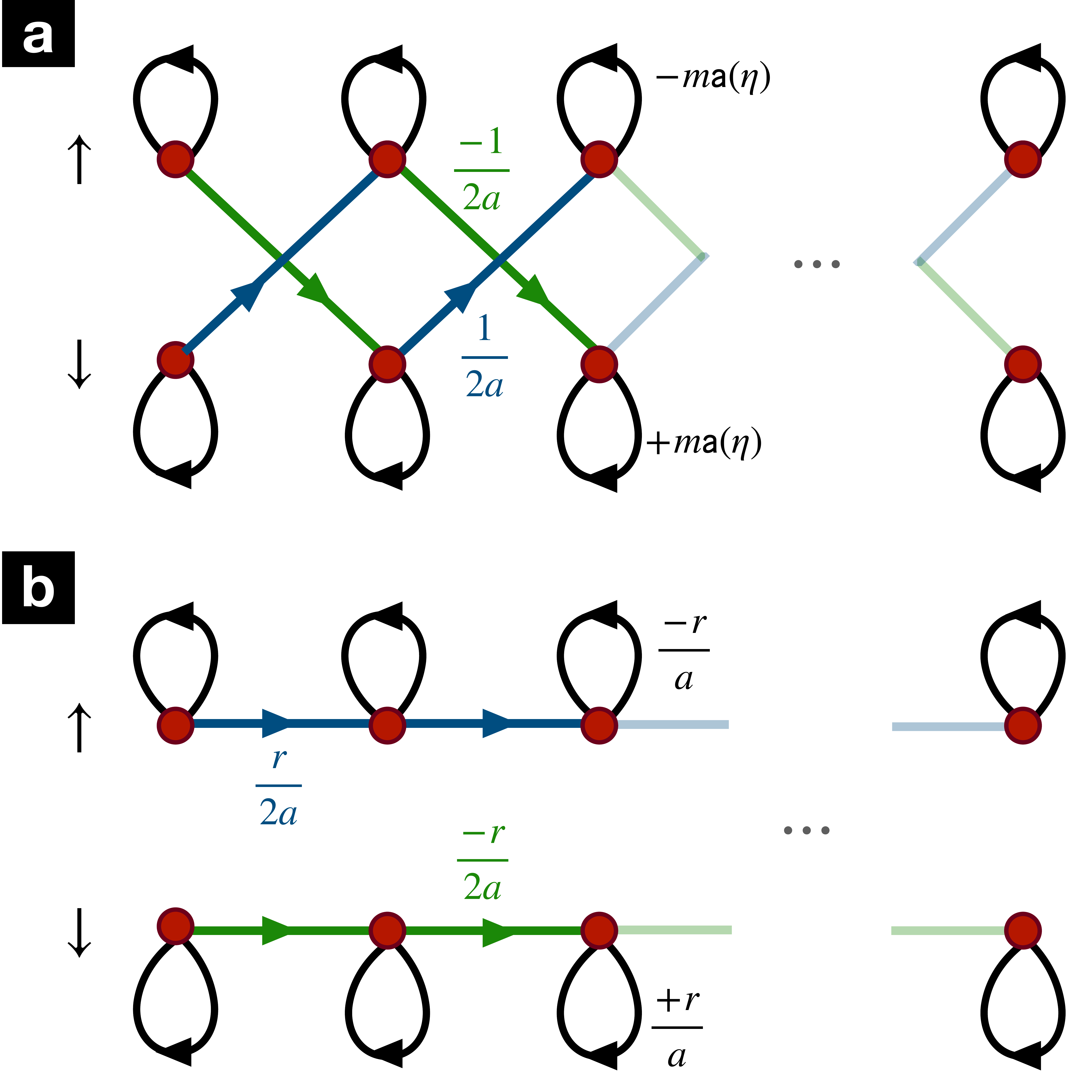}
	\caption{{\bf Visual representation of the lattice field theories:} {\bf (a)} Na\"ive discretization of the Dirac equation, which can be described by a tight-binding model with tunnelings in green and blue that connect neighboring sites and, simultaneously, change the spinor component. Likewise, the bare mass correspond to "self-energy" terms represented by black lines, the amplitude of which changes in time due to the underlying expanding universe. {\bf (b)} In Wilsons discretization, in addition to the tight-binding terms in {\bf (a)}, the Wilson mass term corresponds to additional tunnelings where the spinor component is preserved, again represented by blue and green lines, as well as a constant "self-energy" represented by black lines.} 
\label{fig:diractightbinding}
\end{figure}

Thus, when looking for a single-particle Hamiltonian with OBC, we can assume a chain of $N$ sites and spacing $a$, and express this Hamiltonian as a $2N\times2N$ matrix $H_{n,n'}(\eta)$ considering also the two internal degrees of freedom of the Dirac spinor. Let us note that this matrix is block-tridiagonal, having $2\times2$ matrices in the diagonal that depend on the  bare  and   Wilson mass discretization. The upper and lower blocks correspond to $2\times 2$ matrices that include the na\"ive and Wilson tunnelings, bringing the single-particle excitation to the nearest-neighboring sites.  Then, the procedure is to diagonalize this matrix, obtaining $N$ pairs of positive- and negative-semidefinite eigenenergies, $\omega_{j,\pm}(\eta)$ with  their associated eigenvectors, $\mathsf{v}_j^\pm(\eta)$, where $j$ is an index that labels the eigenvalues, and plays the role of the spatial momentum in the translationally-invariant situation for PBC. After this, the procedure to obtain the Bogoliubov coefficient $\beta_j$ is analogous to that followed in previously. The system of coupled ODEs~\eqref{eoms_naive} now reads
\begin{align}
\begin{split}
&\mathrm{i}\partial_\eta u_{n}=\sum_{n'}H_{n,n'}(\eta)u_{n'},
\end{split}
\end{align}
which must be solved numerically after imposing an initial condition analogous to Eq.~\eqref{eqn:initialconditionoriginal}. In our case, we  consider
\begin{equation}
  u_{j}(\eta_{\rm in})=\mathsf{v}_{j}^+(\eta_{\rm in}),
    \label{eqn:initialcondition}
\end{equation}
$\forall j\in\{1,\cdots, N\}:\omega_{j+}\geq0$. This amounts to an initial situation in which the asymptotic Dirac sea is obtained by filling all the negative-energy solutions and only one of the two topological states of zero energy, which is localized to one edge of the chain. After the de Sitter expansion, we calculate the Bogoliubov coefficient by the analogue of Eq.~\eqref{eq:beta_coeff}, but the matrix vector products are now 
\begin{equation}
\label{eq:beta_coeff_lattice}
|\beta_{j}(\eta_{\rm out})|^2=\sum_{i}| u_j^{\dagger}(\eta_{\rm out})\cdot\mathsf{v}_{i}^-(\eta_{\rm out})|^2, 
\end{equation}
$\forall j\in\{1,\cdots,N\}:\omega_{j-}\leq 0$. This is to be interpreted as the overlap between the evolved states and the eigenstates of the final Hamiltonian. For the propagating modes, this can be understood as an excitation to positive-energy eigenstates, and for the topological modes, as a directed flow from the edge that was initially populated towards the other edge of the chain or bulk states. We thus recover the particle-production spectrum by associating  each $|\beta_j(\eta_{\rm out})|^2 $ parameter with its corresponding energy $\omega_{j+}(\eta_{\rm out}) $. Accordingly, the particle-production spectrum will no loner depend on momentum, but rather on the energy of the particles and antiparticles in the asymptotically-flat spacetimes. Although conceptually equivalent, the process with OBC is  more involved, as it  requires solving $N$ coupled ODEs for each different value of $j$, so a full spectrum calculation requires solving $N$ coupled ODEs $N$ times. This difficulty is due to the lost of periodicity, but it is worthy since  we can now  look for dynamical manifestations of the  SPT phases due to the gravitational production of particles. 

\begin{figure}[t]
	\centering
	\includegraphics[width=0.95\columnwidth]{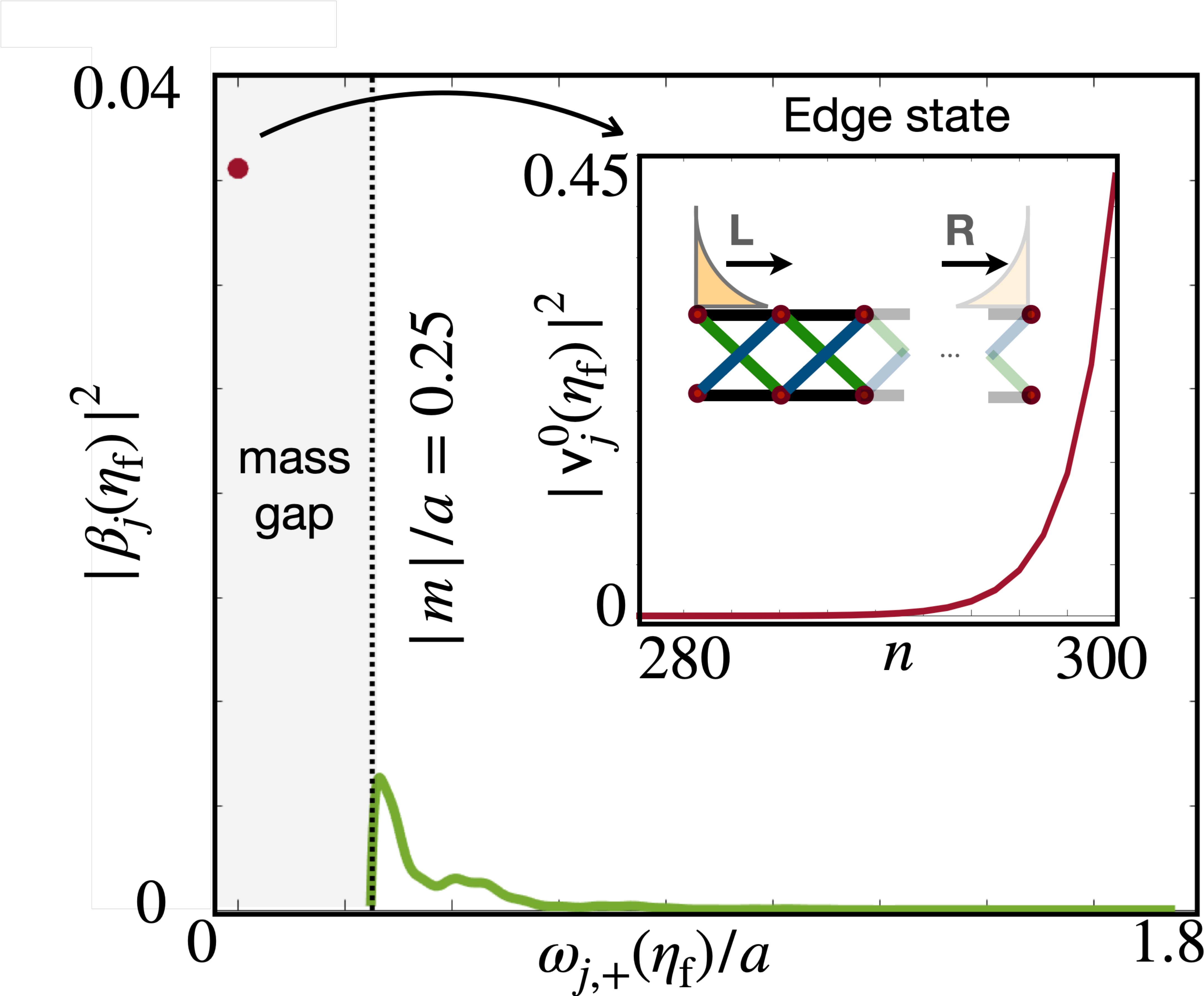}
	\caption{\label{fig:topophasecreation} {\bf Spectral density as a function of the energy for OBC:} We represent the   density of created fermions $|\beta_{j}(\eta_{\rm out})|^2$ for an expansion with time parameters $\eta_{0}=-50a$, $\eta_{\rm in}=-20a$, $\eta_{\rm out}=-10a$ and $\eta_{\rm f}=-0.01a$, as a function of the corresponding bulk energy $\omega_{j,+}$, or the energy of the zero mode. Here, we consider $m=-0.25/a$, $H=0.1/a$,  and set $r=1$ and  $N=300$ sites. The blue dot represents the particle production of topological zero-energy particles. The wavefunction of the zero-mode is shown in the inset figure, where it is clear that the produced particles are exponentially localized in the right boundary, corresponding to the topological state that was initially empty.}
\end{figure}

Since there are zero-energy states exponentially localized to the spatial edges  in the SPT phase, we are interested in distinguishing bulk and boundary contributions to the production caused by the de Sitter expansion. This is automatically allowed by the previous spectrum, as the boundary contribution stemming from the edge states must appear at $\omega_{j+}= 0$ energy. If we solve numerically the system of ODEs, and perform the corresponding numerical diagonalizations, we obtain the result shown in Fig.~\ref{fig:topophasecreation}. In this figure, we see that there is in fact particle creation for this topological zero-modes when we impose OBC, which becomes manifest via a non-zero fermion number inside the energy gap of the spectrum. This is an interesting result that could be observed in a quantum simulation experiment, as discussed in the following section. Since the production of zero-modes is intensive (as can be seen in Fig.~\ref{fig:topophasecreation}, it is inside the energy gap due to the mass), its contribution to the total  density of  particles~\eqref{eq:density_produced} will be negligible in comparison to the contribution of the bulk, which is extensive. On the other hand, if one has frequency resolution, or spatial resolution to localize a probe to the boundaries, the  effect of these topological edge states should be distinguishable from the bulk, as we have just shown. It is  interesting to highlight that, although  there is no particle production for zero-energy bulk propagating modes, as this would require a massless conformally-invariant limit, the edge states change this paradigm allowing for zero-energy particle production.

\section{\bf \label{sec:implementation} Analogue gravity in ultra-cold Fermi gases}

In the previous sections, we have discussed the phenomenon of fermion production during a de Sitter phase of expansion by solving the real-time dynamics of Dirac fermions in a $D=(1+1)$-dimensional Friedmann-Robertson-Walker spacetime. After that, we have regularized the problem on a lattice, which has allowed us to change the topology of the basis domain of the fields from $\mathbb{R}\times\mathbb{R}$ to $\mathbb{R}\times I$, with $I$ a finite subinterval of the real line associated to open boundary conditions. In this case, topological effects in reciprocal space have a boundary manifestation in the form of  zero-energy modes exponentially localized to the boundaries of $I$. We have demonstrated that this topological modes, although having zero energy, can accommodate for particles-antiparticles being created by the de Sitter expansion of the universe, which contrasts with the case of zero-energy massless bulk modes. In this section, we discuss a path for the quantum simulation of the phenomenon of fermion production in expanding universes using table-top experiments, proposing an experimental scheme that employs ultra-cold atomic gases in optical lattices~\cite{RevModPhys.80.885,Bloch2012,Goldman2016,doi:10.1080/00018730701223200}. This would allow us to test the results presented  in the previous sections and, more interestingly, would open new perspectives for the study of non-perturbative effects in real-time dynamics, for example by adding a four-Fermi interaction in the form of a Gross-Neveu-type model which, among other interesting phenomena, can lead to chiral symmetry breaking in this real-time scenario

Before presenting the details of the scheme, let us note that  previous proposals for the quantum simulations of Dirac QFTs in a curved spacetime can be found in the literature, e.g.~\cite{Boada_2011,Minar_2015,PhysRevA.95.013627,Celi2017,10.21468/SciPostPhys.5.6.061}. In this case, there has been a certain focus in spacetimes of reduced dimensionality with spatial inhomogeneities, such as Rindler ones, where a generic formulation based on the previous formalism of vielbeins and the spin connections generally require to implement an inhomogeneous and non-unitary tunneling of the fermionic atoms in the optical lattice. For the specific case of the FRW spacetimes considered in our work, working with conformal time simplifies things considerably, as the inhomogeneities only occur in the temporal direction, and shall amount to a specific real-time modulation of the experimental parameters.  In this sense, our scheme can directly exploit the progress in the quantum simulation of lattice field theories in flat spacetimes reviewed in~\cite{doi:10.1002/andp.201300104,Zohar_2015,Banuls2020,Carmen_Ba_uls_2020,rpa_icfo,Klco_2022}, alleviating some of the difficulties associated to the quantum simulation of  more generic curved spacetimes. In particular, we will show that the schemes presented in~\cite{PhysRevResearch.4.L042012,coldatoms2,PhysRevB.106.045147}, which are based on the idea of Raman optical lattices for implementing an effective spin-orbit coupling~\cite{Zhang_2018,PhysRevLett.110.076401,PhysRevLett.112.086401,PhysRevLett.113.059901}, can be modified minimally, including a real-time modulation of a single experimental parameter,  to allow for a quantum simulation of Dirac fields under the de Sitter expansion. For the sake of completeness, we present a self-contained discussion of the various ingredients of this proposal, emphasizing where the differences with respect to~\cite{PhysRevResearch.4.L042012,coldatoms2,PhysRevB.106.045147} would arise.

\subsection{Raman optical lattices and  expanding spacetimes}

Our objective is to simulate the dynamics of the discretized system shown in Figs.~\ref{fig:diractightbinding} {\bf (a)} and {\bf (b)}. To do so, we consider a gas of fermionic atoms, such as the alkaline-earth ${}^{87}\text{Sr}$ atoms~\cite{Cazalilla_2014}. In this case, the total electronic angular  and spin momentum vanishes  in the ground-state manifold,  which is composed of the Zeeman sub-levels associated to the nuclear spin $F=I=9/2$, so there are 10 Zeeman sub-levels $M_F\in\{-9/2,-7/2,\cdots,9/2\}$, which can be split by applying a weak magnetic field $\bm{B_{\rm ex}}=B\bm{e_z}$ (we choose $\bm{e_z}$ as the quantization direction). Since we are interested in simulating the two spinor states of the Dirac  field,  we focus only on two of those Zeeman sub-levels, which we shall denote $\ket{\uparrow}=\ket{{}^1\text{S}_{0},\,F,\,M_{\uparrow}}$ and $\ket{\downarrow}=\ket{{}^1\text{S}_{0},\,F,\,M_{\downarrow}}$. Note that one  must choose $M_{\uparrow}\text{ and }M_{\downarrow}$ such that the selection rules allow for two-photon Raman transitions between them. The interest of working with these atomic species is that, in addition to the internal $SU(N)$ symmetry of their scattering~\cite{Cazalilla_2014},  they have  ultra-narrow optical transitions that allow to minimize the residual photon scattering associated to these Raman transitions. 

 \begin{figure*}[t]
	\centering
	\includegraphics[width=1.8\columnwidth]{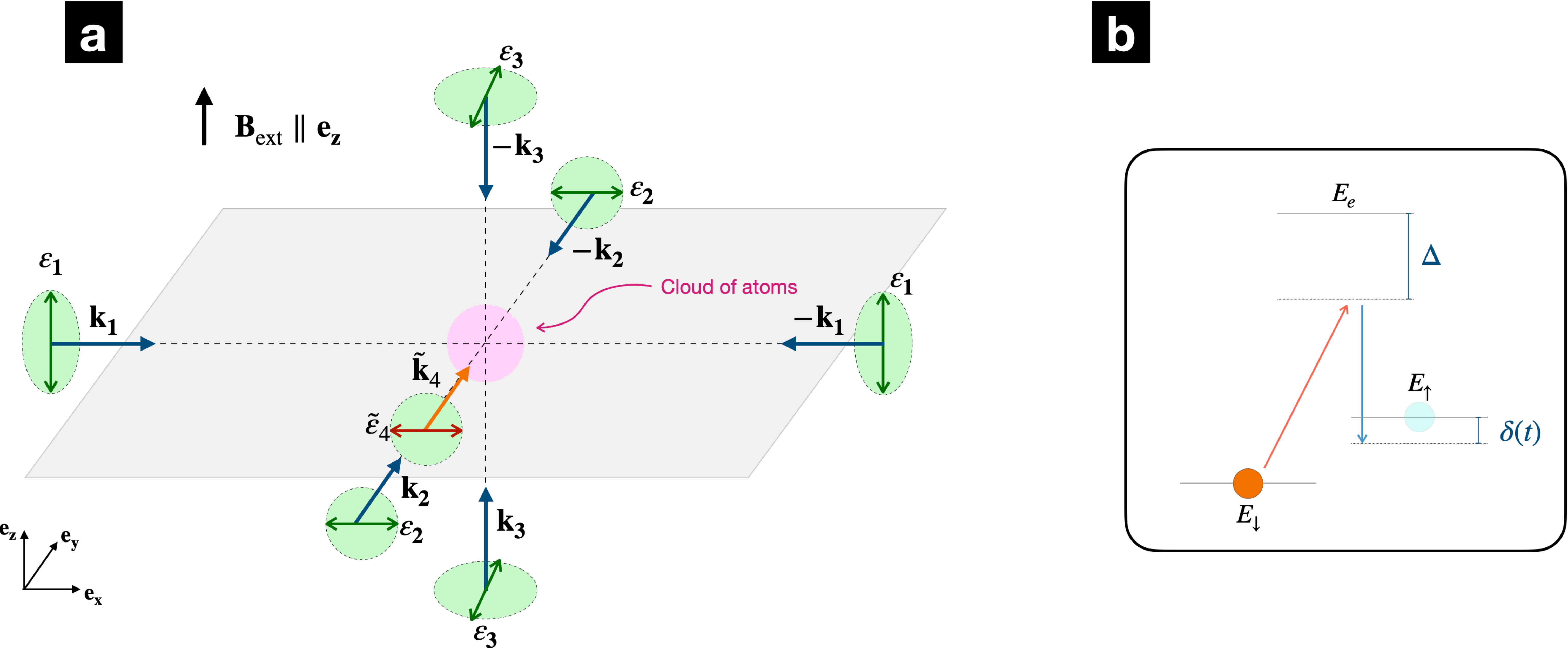}
	\caption{\label{fig:expsetup}\textbf{Scheme for the quantum simulation with ultra-cold atoms in an optical Raman potential:} (a) A cloud of atoms is subjected to three pairs of counter-propagating laser beams in a standing-wave configuration with mutually orthogonal linear polarizations (blue arrows), leading to a cubic  optical lattice. The optical-potential depths in the $y$ and $z$ directions are much larger than in the $x$ direction, leading to effective one-dimensional dynamics. A fourth laser beam in a traveling-wave configuration (orange arrow) is used to induce two-photon Raman transitions in combination with the standing wave along the $x$ axis, which require their beatnote to be tuned close to the resonance   $\delta\omega\approx E_\uparrow-E_\downarrow$, as depicted in \textbf{(b)}. In this way, the Raman terms oscillate with twice the period in comparison to the ac-Stark shifts that lead to the optical lattice.  The expanding background is encoded in the time-dependence of the Raman detuning $\delta(t)$, where  $\delta\omega=E_\uparrow-E_\downarrow-\delta(t)$.}
	
\end{figure*}

This atomic gas is under the influence of three counter-propagating laser beams, as depicted in Fig. \ref{fig:expsetup}. This set-up generates a blue-detuned three-dimensional optical lattice~\cite{RevModPhys.80.885}, with an optical potential of the form 
\begin{align}
\label{eq:optical_lattice}
V_{\rm latt}(\boldsymbol{r})=\sum_jV_{0,j}\cos^2(k_jr_j).
\end{align}
Here, $j\in\{1,\,2,\,3\}\equiv\{x,\,y,\,z\}$ denote each one of the spatial directions, and $\bm{k_{j}}=k_j\bm{e_{j}}$ is the wave-vector of the laser beams, with mutually-orthogonal polarizations $\bm{\varepsilon}_j$, and $V_{0,j}$ is the amplitude of the ac-Stark shift experienced by the states in the groundstate manifold $\ket{{}^1\text{S}_{0},\,F,\,M_{F}}$. We choose $V_{0,1}\ll V_{0,2},\,V_{0,3}$, so that the dynamics are effectively frozen in the directions $y$ and $z$, and the system can simulate our $(1+1)$-dimensional original problem.

We also consider the use of an additional laser beam in a traveling-wave configuration to drive two-photon Raman transitions between $\ket{\uparrow}$ and $\ket{\downarrow}$ by means of off-resonant couplings to states within an excited-state manifold $\ket{e}\in\{\ket{{}^3\text{P}_{1},\,F',\,M_{F}'}\}$, and will thus be referred to as the Raman beam. The wavevector of this extra laser beam will be denoted by $\bm{\tilde{k}}_4$, and its polarization by $\bm{\tilde{\varepsilon}}_4$. We choose them to satisfy $\bm{\tilde{k}}_4\cdot\bm{k}_1=0$ and $\bm{\tilde{\varepsilon}}_4\cdot\bm{\varepsilon}_1=0$ (e.g. $\boldsymbol{k}_1=k_1\textbf{e}_1,\boldsymbol{k}_4=k_4\textbf{e}_2$), adjusting the polarizations in a way that the respective selection rules allow for two-photon transitions between the two Zeeman sub-levels. By virtually populating the excited states $\{\ket{e}\}$, transitions between $\ket{\uparrow}$ and $\ket{\downarrow}$ can occur, which will be used to simulate the spin-dependent tunnelings of Figs.~\ref{fig:diractightbinding} {\bf (a)} and {\bf (b)}. We consider  large detunings from these excited states which, in addition to the narrow linewidth of these  transitions, allow to minimize the heating mechanism due to the residual spontaneous photon emission. Moreover, the additional ac-Stark effect induces a non-linear shift with respect to
the Zeeman quantum number, which can be exploited to set the laser-beam frequencies such that the two-photon processes only involve the states $\ket{\uparrow}$ and $\ket{\downarrow}$~\cite{https://doi.org/10.48550/arxiv.2109.08885}, provided that the Fermi gas has been polarized to one of those states using a preliminary stage of optical pumping. We note that, due to the traveling-wave configuration of the Raman beam, the two-photon processes that absorb a photon from the weaker standing wave and emit it in the Raman beam (and viceversa), lead to an optical Raman potential~\cite{Zhang_2018} with a  period that is twice the one of the optical-lattice potential
\begin{align}
\label{eq:Raman_potential}
V_{\rm Ram}(\boldsymbol{r})=\frac{\tilde{V}_{0}}{2}\cos({k_1}r_1)\ee^{\text{i}\left(\tilde{k}_4r_2-\Delta\omega t\right)}\sigma^{+}+\rm{h.c.},
\end{align}
where $\tilde{V}_0$ is the amplitude of the two-photon Raman transition, and we have introduced $\sigma^{+}=\ket{\uparrow}\bra{\downarrow}$, and $\Delta\omega=\omega_4-\omega_1$. When tuned close to the resonance, i.e. $\Delta\omega\approx\omega_0=E_{\uparrow}-E_{\downarrow}$, this term can drive the aforementioned spin-flip transitions by virtually populating the excited state. This process involves absorbing a photon from the traveling-wave Raman beam and re-emitting it to the standing wave, while simultaneously exciting the atom. However, since the period of the Raman potential~\eqref{eq:Raman_potential} is exactly twice the one of the optical-lattice potential~\eqref{eq:optical_lattice}, atoms standing in the minima of the latter see a vanishing Raman-beam intensity and, therefore, no local spin-flips are driven. Instead, only spin-flipping tunnelings are induced by the Raman potential, which can be used to generate spin-orbit coupling~\cite{Zhang_2018,PhysRevLett.110.076401,PhysRevLett.112.086401,PhysRevLett.113.059901}, as demonstrated in  landmark experiments with bosonic and fermionic gases~\cite{doi:10.1126/science.aaf6689,PhysRevLett.121.150401,doi:10.1126/sciadv.aao4748,https://doi.org/10.48550/arxiv.2109.08885}. In~\cite{PhysRevResearch.4.L042012,coldatoms2,PhysRevB.106.045147}, slight modifications of this Raman lattice  scheme were considered for the quantum simulation of relativistic Dirac QFTs in flat spacetimes with four-Fermi interactions. Let us now discuss how, using  conformal time~\eqref{eqmetric}, the quantum simulation of Dirac fields in an FRW spacetime describing an expanding universe can also be realised with specific modifications of the scheme.

The Hamiltonian field theory of the  ${}^{87}\text{Sr}$ Fermi gas reads, in second quantization~\cite{doi:10.1080/00018730701223200}, as follows
\begin{align}
\nonumber
\hat{H}=&\!\!\int\!\! {\rm d}^3\bm{r}\sum_{\sigma,\,\sigma '}{\hat{\Phi}_\sigma^{\dagger}}(\bm{r})  \left(-\frac{{\nabla}^2}{2m}+\bra{\sigma}V(\bm{r})\ket{\sigma'}\right)  {\hat{\Phi}_{{\sigma '}}(\bm{r})}\\
+&\!\!\int\!\! {\rm d}^3\bm{r}\!\!\int\!\! {\rm d}^3\bm{r'}\!\!\sum_{\sigma,\,\sigma'}\!\!{\hat{\Phi}_{\sigma}^{\dagger}}(\bm{r}){\hat{\Phi}_{\sigma'}^{\dagger}}(\bm{r'})\frac{2\pi a_s}{m}\nonumber\\\cdot&\delta(\bm{r}-\bm{r'}){\hat{\Phi}_{\sigma'}}(\bm{r'}){\hat{\Phi}_{\sigma}}(\bm{r}),
\label{eq:Ham_continuum}
\end{align}
where $\hat{\Phi}_\sigma^{\dagger}(\bm{r})$ and $\hat{\Phi}_\sigma(\bm{r})$ are  fermionic field operators that create/annihilate an atom at   position $\bm{r}$ in one of the two selected  Zeeman sub-levels $\sigma\in\{\uparrow,\downarrow\}$, $V(\bm{r})=V_{\rm latt}(\bm{r})+V_{\rm Ram}(\bm{r})$, $m$ is the mass of the atoms, and $a_s$ is the $s$-wave scattering length for collisions, which can be controlled via optical Feshbach resonances~\cite{Cazalilla_2014}. Although these can interfere with the $SU\!(N)$ symmetry of  interactions, we note that only a couple of levels is required  in this work, so there is no problem in controlling the scattering by admixing with excited states that  have a non-vanishing electronic angular momentum.

We want to map this Hamiltonian to the lattice model in Eqs.~\eqref{eq:naive_hamiltonian} and~\eqref{eq:Wilson_H}. First, in order to obtain a lattice description of the ultra-cold atomic model, we consider the regime of deep optical lattices by imposing $|V_{0,j}|\gg E_{R,j}$, where $E_{R,j}=k_{j}^2/2m$ is the recoil energy. In this limit, atoms are tightly confined to the minima of the cubic optical lattice, which forms a periodic crystal at $\bm{r}_{\bm{n}}^0=\sum_j \frac{\lambda_j}{2}(n_j+1/2)\bm{e}_j$, with $n_j\in\mathbb{Z}_{N_j}$, and $\lambda_j=2\pi/k_j$ being the laser wavelengths. A better description of this lattice model is obtained by working in the so-called Wannier basis~\cite{PhysRevLett.81.3108,PhysRevLett.89.220407}, which leads to the transformation
\begin{align}
\label{eq:wannier_basis}
    \hat{\Phi}_\sigma(\bm{r})=\sum_{\bm{n}}w(\bm{r}-\bm{r_n}^0)\hat{f}_{\bm{n},\,\sigma}.
\end{align}
Here, $w(\bm{r}-\bm{r_n^0})$ represents the Wannier function localized around $\bm{r_n}^0$, and $\hat{f}^{\phantom{\dagger}}_{\bm{n},\sigma},\hat{f}_{\bm{n},\sigma}^{\dagger}$ are dimensionless  creation-annihilation operators on the corresponding lattice site,  which satisfy a fermionic algebra $\{\hat{f}^{\phantom{\dagger}}_{\bm{n}\phantom{'}\!\!,\sigma},\hat{f}_{\bm{n}',\sigma'}^{\dagger}\}=\delta_{\bm{n},\bm{n}'}\delta_{\sigma,\sigma'}$. Using this basis, the Hamiltonian~\eqref{eq:Ham_continuum} can be split into a spin-conserving term and a spin-flipping one. Since the resulting microscopic parameters will be expressed as integrals of the Wannier functions, which are tightly confined to the lattice sites, we can treat the problem as a lattice model with only nearest-neighbor couplings. The spin-conserving terms of the Hamiltonian read
\begin{align}
\label{eq:tunneling_sc}
 \nonumber   \hat{H}_{\rm sc}=\sum_{\bm n}&\biggl(\sum_{\sigma,\,j}\left(-t_j \hat{f}_{\bm{n},\sigma}^{\dagger} \hat{f}_{\boldsymbol{ n}+{\bf e}_j,\sigma}^{\phantom{\dagger}}+\rm{h.c.}\right)\\&+U\hat{n}_{\bm{n},\uparrow}\hat{n}_{\bm{n},\downarrow}\biggl),
\end{align}
where $\hat{n}_{\bm{n},\sigma}=\hat{f}_{\bm{n},\sigma}^{\dagger} \hat{f}_{\bm{n},\sigma}$ is the number operator, and we have introduced $t_j$  as the tunneling strength along the ${\bf e}_j$ direction, and $U$ as the Hubbard interaction. The explicit expression of these parameters is obtained by performing the corresponding integrals of Wannier functions \cite{RevModPhys.80.885} and, assuming $\lambda_j=\lambda=\frac{2\pi}{k}\;\forall j$, they read
\begin{align}
\label{eq:Hubbard_param}
    t_j=\frac{4}{\sqrt{\pi}}E_{R}\left(\frac{V_{0,\,j}}{E_R}\right)^{\frac{3}{4}}\ee^{-2\sqrt{\frac{V_{0,\,j}}{E_R}}},\\U=\sqrt{\frac{8}{\pi}}k a_s E_R\left(\frac{\prod_{j}V_{0,\,j}}{E_R^3}\right)^{\!\!\frac{1}{4}}.
    \label{eq:Hubbard_param_interaction}
\end{align}
As aforementioned, by setting the optical-lattice depths $V_{0,\,1}\gg V_{0,\,2},\,V_{0,\,3}$, the tunnelings along the $y$ and $z$ directions become negligible compared to that along the $x$ direction, $t_1\gg t_2, t_3$. Moreover, we will  also consider that $t_1\gg U$, and neglect the effect of the Hubbard interactions. In the following, we will show how the effect of the additional Raman potential can modify the (1+1)-dimensional tight-binding model in a way that connects to the two different discretizations of Dirac fields in FRW spacetimes discussed in the previous section.

We now discuss how the spin-flipping tunnelings of Fig.~\ref{fig:diractightbinding} {\bf (a)} can be mediated by the Raman potential. One proceeds with the second-quantized Raman term~\eqref{eq:Ham_continuum} in a similar way, expanding the fields in the Wannier basis~\eqref{fig:diractightbinding}, and performing the corresponding overlap integrals. As advanced previously, since the Raman potential vanishes at each minima of the lattice, the contributions to  local spin-flipping terms vanish. To be more precise, the doubled period of the Raman potential makes the integrand of the corresponding Wannier integral  to be an odd function over a symmetric interval of integration, which must thus vanish.  The leading-order contributions are then nearest-neighbor laser-assisted tunnelings of strength $\tilde{t}$ along the direction of the standing wave which, simultaneously,  change the Zeeman sub-level. Let us now discuss the connection to the two possible discretizations:

{\it (i) Wilson-fermion scheme.--} In this case, we also need the spin-dependent tunnelings of Fig.~\ref{fig:diractightbinding} {\bf (b)}. The idea is to 
allow for a  detuning $\delta$ in the Raman beam, such that it drives slightly off-resonant two-photon transitions between $\ket{\uparrow}$ and $\ket{\downarrow}$, such that the beatnote frequency in Eq.~\eqref{eq:Raman_potential} is set to 
\beq
\label{eq:raman_detuning_wilson}
\Delta\omega=\omega_0-\delta,
\eeq
 with $\delta\ll\omega_0$. Moving into a rotating frame, the detuned Raman potential contributes to the lattice Hamiltonian with
\begin{align}
\nonumber
    \hat{H}_{\rm sf}&=\sum_{{n}_1}\tilde{t}\Big(\ee^{\ii\pi(n_1+1)}\Big(\hat{f}_{{n}_1,\uparrow}^{\dagger} \hat{f}_{{n}_1+1,\downarrow}^{\phantom{\dagger}}-\hat{f}_{{n}_1,\uparrow}^{\dagger} \hat{f}_{{n}_1-1,\downarrow}^{\phantom{\dagger}}\Big)\\&+{\rm H.c.}\Big)+\sum_{{n}_1}\frac{\delta}{2}\left(\hat{f}_{{n}_1,\uparrow}^{\dagger} \hat{f}_{{n}_1,\uparrow}^{\phantom{\dagger}}-\hat{f}_{{n}_1,\downarrow}^{\dagger} \hat{f}_{{n}_1,\downarrow}^{\phantom{\dagger}}\right).
    \label{eq:raman_tunneling}
\end{align}
In this expression,  the approximate form of $\tilde{t}$ after applying a Gaussian approximation around the minima of each optical-lattice minima  reads
\begin{align}
\tilde{t}=\frac{\tilde{V}_0}{2}\ee^{-\frac{\pi^2}{4}\sqrt{\frac{V_{0,\,1}}{E_R}}}.
\end{align}

The missing step is  that, after a $U(2)$ gauge transformation and a rescaling to obtain the correct units for the lattice field operators~\eqref{eq:spinor_lattice_field},  one can map the cold-atom creation-annihilation operators to those of the Dirac spinor field 
\begin{align}
\label{eq:gauge_transformation}
    \hat{f}_{{n}_1,\uparrow}^{\phantom{\dagger}}=\sqrt{a}\ee^{-\ii\pi n_1}\hat{\chi}_{{n}_1,u}^{\phantom{\dagger}},\hspace{1ex} \hat{f}_{n_1,\downarrow}^{\phantom{\dagger}}=\sqrt{a}\hat{\chi}_{n_1,d}^{\phantom{\dagger}},
\end{align}
where we have used the notation $\hat{\chi}_{n_1}=(\hat{\chi}_{n_1,u},\hat{\chi}_{n_1,d})^{\rm t}$ for the lattice spinor field. After this transformation,  one can see that the site-dependent phase of the Raman tunneling~\eqref{eq:raman_tunneling} disappears, and we get exactly the  nearest-neighbor tunneling used for the discretization of the Dirac kinetic term~\eqref{eq:naive_hamiltonian}. In addition, this transformation also affects the spin-conserving tunneling in Eq.~\eqref{eq:tunneling_sc}, turning it into a spin-dependent tunneling that can be mapped exactly onto the Wilson mass term~\eqref{eq:Wilson_H} when considering also the Raman  detuning. In summary, we recover  the lattice field theory in Eqs.~\eqref{eq:naive_hamiltonian} and~
\eqref{eq:Wilson_H} with the correspondence between parameters 
\begin{align}
a=\frac{1}{2\tilde{t}},\hspace{1ex}r=\frac{t_1}{\tilde{t}},\hspace{1ex}m\mathsf{a}(\eta)=-\frac{\delta}{2}-2t_1.
\label{eq:equivalences_coldatoms_GNW}
\end{align}

In comparison to previous schemes for the quantum simulation of Dirac fields in flat spacetimes~\cite{PhysRevResearch.4.L042012,coldatoms2,PhysRevB.106.045147}, we see that the required ingredients when using conformal time for  the Dirac fields under a FRW spacetime are exactly the same. One of the  differences of the mapping is that the transformation in Eq.~\eqref{eq:gauge_transformation} has been modified with respect to those of previous works~\cite{PhysRevResearch.4.L042012,coldatoms2,PhysRevB.106.045147}, which is a consequence of  the different convention of the metric signature. Additionally, this choice also changes the sign of the detuning in Eq.\eqref{eq:equivalences_coldatoms_GNW}, which will require using Raman beams that are blue-detuned with respect to the transition between the Zeeman sub-levels. Finally, the most important difference with respect to the quantum simulation of Dirac fields in flat spacetimes~\cite{PhysRevResearch.4.L042012,coldatoms2,PhysRevB.106.045147} is that simulating  the expansion of the FRW universe requires using a time-dependent detuning of the Raman beam $\delta\mapsto\delta(\eta)$. 

{\it (ii) Naive-fermion scheme.--}  Contrary to what would be expected, finding a quantum simulation scheme for the na\"{i}ve-fermion discretization~\eqref{eq:naive_hamiltonian} requires additional experimental complexity in comparison to the Wilson-fermion one. First of all, the spin-dependent tunneling that lead to the Wilson mass is no longer required. This tunneling can be inhibited~\cite{Jaksch_2003} by exploiting a linear gradient of the on-site energies which can be achieved by lattice acceleration, requiring a linear drift of the optical-lattice beams detuning with time~\cite{PhysRevLett.76.4508}, or the application of a  magnetic-field gradient. In both cases, the effective Hubbard model in Eq.~\eqref{eq:tunneling_sc} receives a correction
\beq
\hat{H}_{\rm sc}\mapsto \hat{H}_{\rm sc}+\sum_{n_1,\sigma}\delta\!\omega_{n_1,\sigma}\hat{f}_{{n}_1,\sigma}^{\dagger} \hat{f}_{{n}_1,\sigma}^{\phantom{\dagger}},
\eeq
where $\delta\!\omega_{n_1,\sigma}=n_1\Delta_{\sigma}$, and $\Delta_{\sigma}$ is the aforementioned gradient that can depend on the internal state if it arises from a magnetic field.
Provided that $|t_1|\ll \Delta_\sigma$, the tunneling mediated by the standing wave becomes energetically penalized, and can be neglected up to leading order. In general, this gradient can also inhibit the Raman-mediated tunneling. However, one can modify the Raman-beam frequency in Eq.~\eqref{eq:raman_detuning_wilson}, such that the Raman beams provide the required energy to overcome the gradient penalty during the tunneling. In particular, if one considers a spin-independent gradient $\Delta_\uparrow=\Delta_\downarrow$, and imposes
\beq
\label{eq:raman_detuning_naive}
\Delta\omega=\omega_0-\Delta -\delta,
\eeq
one obtains a direct mapping to the  na\"{i}ve-fermion discretization~\eqref{eq:naive_hamiltonian} with parameters
\begin{align}
a=\frac{1}{2\tilde{t}},\hspace{1ex}m\mathsf{a}(\eta)=-\frac{\delta}{2}.
\label{eq:equivalences_coldatoms_GN_naive}
\end{align}

\subsection{Conformal time in the laboratory}

In this subsection, we start with a small digression to emphasize the simplifications that arise from using conformal time to describe the properties of the Dirac field, specially in light of the requirements for its quantum simulation. Since quantum simulators require a Hamiltonian formulation, one must be aware  that the canonical quantization of Dirac fields in non-static curved spacetimes can present important subtleties~\cite{PhysRevD.79.024020}. In particular, if we follow the standard quantization route discussed for conformal time  below Eq.~\eqref{eq:Dirac_conformal_time_lagrangian}, but considering now the cosmological time $t$~\eqref{eq:conformal_time}, we would get a canonical momentum ${\Pi}_{\psi}(x)=-\sqrt{-g}\overline{\psi}(x)\tilde{\gamma}^0(x)=\ii\mathsf{a}(t)\psi^{\dagger}(x)$ and, upon quantization of the field and its canonical momentum, arrive directly to the Hamiltonian field theory
\beq
\hat{H}=\int\!{\rm d}{\rm x}\hat{\psi}^{\dagger}\!({\rm x})\!\left(\gamma^5\partial_{\rm x}-\ii m\mathsf{a}(t)\gamma^0+\ii\partial_t\mathsf{a}(t)\right)\hat{\psi}({\rm x}).
\label{eq:non_rescaled_ham}
\eeq
In this expression, one clearly notices that  the last term in brackets leads to a non-Hermitian  operator. 
As discussed in~\cite{PhysRevD.79.024020}, this non-Hermitian contribution is generated precisely by the spin connection~\eqref{eq:spin_connection}, and  is  a generic consequence of  the covariant derivative of fermionic problems in non-static metrics such as the FRW spacetime. Accordingly,  one should be more careful in defining a correct process of canonical quantization, which entails a rescaling of the above field  operators $\hat{\psi}({\rm x}),\hat{\psi}^{\dagger}\!({\rm x})$~\cite{Minar_2015}. This leads to more complicated Hamiltonians that, upon a naive-fermion discretization similar to Eq.~\eqref{eq:naive_hamiltonian}, are  described by tight-binding models with both spin-conserving and spin-flipping tunnelings. The strengths of these tunnelings   will generally depend on the spacetime coordinates~\cite{Minar_2015}. In our specific situation, the  spin-conserving and spin-flipping tunnelings would  depend on the scale factor $\mathsf{a}(t)$, and thus become time-dependent. 

On a technical level, one would have to implement the specific time dependence on both type of tunnelings by modulating the intensity of the optical- and Raman-lattice potentials, which contrast with the  simpler modulation of the detuning that is required when working with conformal time~\eqref{eq:raman_detuning_naive}. On a more fundamental level, we see that the na\"ive discretization using cosmological time already requires a combination of  both spin-conserving and spin-flipping tunnelings, whereas only spin-flipping tunnelings where required in the conformal-time case.~\eqref{eq:naive_hamiltonian}. As a consequence, it is not clear how one would proceed to get a Wilson-fermion discretization which, in the conformal-time case, exploited a momentum-dependent mass term~\eqref{eq:WIlson_mass} that comes from spin-conserving tunneling processes as well. Altogether, performing a cosmological-time quantum simulation of fermion production in the boundary of the FRW spacetime would considerably increase the complexity of the scheme and, quite likely, forbid a direct  application of the Raman-lattice toolbox that is currently being used in several experiments~\cite{doi:10.1126/science.aaf6689,PhysRevLett.121.150401,doi:10.1126/sciadv.aao4748,https://doi.org/10.48550/arxiv.2109.08885}. 

According to our proposal, when performing the quantum simulation, we should interpret the real time of the experiment as representing the conformal time. We recall  that  the conformal time is negative for the de Sitter expansion $(-\infty<\eta<0)$, but the lapse between an initial instant  $\eta_0$ and a final one  $\eta_{\rm f}$ is actually positive.  We should  then  identify the real time of the experiment, which starts as $t=0$,  with the conformal-time lapse  $t=\eta-\eta_{0}$ during which we want to simulate a period of expansion of the universe, leading to a bare mass that increases from its initial value $m\mathsf{a}(\eta_{0})$ to $m\mathsf{a}(\eta_{\rm f})$. This mass  is controlled by the detuning of the Raman beam, and its explicit expression is that of \eqref{eq:equivalences_coldatoms_GNW} for Wilson fermions and \eqref{eq:equivalences_coldatoms_GN_naive} for na\"{i}ve fermions. Accordingly, we must tune the Raman detuning as a function of the experimental time by simply shifting the
desired profile that is set by the scale factor 
\begin{align}
    \delta(t)=
  \left\{ {\begin{array}{ll}
{-}2m\mathsf{a}(t+\eta_0),& {\rm for\, \, \text{na\"{i}ve} \, \,fermions} \\
-4t_1{-}2m\mathsf{a}(t+\eta_0) ,& {\rm for \, \,Wilson \, \,fermions}  \\
  \end{array} } \right.. 
    \label{eqn:evolution_raman_detuning}
\end{align}
In both cases, the time-dependence of the Raman-beam detuning changes in time according to  $\Delta\delta(t)=2m\mathsf{a}(t-\eta_0)$, which depends on the bare mass $m$ and is proportional to the shifted  scale factor $\mathsf{a}(t-\eta_0)$. We recall that, in the numerical simulations, we have considered a particularly-smooth adiabatic switching~\eqref{eq:scalefactor},  but other simpler profiles can also be explored in the laboratory. Finally, in the case of Wilson fermions, the detuning will also incorporate a static part  $\delta_0=4t_1$, which depends on the recoil energy  and  the optical lattice depth~\eqref{eq:Hubbard_param}. 

In summary, the dynamics of a Dirac field in a de Sitter phase of expansion can be simulated through a specific time-dependent control of the  Raman-beam detuning. In recent experiments with Fermi gases in two-dimensional Raman lattices~\cite{https://doi.org/10.48550/arxiv.2109.08885},  the consequences of  changing the value of this  detuning in the quench dynamics of the fermions has been explored, which can actually be used to infer the value of a topological invariant that captures the essence of an SPT groundstate, and even pinpoint the appearance of topological phase transitions. For the situation studied in our work, it is not sufficient to change the value of the detuning prior to the time evolution, but one rather needs to change it as time evolves. This allows to connect to the physics of quantum fields in expanding spacetimes, and could allow for a direct observation of the gravitational analogue of particle production, including the gravitational creation of topological modes in the spacetime boundary. 

It should be noted that we have considered that the Hubbard interactions  $U$~\eqref{eq:Hubbard_param_interaction}  vanish via optical Feshbach resonances, since we have focused on a free LFT in a cosmological background. However, the same experimental scheme opens the path for the study of different interacting QFTs in a variety of background metrics. For instance, one of the most direct extensions of our model would be the Gross-Neveu model~\cite{PhysRevD.10.3235}, which only differs from our original Hamiltonian by a four-Fermi interaction term that connects to the Hubbard interactions in the case of two-component Dirac spinors. This interaction term can be switched on by increasing the value of the scattering length, and would allow us to study non-perturbative phenomena such as chiral symmetry breaking or dynamical mass generation, and their interplay with particle production.

\subsection{Measurement and fermion  production}
\label{subsec:measurements}

The remaining ingredient for the quantum simulation of fermion production in an expanding spacetime  is  to discuss how to measure the key observables (i.e. the spectrum of produced particles) in real time. To do so, one can take advantage of the various   detection methods in ultra-cold atoms~\cite{Bloch_2008}, such as  the so-called time-of-flight (TOF) measurements, and the band-mapping technique. In TOF measurements~\cite{Ketterle2008}, one abruptly removes all the applied fields that trap the atoms, letting the gas to expand freely. After this sudden turn off, if the atoms expand  ballistically, there is a relation between their initial momentum and their final position $\hbar \mathbf{k}=M\mathbf{x}/t$. Accordingly,  after a certain time, absorptive imaging is used to obtain the spatial distribution $n(\mathbf{x})$ of the atoms, which gives information about the momentum distribution prior to the release. In this absorptive imaging, photons from an incoming resonant laser are absorbed by the atoms, which consequently cast a shadow that can be  recorded by a CCD-camera, giving one access to the so-called columnar -integrated- density. This technique can also be done in a spin-resolved manner by using laser beams with different frequencies, addressing each of the internal states. 
On the other hand, the band-mapping technique \cite{2001PhRvL..87p0405G,Kohl2005,1995PhRvL..74.1542K} turns off the external fields adiabatically, such that the band structure of the many-body system is slowly transformed into a free-particle dispersion relation. During this ramping-down, the quasi-momentum is approximately conserved, and Bloch states on the $n$th band are mapped onto free states with linear momentum on the $n$th Brillouin zone, giving thus direct access to the population of the different bands prior to the ramping down of the external fields and to its quasi-momentum distribution.

Following the discussion in the previous sections, the quantum simulator can provide a gravitational-analogue for particle creation by considering an initial half-filling condition that first populates the lower band of the lattice models. Then,  a period of de Sitter expansion between two asymptotic flat vacua is simulated by  the real-time evolution of the Fermi gas loaded in the Raman lattice with a time-dependent Raman detuning~\eqref{eqn:evolution_raman_detuning}. During the expansion, not only the state of the Fermi gas  will change, but also the Hamiltonian itself, and consequently the band structure of the system. Hence, if a fermionic  atom is initially in a specific  Bloch state, after the expansion, it will generally be left in a superposition of Bloch states corresponding to  the two different bands, which is the analogue of Eq.~\eqref{eqn:evolutionb} provided that a particle-hole transformation is applied. The number of  produced particles for a certain quasi-momentum is then given by the probability for an atom with a certain quasi-momentum to get excited to the higher band. The $|\beta_{\rm k}(t+\eta_0)|^2$ coefficient can be thus obtained from this probability. The details of this result can be found in Appendix \ref{app_d}. Since we can measure the population of each band for each quasi-momentum with the band-mapping technique \cite{2012Natur.483..302T}, this Bogoliubov coefficient can thus be  accessed experimentally. Going further, combining the band-mapping technique with spin-resolved measurements~\cite{PhysRevLett.107.235301}, one could even  characterize the band topology of the expanding lattice field theory~\cite{2019RvMP...91a5005C}. 

\section{\bf Conclusions and  outlook}
\label{sec:conclusions}

In this work, we have developed the theory of particle production for a fermionic Dirac field in a (1+1)-dimensional FRW spacetime, both from the perspective of the usual continuous theory and for two types of discretizations on the lattice. We have shown that the phenomenon of fermion production for a de Sitter expansion admits an exact solution in terms of a pair of decoupled Bessel differential equations. To avoid problems with the interpretation of the instantaneous vacua, we added an adiabatic switching that connects the de Sitter expansion to a pair of asymptotic Minkowski spacetimes, both of which have a well-defined notion of vacuum. We have numerically shown that the extra switching periods do not modify the particle production, which can still be  described in terms of the analytically-solvable  Bessel equation. 

To pave the way for a quantum simulation of this phenomenon, we have considered two possible lattice discretizations of the Hamiltonian field theory associated to  Dirac fermions in a curved spacetime: a na\"{i}ve- and a Wilson-fermion discretization, which allow us to discuss universes with spatial boundaries. Focusing on the bulk of the lattice, we have shown that the  na\"{i}ve-fermion discretization reproduces the spectrum of particle production predicted by the continuum QFT  for small momenta. However, as one approaches the edge of the Brillouin zone, the phenomenon of fermion doubling leads to a mirror-image of the spectrum, resulting in a doubled number of the produced fermions with respect to the continuum prediction. Turning into the Wilson-fermion discretization, which sends the fermion doubler to the ultra-violet cutoff of the QFT, we have shown that the spectrum and total number of particles produced at the bulk matches nicely the continuum expressions. 

On the other hand, the asymptotic vacua of the Wilson-fermion lattice field theories can actually correspond to a couple of  SPT groundstates characterized by a non-zero topological invariant, which has a boundary correspondence in terms of the groundstate degeneracy by the appearance of topological zero-energy modes  localized to the spatial boundaries of the FRW universe. 
This has allowed us to study the role of topology in the phenomenon of particle production and, in particular, to show that there can be gravitational particle production of zero-energy fermions localized to the edges of the spacetime. A similar phenomenon would also occur in situations where the mass of the Dirac fermion has some solitonic profile that changes its sign, leading to particle production in the form of domain-wall fermions. By numerically solving a set of coupled differential equations for the case of open boundary conditions, we have calculated the spectrum of fermion production, which is now a function of the energy of the produced particles as momentum is no longer a good quantum number in the absence of translational symmetry. We have shown that this spectrum allows one to identify clearly the production of fermions at the boundary of the expanding FRW universe, as one finds production of particles for energies below the mass gap. 

To conclude, we have shown that current experiments of ultra-cold alkaline-earth Fermi gases in Raman optical lattices would be an ideal platform for the quantum simulation of this curved quantum field theories, provided that one exploits the simplifications that arise when  working with conformal time. In connection to those experiments, our model requires to simulate the expanding spacetime by encoding its effect in a time-dependent mass that depends on the scale factor of the expansion, which corresponds in the experiment to changing the Raman-beam detuning as a function of time. This would allow for an experimental simulation of fermion production in a FRW  spacetime, including its interplay with topology and the boundaries of such an effective universe.

\acknowledgements
We thank S. Hands and C. Sab\'in for useful discussions. We acknowledge support from PGC2018-099169-
BI00 (MCIU/AEI/FEDER, UE), and PID2021- 127726NB-I00 (MCIU/AEI/FEDER, UE), from the Grant IFT Centro
de Excelencia Severo Ochoa CEX2020-001007-S, funded by MCIN/AEI/10.13039/501100011033, from the grant QUITEMAD+ S2013/ICE-2801, and from the CSIC Research Platform on Quantum Technologies PTI-001.

\appendix

\section{\bf \label{app:dirac_qftcs} Quantum field theories of Dirac fermions in  a $(d+1)$-dimensional curved spacetime}

In this Appendix, we review the formulation of Dirac QFTs in a  curved spacetime of $D=d+1$ dimensions. One typically starts from the flat-spacetime limit, in which the events are described by $D$-vectors $x=(t,\boldsymbol{x})$ in a Minkowski spacetime of metric $g_{\mu\nu}(x)\mapsto\eta_{\mu\nu}={\rm diag}(-1,+1,\cdots,+1)$, where $\mu,\nu\in\{0,1,\cdots,d\}$ label the spacetime coordinates. Here,  we have chosen the mostly-plus metric~\cite{weinberg_1995}, which is customarily used in treatments of general relativity~\cite{carroll_2019}. The underlying Poincar\'e symmetry, which consists of spacetime translations and Lorentz transformations, has a specific representation for Dirac fermions that requires  introducing spinor fields $\psi(x)$ and the so-called gamma matrices $\gamma^{\mu}$, which obey a Clifford algebra $\{\gamma^\mu,\,\gamma^\nu\}=2\eta^{\mu\nu}$. Note that the choice of the  mostly-plus  metric exchanges the Hermitian (anti-Hermitian) nature of the temporal (spatial) gamma matrices with respect to  the  mostly-minus metric $\hat{\gamma}^\mu$, which is the typical choice  in particle physics~\cite{Peskin:1995ev}. Both matrices can be related by a simple global phase $\gamma^\mu=\ii\hat{\gamma}^\mu$. In this manuscript, we stick to the mostly-plus metric in which the action of massive  Dirac fields, which  corresponds to the simplest field bi-linear that is a scalar under the Poincar\'e group~\cite{freedman_12},  reads as follows
\begin{equation}
\label{eq:Dirac_action}
    S=\int\!\!{\rm d}^Dx\overline{\psi}(x)\left(-\eta^{\mu\nu}\gamma_\mu{\partial}_{\!\nu}+m\right)\psi(x).
\end{equation}
Here, $\overline{\psi}(x)=\ii\psi^{\dagger}\!(x)\gamma^0$ is the adjoint field, which again differs with the standard choice for the mostly-minus metric~\cite{Peskin:1995ev}, $\partial_\mu=\partial/\partial x^\mu$, and $m$ is the bare mass. In this section, we use  natural units $\hbar=c=1$ and the repeated-index summation.

 When  considering curved spacetimes, the metric is not necessarily flat, and one needs to exchange $\eta_{\mu\nu}\mapsto g_{\mu\nu}(x)$ in the action~\eqref{eq:Dirac_action}. In addition, the integral measure must now include the volume form of the Lorentzian manifold associated to the curved spacetime. Therefore, one must substitute  ${\rm d}^Dx\mapsto{\rm d}^Dx\sqrt{-g}$, where $g={\rm det}(g_{\mu\nu}(x))$, such that different regions of spacetime are    weighted in a way that is invariant under diffeomorphisms.   Finally, the partial derivatives in Eq.~\eqref{eq:Dirac_action}, which connect  fields defined on nearby spacetime points of the flat Minkowski spacetime, must also be generalized in the presence of curvature,  which leads to the covariant derivative $\partial_\mu\mapsto\nabla_{\!\!\mu}$. For Dirac fields, the covariant derivative  contains the generators  of Lorentz transformations $x\mapsto x'=\Lambda x$ on the spinor fields $ \psi(x)\mapsto S[\Lambda]\psi(\Lambda^{-1} x)$, where 
 \beq
 \label{eq:lorentz_spinors}
S[\Lambda]={\ee}^{-\half\Omega^{ab}S_{ab}},\hspace{2ex} S_{ab}=\fourth[\gamma_a,\,\gamma_b].
 \eeq
  Here,   $\Omega^{ab}$ parametrizes the specific Lorentz transformation, where we note that the changes with respect to Ref.~\cite{Peskin:1995ev} are due to the choice of the mostly-plus metric, and the consequent change in the metric and gamma matrices. To connect nearby spinor fields, the covariant derivative also contains a correction due to the spin connection $ \omega^{ab}_{\mu}(x)$, which leads to 
\beq
\label{eq:cov_derivative}
\nabla_\mu\psi(x)=(\partial_\mu+\omega_\mu\!(x))\psi(x),
\eeq
where we have introduced a connection field 
\beq
\label{eq:connection_field}
\omega_\mu\!(x)=\omega^{ab}_{\mu}(x)S_{ab}.
\eeq

The specific form of the spin connection is found by introducing $D$-$beins$~\cite{doi:10.1063/1.1703702,covder}, which form a basis of vector fields with components $e^\mu_a(x)$ that allows us to express the curved metric in terms of the  flat Minkowski one~\cite{Birrell, Mukhanov,Parker}, namely
\beq
\label{eq:vielbein}
 g_{\mu\nu}(x)= \eta_{ab}e_\mu^a(x) e_\nu ^b(x).
 \eeq 
On the one hand, $D$-$beins$ allow one to  generalise the Clifford algebra to the curved spacetime $\gamma^\mu\mapsto\tilde{\gamma}^{\mu}(x)$ 
 \beq
 \{\tilde{\gamma}^\mu(x),\tilde{\gamma}^\nu(x)\}=2g^{\mu\nu}(x),
 \eeq
   where the new set of curved gamma matrices  is
 \beq
 \label{eq:curved_gammas} 
 \tilde{\gamma}^\mu\!(x)=e_a^\mu(x)\gamma^a.
 \eeq
 On the other hand, $D$-$beins$ also play a key role in  the spin connection. Making use of  the Christoffel symbols 
 \beq
 \label{eq:christoffel}
 \Gamma^\nu_{\tau\mu}(x)=\half g^{\nu\sigma}(x)(\partial_\mu g_{\sigma\tau}(x)+\partial_\tau g_{\sigma\mu}(x)-\partial_\sigma g_{\tau\mu}(x))
 \eeq
  for  the curved spacetime, the spin connection becomes 
     \beq
     \label{eq:spin_connection}
     \omega^{ab}_{\mu}(x)=-\frac{1}{2}\!\big(e^{\tau a}(x)e^b_{\nu}(x)\Gamma^{\nu}_{\tau\mu}(x)-e^{\tau a}(x)\partial_\mu e_{\tau}^{ b}(x)\big).
     \eeq
      Here, the overall minus sign with respect to Ref.~\cite{covder} comes from the different signature of the metric in the mostly-plus and -minus conventions. Note that, as customary in the context of curved Dirac fields,  latin indexes $a,\,b$ are raised (lowered) via the flat metric $\eta^{ab}(\eta_{ab})$, whereas greek ones $\mu,\nu$ require  $g^{\mu\nu}(x) (g_{\mu\nu}(x))$.

Equipped with all these tools,   Dirac fermions in curved spacetimes can be finally described by the  action
\begin{equation}
\label{eq:Dirac_action_curved}
    S=\!\!\int\!\!{\rm d}^Dx\sqrt{-g}\,\overline{\psi}(x)\!\left(-g^{\mu\nu}(x) \tilde{\gamma}_\mu\!(x){\nabla}_{\!\nu}+m\right)\!\!\psi(x).
\end{equation}

\section{\bf Dynamical gravity in $(1+1)$ dimensions \label{app:jt_gravity}}

 In this Appendix, we review two formalisms that allow for a dynamical description of gravity in $(1+1)$ dimensions, where Einstein's equations are not dynamical. There exist different alternative formulations for a consistent theory of gravity in $D=1+1$ dimensions, such as the so-called Jackiw-Teitelboim model~\cite{Teitelboim:1983ux,Jackiw:1984je}, which reproduces several phenomena characteristic of the higher-dimensional Einstein gravity.  This model  can also incorporate source terms that induce  spacetime curvature, provided that one adds a term proportional to the trace of the stress-energy tensor in the classical field equations~\cite{Mann:1989gh,AESikkema1991}. The solution of these classical field equations determines the corresponding metric $g^{\mu\nu}(x)$, and can lead to certain analogues of Einstein's gravity, including gravitational collapse, black-hole physics and, more relevant for the topic of this work, analogues of matter- and radiation-dominated FRW spacetimes~\cite{AESikkema1991}. We recall that, in this manuscript, we are concerned with a semi-classical approach that neglects back-action and treats the metric and the spin connection as  classical background fields. At this level, we can use the solutions of JT gravity, and calculate their effect on the properties of the low-dimensional Dirac field.  
 We thus avoid the need of treating the Dirac fields in conjunction with the auxiliary dilaton fields that allow for a covariant action of JT gravity~\cite{Jackiw:1984je,Brown:1988am}.

     Another possibility is to consider that the low-dimensional  QFT arises as an effective field theory in situations in which the  fermions  are forced to propagate only along a two-dimensional section of an underlying curved  spacetime in  $D=3+1$ dimensions~\cite{Sabin_2018,PhysRevD.99.025008}. This can be achieved, for instance,  by considering an anisotropic  mass of the fermion  that is very large along the two remaining spatial directions. Alternatively, one may consider solitonic profiles for the mass, which connect to the aforementioned domain-wall constructions~\cite{Kaplan:2009yg,KAPLAN1992342,GOLTERMAN1993219,10.1143/PTP.73.528}, and can also be used to impose constraints on the propagation of the fermions.    From this  perspective, one can  use the specific time-dependence of the scale factor in a vacuum-dominated FRW spacetime in four dimensions, which are determined by the Friedmann equations~\cite{carroll_2019}, and incorporate it in the $(1+1)$-dimensional QFT of Dirac fermions in a curved metric. Note, however, that the number of spinor components is four in  $D=3+1$, whereas one works with two-component spinors in $D=1+1$. Luckily, when considering a two-dimensional section of the FRW spacetime~\eqref{eq:line_element}, the required gamma matrices and the resulting connection field respect a block structure, such that the 4 spinor components decouple in two disconnected pairs that evolve in time independently. Accordingly, one can use a QFT with two-component Dirac spinors in the FRW spacetime with a time-dependence of the scale factor determined by the higher-dimensional Einstein equations. Both approaches lead to the same de Sitter expansion. We remark that this does not occur for generic metrics and stress-energy tensors.

\section{\bf Analytical solution for a  de Sitter  expansion}
\label{app:exact_solution}

In this Appendix, we present the details of the derivation of Eq.~\eqref{solutions_Hankel}.
From the single-particle Dirac Hamiltonian (\ref{hamdS}), we have 
\begin{subequations}
\begin{align}
&\mathrm{i}\dot{u}_1=\frac{m}{H\eta}u_1-\mathrm{i}\mathrm{k}u_2,\label{eom1}\\
&\mathrm{i}\dot{u}_2=\mathrm{i}\mathrm{k}u_1-\frac{m}{H\eta}u_2\label{eom2},
\end{align}
\end{subequations}
where we omit the subscript $\rm k$ in the mode functions to ease the notation, and use dots for the derivatives with respect to conformal time.
Taking derivatives on the first equation again 
\begin{equation}
    \mathrm{i}\ddot{u}_1=-\frac{m}{H\eta^2}u_1+\frac{m}{H\eta}\dot{u}_1-\mathrm{i}\mathrm{k}\dot{u}_2,
    \label{33}
\end{equation}
such that one can now insert Eq.(\ref{eom2}) in (\ref{33}) to get
\begin{equation}
    \mathrm{i}\ddot{u}_1=-\frac{m}{H\eta^2}u_1+\frac{m}{H\eta}\dot{u}_1-\mathrm{i}\mathrm{k}^2u_1+\frac{mk}{H\eta}u_2.
    \label{34}
\end{equation}

We can find $u_2$ as a function of $u_1$ from Eq.~(\ref{eom1})
\begin{equation}
    u_2=\frac{1}{\mathrm{k}}\left(-\dot{u}_1-\mathrm{i}\frac{m}{H\eta}u_1\right),
    \label{35}
\end{equation}
which, after substitution  in Eq.~(\ref{34}), leads to
\begin{equation}
    \ddot{u}_1+\left(\mathrm{k}^2+\frac{m^2}{H^2\eta^2}-\mathrm{i}\frac{m}{H\eta^2}\right)u_1=0.
\end{equation}

Repeating this with $u_2$ we arrive at an analogous  equation
\begin{equation}
    \ddot{u}_2+\left(\mathrm{k}^2+\frac{m^2}{H^2\eta^2}+\mathrm{i}\frac{m}{H\eta^2}\right)u_2=0.
\end{equation}

We can solve these equations by making the replacement $u_{1,2}(\eta)=\sqrt{\eta} f_{1,2}(\eta)$. The functions $f_{(1,2)}$ satisfy the ODEs
\begin{align}
    &\eta^2\ddot{f}_1+\eta\dot{f}_1+\left(\mathrm{k}^2\eta^2-\left(\half+\mathrm{i}\frac{m}{H}\right)^2\right)f_1=0,\\&\eta^2\ddot{f}_2+\eta\dot{f}_1+\left(\mathrm{k}^2\eta^2-\left(\half-\mathrm{i}\frac{m}{H}\right)^2\right)f_2=0.
\end{align}
Finally, after performing the change of variable $\eta\rightarrow z=\mathrm{k}\eta$, we arrive at
\begin{align}
    &z^2f''_1+zf'_1+\left(z^2-\left(\half+\mathrm{i}\frac{m}{H}\right)^2\right)f_1=0,\label{41}\\\label{42}&z^2f''_2+zf'_1+\left(z^2-\left(\half-\mathrm{i}\frac{m}{H}\right)^2\right)f_2=0,
\end{align}
where the prime denotes differentiation with respect to $z$. We have finally  arrived at a pair of decoupled   Bessel equations for each of the spinor components. Defining  $\nu_+=\half+\mathrm{i}\frac{m}{H}$ for (\ref{41}) and $\nu_-=\half-\mathrm{i}\frac{m}{H}$ for (\ref{42}), a well-known basis of solutions  is that of the so-called Hankel functions~\cite{gradshteyn2007}
\begin{align}
\label{eq:solution}
    &f_1=C_{(1,1)}H^{(1)}_{\nu_+}(\mathrm{k}\eta)+C_{(1,2)}H^{(2)}_{\nu_+}(\mathrm{k}\eta),\\
    \label{eq:solution_2}
    &f_2=C_{(2,1)}H^{(1)}_{\nu_-}(\mathrm{k}\eta)+C_{(2,2)}H^{(2)}_{\nu_-}(\mathrm{k}\eta).
\end{align}
We choose this basis instead of, for example, the more commonly used Bessel functions of the first- $J_{\nu}(z)$ and second-kind $Y_{\nu}(z)$, because of their asymptotic behaviour. In particular, the Hankel functions behave asymptotically  as plane waves for $z\rightarrow\infty$, which will be useful for us when connecting to the vacuum of the remote past. As a consistency check, let us note that our solution can be shown to be  equivalent to the one found by V. M. Villalba~\cite{victorvillalba}, which was expressed in terms the so-called cylinder functions. Although not apparent at first sight, this equivalence can be proven by using the properties of the Hankel functions $H^{(1)}_{-\nu}(z)=\ee^{\ii\nu\pi}H^{(1)}_\nu(z)$, $H^{(2)}_{-\nu}(z)=\ee^{-\ii\nu\pi}H^{(2)}_\nu(z)$, and their relation to the aforementioned cylinder functions $Z_\nu(z)=aH^{(1)}_{\nu}(z)+bH^{(2)}_{\nu}(z)$,  $Z_{\nu-1}(z)=cH^{(1)}_{\nu-1}(z)+dH^{(2)}_{\nu-1}(z)$, where $a,\,b,\,c,\,d$ are related to the constants  of integration $C_{(i,\,j)}$ in~\eqref{eq:solution}-\eqref{eq:solution_2}.

\section{\bf Particle production for  infinitely-long expansions}
\label{app_b}

In this Appendix, we present the details of the derivation of the Bogoliubov parameter in Eq.~\eqref{eqn:thermic}.
Let us  begin from the limiting form of the instantaneous eigenstates  $\mathsf{v}_k^-(\eta_{\rm out}\rightarrow0^-)=(0,\,1)^{\rm t}$, and use the asymptotic solution of the Dirac spinor in Eq.~(\ref{eqn:solutioninfpast}). We have that $|\beta_{\mathrm{k}}(\eta)|^2=|u_k^{\rm t}(\eta)\cdot\mathsf{v}_k^{-}(\eta)|^2=|C_{\mathrm{k}}\sqrt{\eta}\ee^{\mathrm{i}\pi\nu_{-}}H_{\nu_{-}}^{(1)}(k\eta)|^2$, which is only valid when $m\neq0$ and for $\eta\approx0^-$, since we used the limit $\eta\rightarrow0^{-}$. Let us recall that $C_{\mathrm{k}}=\frac{\sqrt{\pi \mathrm{k}}}{2}\ee^{-\mathrm{i}(\mathrm{k}\eta_{\rm in}+\frac{\pi\nu_{-}}{2}-\frac{\pi}{4})}$, and that the expansion of the Hankel function is $H_{\nu}^{(1)}(z)\to-\frac{1}{\pi}\Gamma\left(\nu\right)\left(\half z\right)^{-\nu}$ as $z\to 0$. Using this, and substituting $\nu_{-}=\half-\ii\frac{m}{H}$, we find
\begin{align}
    |\beta_{\rm k}(\eta)|^2=\frac{1}{2\pi}e^{-\frac{\pi m}{H}}\left|\Gamma\left(\half-\mathrm{i}\frac{m}{H}\right)\right|^2\theta(m),
\end{align}
where $\theta(m)$ is the Heaviside's step function and takes account for the requirement $m\neq0$.
Using the property of the gamma function $\left|\Gamma\left(\half+\mathrm{i}b\right)\right|^2=\frac{\pi}{\text{cosh}(\pi b)}$, we finally arrive at
\begin{align}
    |\beta_{\rm k}(\eta)|^2&=\frac{e^{\frac{-\pi m}{H}}}{2\pi}\frac{\pi}{\text{cosh}\left(\frac{\pi m}{H}\right)}\theta(m)=\frac{1}{1+\ee^{\frac{2\pi m}{H}}}\theta(m),
\end{align}
which is the result that we stated in Eq. (\ref{eqn:thermic}).
Our derivation differs from that found in \cite{victorvillalba}. There, no restriction on the mass was assumed, which would imply particle production in the massless conformal limit $m=0$. Additionally, our derivation  also differs from that of \cite{Haro_2008}, which was only valid for $m/H\gg0$. We note that, for the massless case $m=0$, the previous asymptotic eigenvector  is no longer valid, and must be substituted by $\mathsf{v}_{\mathrm{k}}^-(\eta\rightarrow0^-)=\left(\mathrm{i},\,1\right)^{\rm t}/\sqrt{2}$. The corresponding Bogoliubov parameter vanishes, as it should be, since in the massless case the fermionic field is conformally coupled, and hence there is no particle production. 
  
  \section{\bf Mode expansion and \mbox{Bogoliubov} transformations}
  \label{app_c}
  
  In this Appendix, we present some details on the mode expansion of the fermionic field~\eqref{eq:normalizedmodeexpansion}, and its relation with the   particle creation and Bogoliubov transformations. 
Working within the Heisenberg picture for the rescaled field operator $\hat{\chi}(\eta,\,x)$, the time evolution dictated by  the Hamiltonian~\eqref{eq:Dirac_hamiltonian_curved_rescaled} is given by  the Heisenberg equation
\begin{align}
    \ii\frac{d}{d\eta}\hat{\chi}(\eta,\,{\rm x})=-\left[\hat{H}(\eta),\,\hat{\chi}(\eta,\,{\rm x})\right]. 
    \label{eqn:app_Heisenberg_eq}
\end{align}

The effect of particle production arises when expanding the field operator in modes. As discussed below Eq.~\eqref{eqn:decompositionmodefunctions} of the main text, in flat spacetimes and for inertial observers, there is a natural choice of these modes as positive- and negative-frequency plane waves. Conversely, in curved spacetimes, this choice is not unique, and  must be discussed in detail. In our case, we choose the convention of  \cite{Grib:1976pw, Shtanov:1994ce,Haro_2008}, which amounts to taking these modes as the instantaneous eigenstates of the single-particle Hamiltonian, ${\rm v}^\pm_{\rm k}(\eta)$, such that
\begin{align}
    \hat{\chi}(\eta,\,{\rm x})=\int\frac{d{\rm k}}{2\pi}\left(\hat{a}_{\rm k}(\eta) {\rm v}^{+}_{\rm k}(\eta)+\hat{b}_{\rm -k}^\dagger(\eta){\rm v^-_{\rm k}}(\eta)\right)\ee^{\ii{\rm k} {\rm x}}.
    \label{eqn:app_diagon_method}
\end{align}
This choice diagonalizes the Hamiltonian field theory
\begin{align}
\hat{H}=\int\frac{d{\rm k}}{2\pi}\;\omega_k(\eta)\left(\hat{a}_{\rm k}^\dagger(\eta)\hat{a}_{\rm k}(\eta)-\hat{b}_{\rm -k}(\eta)\hat{b}_{\rm -k}^\dagger(\eta)\right)e^{\rm ikx},
\end{align}
and is thus known as the diagonalization method, which has been  widely developed for both the scalar~\cite{Grib:1976pw, Shtanov:1994ce} and the fermionic cases~\cite{Haro_2008}. Within this method, the instantaneous  groundstate $|0_{\star}\rangle$ at a certain time $\eta=\eta_{\star}$ is the state that is annihilated by the annihilation operators defined by the expansion \eqref{eqn:app_diagon_method}. The time-dependence of the operators $\hat{a}_{\rm k}(\eta)/\hat{b}_{\rm k}(\eta)$ is obtained by using  equation~\eqref{eqn:app_Heisenberg_eq} on the field $\hat{\chi}(\eta,\,x)$. Note that, within this convention, not only the creation-annihilation operators are dynamical, but there is also a parametric time-dependence on  the mode functions. 

From an operational point of view, however, it is  easier to proceed in two separate steps: \textit{(i)} We decompose the field assuming time-dependent mode functions, and time-independent creation-annihilation operators, 
\begin{align}
\label{eq:static_creation_ann}
\hat{\chi}(\eta,\,{\rm x})=\int\frac{d{\rm k}}{2\pi}\left(\hat{a}_{\rm k} u_{\rm k}(\eta)+\hat{b}_{\rm -k}^\dagger v_{\rm -k}(\eta)\right)\ee^{\ii{\rm k} {\rm x}}.
\end{align}
The equations of motion that follow from Eq.~\eqref{eqn:app_Heisenberg_eq} are
\begin{align}
&\dot{\hat{a}}_{\rm k}=\dot{\hat{b}}_{\rm k}=0,\\
&\ii \dot{u}_{\rm k}(\eta)=H_{\rm k}(\eta)u_{\rm k}(\eta),\label{eqn:app_schrodinger_u}\\
&\ii\dot{v}_{\rm -k}(\eta)=H_{\rm k}(\eta)v_{\rm -k}(\eta)\label{eqn:app_schrodinger_v},
\end{align}
where the dot denotes differentiation with respect to the conformal time, and $H_{\rm k}(\eta)$ is the single-particle Hamiltonian corresponding to Eq.~\eqref{hamdS}  for our problem. This expansion does not assume any notion of particle, and this must be imposed as a second step, such that the picture in \eqref{eqn:app_diagon_method} is recovered. \textit{(ii)} At the end of the expansion phase, we obtain particle production by decomposing the mode solutions $u_{\rm k}(\eta)$ and $v_{\rm k}(\eta)$ in the basis of instantaneous eigenvectors, 
 \begin{align}
 &u_{\rm k}(\eta_{\rm f})=\alpha_{\rm k}(\eta_{\rm f}){\rm v}^+_{\rm k}(\eta_{\rm f})+\beta_{\rm k}^{*}(\eta_{\rm f}){\rm v}_{\rm k}^{-}(\eta_{\rm f}), \label{eqn:app_modes_decomposition_u}\\
 &v_{-\rm k}(\eta_{\rm f})=-\beta_{\rm k}(\eta_{\rm f}){\rm v_k^+}(\eta_{\rm f})+\alpha_{\rm k}^{*}(\eta_{\rm f}){\rm v}_{\rm k}^{-}(\eta_{\rm f}).
 \label{eqn:app_modes_decomposition_v}
 \end{align}
 recovering thus, at each time, the decomposition in \eqref{eqn:app_diagon_method}. To preserve orthonormality, the coefficients in this decompositions must obey the constraint $|\alpha_{\rm k}(\eta)|^2+|\beta_{\rm k}(\eta)|^2=1$.

With these two steps, one can easily obtain the production of particles. Firstly, one sets as the initial conditions for the mode functions as follows
\begin{align}
&u_{\rm k}(\eta_0)={\rm v}_{\rm k}^+(\eta_0),\\
&v_{\rm -k}(\eta_0)={\rm v}_{\rm k}^-(\eta_0).
\end{align}
Therefore, at $\eta=\eta_0$, we have
\begin{align}
\hat{\chi}(\eta_0,\,{\rm x})=\int\frac{d{\rm k}}{2\pi}\left(\hat{a}_{\rm k}^0(\eta_0){\rm v}_{\rm k}^+(\eta_0)+\hat{b}_{\rm -k}^{0\dagger}(\eta_0){\rm v}_{\rm k}^-(\eta_0)\right)\ee^{{\rm ik}{\rm x}},
\end{align}
where we have used the superscript $0$ in the operators to note that they correspond to the correct notion of particle only at $\eta=\eta_0$, corresponding to the asymptotically-flat remote past. 
Then, one solves the equations of motion for the mode functions \eqref{eqn:app_schrodinger_u}-\eqref{eqn:app_schrodinger_v}, so that at $\eta=\eta_{\rm f}$, one has 
\begin{align}
\hat{\chi}(\eta_{\rm f},\, {\rm x})=\int\frac{d{\rm k}}{2\pi}\left(\hat{a}_{\rm k}^0(\eta_0)u_{\rm k}(\eta_{\rm f})+\hat{b}_{\rm -k}^{0\dagger}(\eta_0)v_{\rm -k}(\eta_{\rm f})\right)\ee^{{\rm ik}{\rm x}}.
\label{eqn:app_evolved_field}
\end{align}

In \eqref{eqn:app_evolved_field}, the non-adiabatic dynamics induced by the expansion of the universe implies that these mode solutions depart from the instantaneous eigenvectors at latter times $\eta>\eta_0$. Introducing the decomposition  \eqref{eqn:app_modes_decomposition_u}-\eqref{eqn:app_modes_decomposition_v} at $\eta=\eta_{\rm f}$, we arrive at
\begin{align}
\nonumber\hat{\chi}(\eta_{\rm f},\,{\rm x})
=\int\frac{d{\rm k}}{2\pi}\Big[&\underbrace{\{\alpha_{\rm k}(\eta_{\rm f})\hat{a}_{\rm k}^0-\beta_{\rm k}(\eta_{\rm f})\hat{b}_{\rm -k}^{0\dagger}\}}_{\hat{a}_{\rm k}^{\rm f}}{\rm v}_{\rm k}^{+}(\eta_{\rm f})\nonumber \\
+&\underbrace{\{\beta_{\rm k}^{*}(\eta_{\rm f})\hat{a}_{\rm k}^0+\alpha_{\rm k}^{*}(\eta_{\rm f})\hat{b}_{\rm k}^{\rm f\, \dagger}\}}_{\hat{b}_{\rm -k}^{\rm f\, \dagger}}{\rm v}_{\rm k}^{-}(\eta_{\rm f})\Big]\ee^{{\rm i k}{\rm x}}.
\end{align}
Thus, the manifestation of \eqref{eqn:app_modes_decomposition_u}-\eqref{eqn:app_modes_decomposition_v} at the level of operators leads to the so-called Bogoliubov transformations
\begin{align}
&\hat{a}_{\rm k}^{\rm f}=\alpha_{\rm k}(\eta_{\rm f})\hat{a}_{\rm k}^0-\beta_{\rm k}\hat{b}_{\rm -k}^{0\,\dagger},\\
&\hat{b}_{\rm -k}^{\rm f\,\dagger}=\beta_{\rm k}^*(\eta_{\rm f})\hat{a}_{\rm k}^0+\alpha_{\rm k}^*\hat{b}_{\rm -k}^{0\,\dagger}.
\end{align}
These operators define the vacuum state $|0_{\rm f}\rangle$ at $\eta=\eta_{\rm f}$, which corresponds to the asymptotically-flat vacuum of the distant future.
The expectation value of the number operator of particles/antiparticles at $\eta=\eta_{\rm f}$ for a particular mode is given by
\begin{align}
\langle0_0|\hat{a}_{\rm k}^{\rm f\,\dagger}\hat{a}_{\rm k}^{\rm f}|0_0\rangle=\langle0_0|\hat{b}_{\rm -k}^{\rm f\,\dagger}\hat{b}_{\rm -k}^{\rm f}|0_0\rangle=|\beta_{\rm k}(\eta_{\rm f})|^2.\label{eqn:app_number_op}
\end{align}

From this discussion, we conclude that the production of particles is given by the $\beta$ coefficient of the Bogoliubov tansformation
\begin{align}
|\beta_{\rm k}(\eta_{\rm f})|^2=\left|\left({\rm v}_{\rm k}^{-}(\eta_{\rm f})\right)^\dagger\cdot u_{\rm k}(\eta_{\rm f})\right|^2.
\end{align}
This way of proceeding ensures that the Bogoliubov coefficients, which describe the gravitational production of particles, incorporate the correct notion of particles/antiparticles both at $\eta=\eta_{\rm 0}$ and at $\eta=\eta_{\rm f}$. 

\vspace{0.4cm}
\section{\bf Excitation probability in the experimental analogue  \label{app_d}}
  
  In this Appendix, we present the details of the scheme for the cold-atom measurement of the production of particles, given by $|\beta_{\rm k}(\eta_{\rm f})|^2$, in the analogue-gravity experiment discussed in Subsection \ref{subsec:measurements}. 
  
  The problem of particle production in curved spacetimes is generally treated within the Heisenberg picture, where the field operator dynamically evolves according to \eqref{eqn:app_Heisenberg_eq}. However, from the point of view of the analogue experiment, and the measurements that must be performed, it is more natural to work in the picture where the expansion of the field is expressed in terms of the instantaneous eigenvectors~\eqref{eqn:app_diagon_method} but with time-independent creation/annihilation operators.
 
 In this picture, the dynamics induced by the expanding background which generates particle production are encoded in the state of the system. Thus, the state, which is initially prepared to be the vacuum state, will experience an evolution
  \begin{align}
  |0\rangle\xrightarrow{\eta_0\rightarrow\eta_{\rm f}}|\psi_{\rm S}\rangle.
  \end{align}

  In order to obtain an explicit expression for $|\psi_{\rm S}\rangle$, first we note that in the Heisenberg picture, the Bogoliubov transformations \eqref{eqn:evolutiona}-\eqref{eqn:evolutionb} are induced by   a so-called two-mode fermionic squeezing operator~\cite{PhysRevLett.65.3341}, which is the unitary operator
  \begin{align}
  \hat{U}_{\rm k}(\zeta_{\rm k})=\exp\left(\zeta_{\rm k}\hat{a}^\dagger_{\rm k}(\eta_0)\hat{b}^\dagger_{-\rm k}(\eta_0)-\zeta_{\rm k}^{*}\hat{b}_{\rm -k}(\eta_0)\hat{a}_{\rm k}(\eta_0)\right),
  \label{eqn:app_d_squeezing_operator}
  \end{align}
  where $\zeta_{\rm k}=r_{\rm k}\ee^{{\rm i}\theta_{\rm k}}$ is the squeezing parameter. This operator satisfies 
  \begin{align}
  &\hat{U}_{\rm k}(\zeta_{\rm k})\hat{a}_{\rm k}(\eta_0)\hat{U}_{\rm k}^\dagger(\zeta_{\rm k})=\hat{a}_{\rm k}(\eta_{\rm f}),\label{eqn:app_d_squeezing_a}\\
  &\hat{U}_{\rm k}(\zeta_{\rm k})\hat{b}_{\rm -k}(\eta_0)\hat{U}_{\rm k}^\dagger(\zeta_{\rm k})=\hat{b}_{\rm -k}(\eta_{\rm f}),\label{eqn:app_d_squeezing_b}
  \end{align}
  as long as one identifies 
  \begin{align}
  &\alpha_{\rm k}(\eta_{\rm f})=\cos(r_{\rm k}),\\
  &\beta_{\rm k}(\eta_{\rm f})=\ee^{{\rm i}\theta_{\rm k}}\sin(r_{\rm k}).
  \end{align}
  Then, we note that the equivalence between the Schr\"odigner and Heisenberg pictures implies that
  \begin{align}
  \langle0|\hat{a}_{\rm k}^\dagger(\eta_{\rm f})\hat{a}_{\rm k}(\eta_{\rm f})|0\rangle=\langle\psi_{\rm S}|\hat{a}_{\rm k}^\dagger(\eta_0)\hat{a}_{\rm k}(\eta_0)|\psi_{\rm S}\rangle
  \label{eqn:app_d_equivalence_pictures}
  \end{align}
  
  Introducing in the left hand side of \eqref{eqn:app_d_equivalence_pictures} the transformation \eqref{eqn:app_d_squeezing_b}, we arrive at
  \begin{align}
  \nonumber
   \langle0_0|U_{\rm k}(\zeta_{\rm k})\hat{a}_{\rm k}^\dagger(\eta_0)\hat{a}_{\rm k}(\eta_0)U^\dagger_{\rm k}(\zeta_{\rm k})|0_0\rangle=&\\=\langle\psi_{\rm S}|\hat{a}_{\rm k}^\dagger(\eta_0)\hat{a}_{\rm k}(\eta_0)|\psi_{\rm S}\rangle.&
  \end{align}
  
  From this expression, one readily concludes that
  \begin{align}
  |\psi_{\rm S}\rangle=U^\dagger_{\rm k}(\zeta_{\rm k})|0\rangle.
  \label{eqn:app_d_expr_psi_symb}
  \end{align}
  
  The explicit expression of $|\psi_{\rm S}\rangle$ in terms of the Bogoliubov parameters can be obtained by noting that the fermionic operators in \eqref{eqn:app_d_squeezing_operator} realize an $\mathfrak{su}(2)$ algebra, and so we can use the disentangled form of the two-mode squeezing operator \cite{Ban:93}
  
  \begin{align}
  \nonumber
  U&_{\rm k}(\zeta_{\rm k})=\exp\left[\ee^{{\rm i}\theta_{\rm k}}\tan(r_{\rm k})\hat{a}_{\rm k}^\dagger(\eta_0)\hat{b}_{\rm -k}^\dagger(\eta_0)\right]\\\cdot&\nonumber\exp\left[-\ln(\cos(r_{\rm k}))\left(\hat{n}^{a}_{\rm k}(\eta_0)+\hat{n}^{b}_{\rm -k}(\eta_0)-1\right)\right]\\\cdot&\exp\left[\ee^{-{\rm i}\theta_{\rm k}}\tan(r_{\rm k})\hat{a}_{\rm k}(\eta_0)\hat{b}_{\rm -k}(\eta_0)\right],
  \end{align}
  where $\hat{n}^{a}_{\rm k}(\eta_0)=\hat{a}_{\rm k}^\dagger(\eta_0)\hat{a}_{\rm k}(\eta_0)$ and $\hat{n}^{b}_{\rm -k}(\eta_0)=\hat{b}_{\rm -k}^\dagger(\eta_0)\hat{b}_{\rm -k}(\eta_0)$.  Inserting this expresion in \eqref{eqn:app_d_expr_psi_symb} and recalling that the Schr\"odinger and Heisenberg creation/annihilation operators, as well as the vacuum states, coincide at $\eta=\eta_0$, prior to any dynamical evolution, one arrives at
  \begin{align}
  |\psi_{\rm S}\rangle=\bigotimes_{\rm k}\left(\alpha_{\rm k}(\eta_{\rm f})-\beta_{\rm k}(\eta_{\rm f})\hat{a}_{\rm k}^\dagger\hat{b}_{\rm -k}^\dagger\right)|0\rangle, 
  \label{eqn:app_d_psi_explicit}
  \end{align}
  where the interpretation of particle creation is manifest: particles and antiparticles are created in pairs with oposite momentum as a consequence of the dynamics induced by the expansion of the universe. Note that if the adiabatic theorem holds, $\beta_{\rm k}(\eta_{\rm f})$ would be negligible and thus the evolution would keep the state of the system in the instantaneous groundstate, $|\psi_{\rm S}\rangle\sim|0\rangle$.
  
  This picture is particularly useful for the analysis of the analogue experiment with ultra-cold atoms. The experimental set-up described in Section \ref{sec:implementation} comprises a positive- and a negative-energy band which are symmetric around zero energy. The groundstate of this system is obtained by filling the lower band with atoms, and leaving the upper band empty. This situation, upon a particle-hole transformation, represents the vacuum state $|0\rangle$ of the Dirac QFT. On the other hand, a situation where all the atoms are excited within the upper band represents a state with maximal content of pairs of particle/antiparticle, $\bigotimes_{\rm k}|1_{\rm k},\,\bar{1}_{\rm -k}\rangle$, where $|1_{\rm k},\,\bar{1}_{\rm -k}\rangle\equiv\hat{a}_{\rm k}^\dagger\hat{b}_{\rm -k}^\dagger|0\rangle$, and $|0\rangle\equiv\bigotimes_{\rm k}|0_{\rm k}\rangle$, such that $\hat{a}_{\rm k}|0_{\rm k}\rangle=\hat{b}_{\rm -k}|0_{\rm k}\rangle=0$.
  
  Thus, to simulate the phenomenon of particle production, the system must be initially prepared at half-filling in the groundstate, with all atoms within the lower band, so the initial state is $|0\rangle$. Then, the dynamics induced by the expansion of the universe are simulated by means of the optical potential, as described in Section \ref{sec:implementation} and, consequently, at the end of the expansion atoms will no longer be in a well-defined Bloch state, but rather delocalized in both bands according to \eqref{eqn:app_d_psi_explicit}. Conversely, quasi-momentum is conserved. Since we can access to the energy band of each atom by means of the band-mapping technique, the coefficients $|\alpha_{\rm k}(\eta_{\rm f})|^2$ and $|\beta_{\rm k}(\eta_{\rm f})|^2$ will be given by the probability of finding an atom with quasi-momentum $\rm k$ in a certain Bloch state. In particular, for a certain quasi-momentum, the probability of finding it in the lower ($\downarrow$) or in the upper ($\uparrow$) band will be given by
  \begin{align}
 &{\rm p}_{{\rm k},\downarrow}=|\langle0_{\rm k}|\psi_{\rm S}\rangle|^2=|\alpha_{\rm k}(\eta_{\rm f})|^2,\\
 & {\rm p}_{{\rm k},\uparrow}=|\langle1_{\rm k},\,\bar{1}_{\rm -k}|\psi_{\rm S}\rangle|^2=|\beta_{\rm k}(\eta_{\rm f})|^2.
  \end{align}
  
  Thus, the relation between the Bogoliubov parameters and the probability of excitation of the atoms is made explicit within the Schr\"odinger picture.   
  
\bibliographystyle{apsrev4-1}
\bibliography{apssamp.bib}

\end{document}